\documentclass[twocolumn]{aastex631}

\usepackage{amsmath}
\usepackage{multirow}
\usepackage{enumitem}
\usepackage{xspace}  

\newcommand{\Mone}{$\rm{M_{1kpc}/M_*}$\xspace}

\newcommand{\Se}{$\Sigma_e$\xspace}
\newcommand{\Sone}{$\Sigma_{\rm{1kpc}}$\xspace}

\newcommand{\re}{$R_{\rm{e}}$\xspace}
\newcommand{\n}{$n$\xspace}

\newcommand{\zf}{$z_{\rm{form}}$\xspace}

\newcommand{\zobs}{$z_{\rm{obs}}$\xspace}
\newcommand{\tage}{$t_{\rm{age}}$\xspace}

\newcommand{\tausf}{$\tau_{\rm{SF}}$\xspace}
\newcommand{\tauq}{$\tau_{\rm{Q}}$\xspace}
\newcommand{\fmulsf}{$\mathcal{F}_{\rm{multi-SF}}$\xspace}

\def\B{\ensuremath{B_{435}}\xspace}
\def\V{\ensuremath{V_{606}}\xspace}
\def\I{\ensuremath{I_{814}}\xspace}
\def\Y{\ensuremath{Y_{105}}\xspace}
\def\J{\ensuremath{J_{125}}\xspace}

\def\H{\ensuremath{H_{160}}\xspace}

\newcommand{\jwst}{\textit{JWST}\xspace}

\newcommand{\hst}{\textit{HST}\xspace}

\newcommand{\galfit}{{\sc Galfit}\xspace}
\newcommand{\sextractor}{{\sc SExtractor}\xspace}
\newcommand{\prospector}{{\sc Prospector}\xspace}

\newcommand{\sersic}{S\'{e}rsic\xspace}
\newcommand{\dms}{$\rm{\Delta MS}$\xspace}

\begin{document}

\defcitealias{Ji2022}{Paper~I}

\title{Reconstructing the Assembly of Massive Galaxies. II. Galaxies Develop Massive and Dense Stellar Cores as They Evolve and Head Toward Quiescence at Cosmic Noon} 

\correspondingauthor{Zhiyuan Ji}
\email{zhiyuanji@astro.umass.edu}

\author[0000-0001-7673-2257]{Zhiyuan Ji}
\affiliation{University of Massachusetts Amherst, 710 North Pleasant Street, Amherst, MA 01003-9305, USA}
\author[0000-0002-7831-8751]{Mauro Giavalisco}
\affiliation{University of Massachusetts Amherst, 710 North Pleasant Street, Amherst, MA 01003-9305, USA}

\begin{abstract}

We use the SED-fitting code \prospector to reconstruct the nonparametric star formation history (SFH) of massive ($\log M_*>10.3$) star-forming galaxies (SFGs) and quiescent galaxies (QGs) at redshift \zobs$\sim2$ to investigate the joint evolution of star-formation activity and structural properties. Compared to extended SFGs, compact SFGs are more likely to have experienced multiple star-formation episodes, with the fractional mass formed during the older ($\ge1$ Gyr) episode being larger, suggesting that high-redshift SFGs assembled their central regions earlier and then kept growing in central mass as they become more compact. The SFH of compact QGs does not significantly differ from the average for this category, and shows an early burst followed by a gradual decline of the star formation rate. The SFH of extended QGs, however, is similar to that of post-starburst galaxies and their morphology is also frequently disturbed. Knowledge of the SFH also enables us to empirically reconstruct the structural evolution of individual galaxies. While the progenitor effect is clearly observed and accounted for in our analysis, it alone is insufficient to explain the observed structural evolution. We show that, as they evolve from star-forming phase to quiescence, galaxies grow massive dense stellar cores. Quenching begins at the center and then propagates outward to the rest of the structure. We discuss possible physical scenarios for the observed evolution and find that our empirical constraints are in good quantitative agreement with the model predictions from dissipative accretion of gas to the center followed by massive starbursts before final quiescence (wet compaction).

\end{abstract}

\keywords{Galaxy formation(595); Galaxy evolution(594); Galaxy structure(622); High-redshift galaxies(734); Galaxy quenching (2040)}

\section{Introduction} \label{sec:intro}
Observations over a wide range of cosmic time, from the present day up to the so-called epoch of ``cosmic noon" at $1<z<4$ \citep{Madau2014}, have shown that galaxies are characterized by a bimodality of colors at UV through NIR wavelengths (e.g. see \citealt{Baldry2004, Bell2004,Brammer2009,Williams2009}, among many other). Blue galaxies on average are actively forming stars, and have higher star formation rates (SFRs) than red galaxies which generally have substantially lower levels of star formation activity, leading to a natural classification into the star-forming galaxies (SFGs) and the quiescent galaxies (QGs). Understanding the transition from SFGs to QGs, a process referred to as galaxy quenching, remains a key missing piece toward a complete picture of galaxy evolution.

Broadly, two general categories of quenching mechanisms -- mass quenching and environmental quenching -- have been identified \citep{Peng2010}. Environmental quenching is associated with the external environment of a galaxy, and it might be able to alter the structural evolution of galaxies \citep[e.g.,][]{Valentinuzzi2010,Papovich2012,Cappellari2013,HuertasCompany2013,Strazzullo2013,Newman2014,Matharu2019}. Observations generally show that the environmental effect on galaxy evolution primarily happens in relatively low-mass (stellar mass $\log M_*<10$), low-redshift ($z<1.5$) galaxies in dense environments such as groups/clusters of galaxies \citep{Peng2010,Guo2017,Kawinwanichakij2017, Ji2018}, which, however, are not the subject of the investigation presented here. In this work, we will specifically focus on massive galaxies at cosmic noon epoch, defined here as having stellar mass $\log M_*\ge 10.3$ and in the redshift range of $1<z<4$ (see section \ref{sec:sample}). Observations show that at this epoch massive galaxies begin to quench in large numbers \citep[e.g.,][]{Muzzin2013} and suggest that the quenching should be mainly due to physical processes internal to the galaxies themselves, i.e., mass quenching (see a recent review by \citealt{Man2018} and references therein).

Strong correlations are observed between the star-formation activity and the structural properties of galaxies, such as size, light profile and central mass density. It is now well-established that these correlations persist at least out to $z\sim3$. These include the emergence of the Hubble Sequence at $z>2$ \citep[e.g.][]{Franx2008,Kriek2009,Wuyts2011}, and the dependence of the mass-size relation and \sersic index $n$ on specific star formation rate (sSFR, e.g. \citealt{Williams2010,Wuyts2011,Newman2012,Barro2013,Patel2013,vanderWel2014,Shibuya2015}). On average, QGs have much smaller sizes, steeper light profiles (i.e. larger $n$) and larger central mass densities than SFGs of similar mass and redshift. The mass-size relation of QGs also is substantially steeper than that of SFGs. Phenomenologically, these correlations could be interpreted as evidence of a causal link between the structural transformations and the quenching of galaxies, but because the relative timing sequence of these two events remains  empirically unconstrained, whether the causality is real remains unknown. In fact, some evidence suggests that at $z\sim2$ galaxies quench first and then secularly transform their structures later. This is based on the observations that (1) a substantial amount of massive QGs at redshift $z= 2\sim3$ appear to have a disk-like morphology \citep[e.g.][]{McGrath2008,Bundy2010,vanderWel2011,Bruce2012,Chang2013}, and (2) the spatially resolved spectra of a handful of strongly lensed QGs show disk-like kinematics, namely systems that are predominantly supported by rotation although with a larger velocity dispersion than modern disks \citep{Newman2015,Toft2017,Newman2018}.

Imaging observations, especially those with high-angular resolution taken with the \textit{Hubble Space Telescope} ({\it HST}), have unveiled what appears to be a threshold of central stellar-mass surface density ($\Sigma$) above which galaxies are found to be predominately quiescent. This observational finding seemingly suggests that the presence of dense central core in massive galaxies might be a necessary condition for, or a consequence of quenching  \citep[e.g.][]{Kauffmann2003,Franx2008,Cheung2012,Fang2013,Lang2014,Whitaker2017,Barro2017,Lee2018}, and this consideration motivated a number of theoretical studies to explore possible physical processes that would result in such a phenomenology. For example, one such mechanism is actually a class of processes, generically referred to as ``wet compaction" whose main feature is highly dissipative gas accretion toward the central regions of a galaxy which, in turn, promotes enhanced activity of star formation at a higher rate relative to the average SFR \citep{Dekel2009}. Specifically, the larger gas mass fraction of galaxies at high redshifts compared to local ones (\citealt{Tacconi2020} and references therein) makes the early galactic disks gravitationally unstable on the scale of  $\lesssim1$ kpc \citep{Toomre1964}. The disks can thus fragment into star-forming clumps \citep[e.g.][]{Genzel2008,Guo2015} which then migrate toward the center of galaxies, trigger central starbursts and form dense central cores. In the absence of further gas inflow, these lead to the central gas depletion and finally the quenching of central star formation \citep{Ceverino2010,Dekel2014,Wellons2015,Zolotov2015,Tacchella2016}. Simulations suggests that any mechanism  that can cause a drastic dissipative loss of angular momentum would lead to hydrodynamical instabilities and the compaction, such as wet major mergers \citep{Zolotov2015,Inoue2016} or collisions of counter-rotating streams that feed the galaxy \citep{Danovich2015}. 

An apparent correlation between quiescence and central density, including the the apparent $\Sigma$ threshold, however, does not necessarily imply any physical causality between the two properties, because it can also be the result of the ``progenitor effect" \citep{Lilly2016,Ji2022}. In an expanding universe, bound systems should keep memory of the cosmic density at the time when the halo collapses, implying that halos formed earlier should have higher densities than those formed later \citep{Mo2010}. Therefore, at a fixed epoch QGs are expected to be smaller and denser than SFGs, as the star-forming progenitors of what are being observed as QGs have necessarily formed earlier. The implication is that the apparent correlations between structural and star-formation properties reflect the collective effects of the interplay among different physical mechanisms, e.g., the progenitor effect vs. the physical compaction. This makes the empirical investigation of the relative contribution from each of them critical for  understanding the structural transformations of galaxies and their relationships with quenching. 

A substantial body of existing observational studies has investigated the redshift evolution of the mass-size and mass-$\Sigma$ relations of SFGs and QGs. Both relations contain, \textit{in an integral form}, information of the structural evolution. This means that a certain galaxy population, observed at redshift \zobs, includes galaxies that have formed at all different epochs of $z>$ \zobs. The progenitor effect, therefore, is inherently mixed with whatever other physical mechanism (e.g., wet compaction) is at play. This makes the interpretation of the observations model-dependent (see e.g., section 3.3 of \citealt{Barro2017}). 

To account for the progenitor effect on the apparent structural evolution, the key is to identify galaxies that formed at different epochs. In this way we can then avoid aggregating all galaxies together. Some early attempts achieved this by selecting galaxies with a constant number density \citep[e.g., ][]{vanDokkum2010,Patel2013}. This is based on the idea of selecting dark matter halos of the same mass, which at high redshifts would be roughly coeval. In our view, a further step forward is to utilize the star-formation history (SFH) of galaxies, i.e. their full stellar-mass assembly histories, and to select galaxies following a common evolutionary path, namely formed at a similar epoch, have a similar stellar mass and are observed at a similar evolutionary stage. In this way any form of the progenitor effect is naturally taken into account. We can then statistically reconstruct the structural evolution of individual galaxies. And by looking at how the average properties of galaxies that formed at the same time and are observed at the same evolutionary stage change as a function of time, we can empirically constrain the physics behind the structural transformations of galaxies.    

With the modelling of Spectral Energy Distributions (SEDs) growing in accuracy and sophistication, and with the availability of high-quality data that cover truly  panchromatic swathes of wavelengths, significant progress has been made to reconstruct the SFH of galaxies at high redshifts, both in parametric \citep[e.g.][]{Maraston2010,Papovich2011,Ciesla2016,Lee2018,Carnall2018,Carnall2019} and nonparametric forms \citep[e.g.][]{Ocvirk2006,Tojeiro2007,Pacifici2012,Leja2017,Leja2019,Belli2019,Iyer2019}. The flexibility of the nonparametric form, in particular, is critical for not only an unbiased inference of physical parameters such as stellar mass, SFR and stellar age \citep[e.g][]{Leja2019,Lower2020}, but also for the fidelity of the reconstruction of the SFH itself, which can have arbitrarily complex forms generally not captured by parametric models \citep{Leja2019, Johnson2021, Tacchella2022}.

In the first paper of this series (\citealt{Ji2022}, hereafter \citetalias{Ji2022}), we have utilized the fully Bayesian SED fitting code \prospector \citep{Leja2017,Johnson2021} to reconstruct the nonparametric SFH of a sample of massive QGs at $z\sim2$ to quantify the progenitor effect. We found that the progenitor effect is strong in QGs with $\log_{10}(M_*/M_\sun)=10.3\sim11$, while much milder in more massive QGs, implying that the post-quenching mass and size growths of the latter happen via mergers, which reduce the memory of the time of their formation in their structural properties. What remains to be explored is if, in addition to the progenitor effect, the central regions of galaxies grow denser and more massive relative to the outskirts as they evolve toward quenching. In this second paper, we expand our investigation to all massive galaxies at $z\sim2$, selected from the CANDELS/COSMOS and GOODS fields, regardless of their star-formation activity. We use the SFHs to empirically and quantitatively estimate the contributions from different physical mechanisms to the apparent correlations between star-formation and structural properties. Throughout this paper, we adopt the AB magnitude system and the $\Lambda$CDM cosmology with parameters $\Omega_m = 0.3$, $\Omega_\Lambda = 0.7$ and $\rm{h = H_0/(100 kms^{-1}Mpc^{-1}) = 0.7}$.

\section{The sample} \label{sec:sample}

The parent sample considered in this work is extracted from the portions of the COSMOS, GOODS-South and GOODS-North fields observed by the CANDELS program  \citep{Grogin2011,Koekemoer2011}. The original CANDELS photometric catalogs in each field, which already include \hst, {\it Spitzer} and ground-based broadband data, and have been augmented by additional narrow- and medium-band data from deep ground-based imaging. 

The photometric catalogs used in this work are the same as those in \citetalias{Ji2022}. In brief, for the CANDELS/COSMOS field, we use the photometric catalog from \citet{Nayyeri2017} which includes the intermediate- and narrow-band optical photometry from Subaru \citep{Taniguchi2015} and medium-band NIR photometry from Mayall NEWFIRM \citep{Whitaker2011}. For the GOODS-North field, we use the photometric catalog from \citet{Barro2019} which includes twenty-five medium-band photometry at the optical wavelengths acquired during the SHARDS survey \citep{PerezGonzalez2013}. Finally, for the GOODS-South field, we use the latest photometric catalog from the ASTRODEEP project (\citealt{Merlin2021}). This expands the original \citet{Guo2013} catalog in the GOODS-South with eighteen new medium-band photometric data obtained with the Subaru SuprimeCAM \citep{Cardamone2010}, and another five NIR photometric bands observed with the Magellan Baade FourStar during the ZFOURGE survey \citep{Straatman2016}. In all cases the photometry has been aperture-matched where low angular-resolution images were de-blended based on positional priors from high angular-resolution ones. Overall, densely sampled ($\approx$ 40 bands), deep photometry covering the spectral range from the rest-frame UV to NIR for galaxies at $z\sim1-4$, is available in all the three fields considered here. 

Starting from the CANDELS photometric catalog, we first impose a cut on the integrated isophotal signal-to-noise ratio, S/N $\ge10$, in the \hst/WFC3 \H band to ensure good-quality photometry and morphology. We then remove galaxies that have clear evidence of AGN activity using the \texttt{AGNFlag} in the aforementioned catalogs. Because the focus of this work is on massive galaxies at cosmic noon, we then select galaxies with redshift $1.2<$ \zobs $<4$ and stellar mass $\log_{10}(M_*/M_\odot)>10$ using the existing measurements of \citet{Nayyeri2017} for the COSMOS field and of \citet{Lee2018} for the GOODS fields. This initial cut on stellar mass also ensures high stellar-mass completeness in all fields considered in this work, namely $>80\%$, see \citealt{Ji2018} for the CANDELS/GOODS fields and \citealt{Nayyeri2017} for the CANDELS/COSMOS field. In principle, these cuts on \zobs and $M_*$ would not be needed, because we could use the results from our new SED modeling with \prospector (section \ref{sec:prosp}) from the upstart. In practice, however, running \prospector is computationally very intensive\footnote{In our case, typical running time for a single galaxy is $\approx10-20$ hours}, making it impractical to run it on the entire CANDELS catalogs for the primary selection. Thus, we resort to first make the cuts on \zobs and $M_*$ from previous measurements, and then run \prospector to derive the SFH of a substantially smaller sample. 

\begin{figure}
    \centering
    \includegraphics[width=0.47\textwidth]{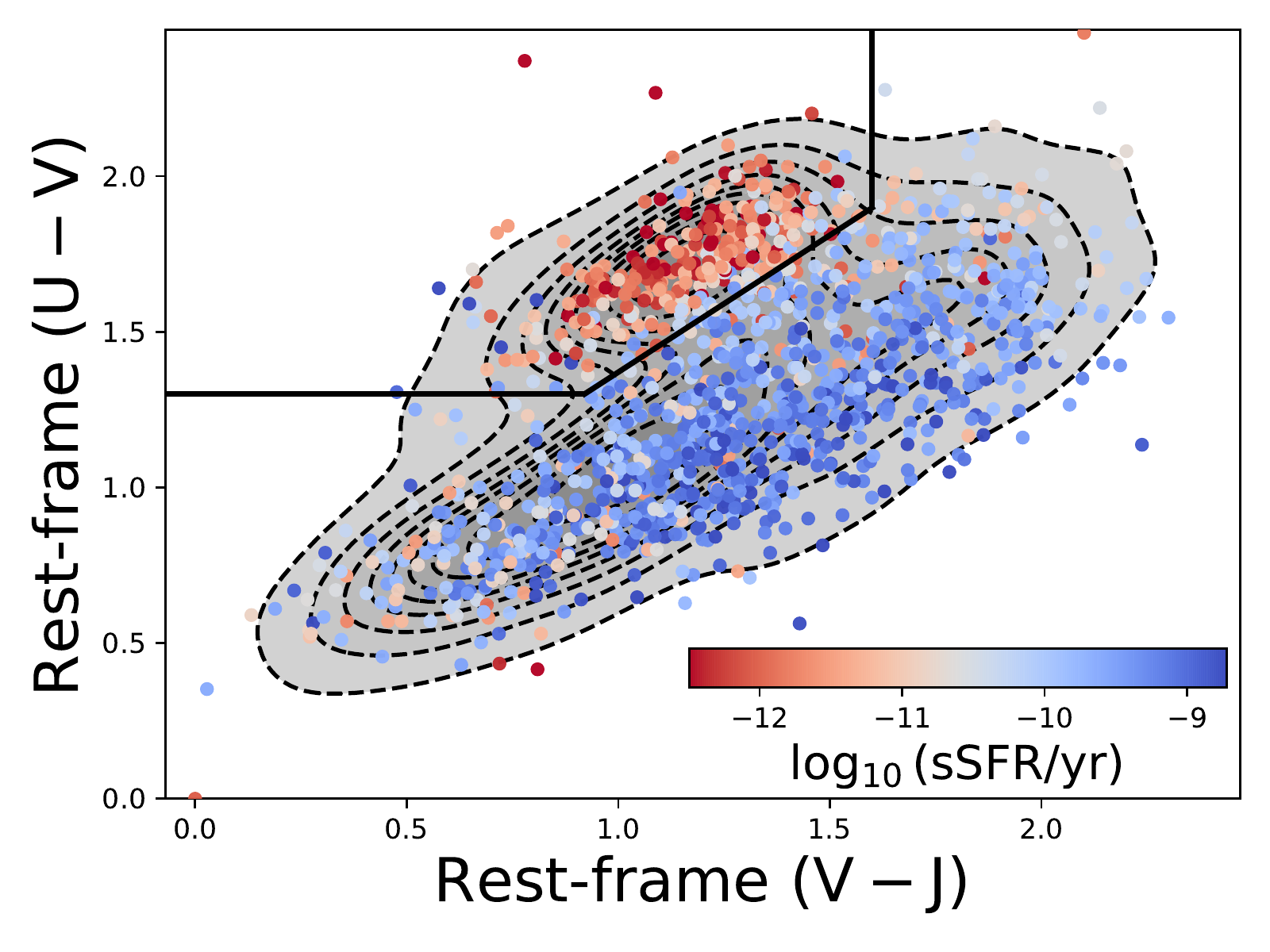}
    \caption{The rest-frame UVJ-color diagram. Background grey contours show the distribution of galaxies in the CANDELS/COSMOS field, with $\log_{10}(M_*/M_\sun)>10$, $1.2<$ \zobs $<4$ and S/N $>10$ in the \H band. Black solid lines show the SFG-QG separation boundary of \citet{Schreiber2015}. The final sample of this work (section \ref{sec:sample}) is shown as individual points that are color-coded according to sSFR derived from our new SED fitting with \prospector (section \ref{sec:prosp}). }
    \label{fig:uvj}
\end{figure}

Following \citetalias{Ji2022}, to ensure the quality of the adopted de-blending procedures of low-resolution images and hence the reliability of the photometry, we have visually inspected the galaxies of our sample in all bands in order to select galaxies with high-quality photometry (see Figure 2 of \citetalias{Ji2022}). We retain in the sample only galaxies with \texttt{GALFITFlag = 0} from \cite{vanderWel2012} to ensure that morphological measurements are reliable (section \ref{sec:morph}). Finally, using the more refined measures of the stellar mass  from our \prospector fitting (section \ref{sec:prosp}), we adjust the stellar-mass cut to $\log_{10}(M_*/M_\sun)>10.3$ such that the environmental effect should only play a relatively minor role in the evolution and quenching of galaxies \citep[e.g.,][]{Ji2018}. Our final sample contains 1317 galaxies in total, with 492 located in the COSMOS, 448 in the GOODS-South and 377 in the GOODS-North. In Figure \ref{fig:uvj}, we show the distribution of the final sample in the rest-frame UVJ diagram \citep{Williams2009}.

\section{Measurements}

\subsection{SED Fitting with \prospector}\label{sec:prosp}

We have derived the integrated properties of the stellar populations of the galaxies in our sample by modeling their SEDs with the \prospector package \citep{Leja2017, Johnson2021}. 

\prospector is built upon a fully Bayesian framework, and is ideally positioned to exploit the vast improvement in quality, quantity and wavelength coverage of modern (mostly) photometric data accumulated over the last two decades in legacy fields, such as the three fields that are subjects of this work, i.e. the COSMOS and the two GOODS fields. Packages such as \prospector now make it possible to fit the SED of high-redshift galaxies with advanced, complex stellar population synthesis models that provide robust constraints on the stellar age of galaxies. One key feature of \prospector is that it allows flexible parameterizations of galaxies' SFHs, including a piece-wise, nonparametric form which we have elected to adopt in this work. 

Adopting a nonparametric SFH means that the assembly history of galaxies can be arbitrarily complex and yet the code yields an unbiased inference of the properties of stellar populations. Despite the reconstruction of SFHs is still a relatively new development, several recent studies have tested its robustness with \prospector using synthetic observations generated from cosmological simulations \citep{Leja2019, Johnson2021,Tacchella2022, Ji2022}. These tests have demonstrated that, at least in the case of the SFHs of the synthetic galaxies encountered in the simulations, \prospector is capable of recovering the nonparametric SFHs with high fidelity, if high-quality, panchromatic data coverage of the SED is available.

In this work, we assume the same model priors and settings of \prospector as we did in \citetalias{Ji2022}, and we refer readers to that paper for a detailed description and the quality tests that we have already conducted. Here we only briefly summarize the key assumptions. For the basic setup of the fitting procedure, we adopt the Flexible Stellar Population Synthesis (FSPS) code \citep{Conroy2009,Conroy2010} where the stellar isochrone libraries MIST \citep{Choi2016,Dotter2016} and the stellar spectral libraries MILES \citep{Falcon-Barroso2011} are used. We assume the \citealt{Kroupa2001} initial mass function (IMF), the \citealt{Calzetti2000} dust attenuation law and fit the V-band dust optical depth with a uniform prior $\tau_V\in(0,2)$, and the \citealt{Byler2017} nebular emission model. During the SED modeling, we assume that all stars in a galaxy have the same metallicity $Z_*$, and fit it with a uniform prior in the logarithmic space $\log_{10}(Z_*/Z_\sun)\in(-2,0.19)$, where $Z_\sun=0.0142$ is the solar metallicity, and the upper limit of the prior is chosen because this is the highest value of metallicity that the MILES library has. We do not keep track of the time evolution of the metallicity. Finally, we also fit the redshift \zobs as a free parameter but we use a strong prior in the form of a normal distribution centered at the best available redshift value from the CANDELS catalogs, either photometric or spectroscopic redshift when available, and with a width equal to the corresponding redshift uncertainty, i.e. either the spectroscopic redshift error bar or the width of the posterior distribution function of photometric redshifts. 

Following recent studies \citep{Leja2017,Leja2019, Leja2021, EstradaCarpenter2020, Tacchella2022, Ji2022}, we model the nonparametric SFH with a piece-wise step function with nine bins of lookback time. Specifically, the first and second bins are fixed to be 0 - 30 and 30 - 100 Myr, the last bin is assumed to be 0.9$\rm{t_H}$ - $\rm{t_H}$ where $\rm{t_H}$ is the Hubble Time of \zobs, and the remaining six bins are evenly spaced in  logarithmic lookback time intervals between 100 Myr and 0.9$\rm{t_H}$. Also, as done in \citetalias{Ji2022}, our fiducial measures throughout the paper have been obtained assuming the nonparametric SFH with the Dirichlet prior that (1) results in a symmetric prior on stellar age and sSFR, and a constant SFH with $\rm{SFR(t) = M_* /t_H}$; (2) has been tested with nearby galaxies across all different types \citep{Leja2017,Leja2018}; and (3) has been validated with simulated observations of synthetic galaxies from cosmological simulations \citep[][]{Leja2019,Ji2022}. However, we stress that the choice of the prior for the  nonparametric SFH still remains somewhat arbitrary and the validity of each assumption still needs to be further investigated. For example, some other recent work \citep[e.g][]{EstradaCarpenter2020,Tacchella2022} adopted a so-called Continuity prior that is very similar to the Dirichlet prior, except it strongly disfavours sudden changes of SFRs in adjacent time bins, such as those resulting from powerful bursts of star formation. Which prior should be used for high-redshift galaxies is awaiting to be tested with future spectroscopic data that can provide independent measures of parameters such as stellar ages and timescales of star formation and of quenching. It is very likely that we will need a more complex prior than the aforementioned two to better capture different episodes of galaxies' assembly histories \citep{Suess2022}, and the choice of such priors might also depend on galaxy types. Here, without further constraints from other independent measurements, we decide to use the well-tested Dirichlet prior as our fiducial one. But we have also run another set of \prospector fittings for the entire sample with the Continuity prior, to check whether the choice of prior introduces significant differences in our results and conclusions. As Appendix \ref{app:cont_diri} shows, we find that the choice of the nonparametric SFH prior leaves the key conclusions of this work unchanged. In the following all the results presented, therefore, are the ones obtained using the Dirichlet prior.

Finally, once we have obtained the SFHs for the individual galaxy in our sample, we use a number of metrics, which we have introduced in \citetalias{Ji2022}, to characterize the overall shape of the SFHs, including different relevant timescales. We refer readers to \citetalias{Ji2022} for the  definitions and characteristics of the metrics. Here we only briefly outline the ones discussed in the main text of this work, i.e.
\begin{itemize}
    \item Mass-weighted age \tage
    \item Formation redshift \zf: the redshift corresponds to when the time interval between \zf and \zobs equals to \tage
    \item Asymmetry \tausf/\tauq (see equation 5 in \citetalias{Ji2022}): the ratio between the mass-weighted time widths of the period between \tage $\rm{<t<t_H}$ (\tausf) divided by that of the period between $0<t<$ \tage (\tauq)
\end{itemize}

\subsubsection{Comparing with Existing Measurements} \label{sec:prosp_compare}

\begin{figure*}
    \centering
    \includegraphics[width=1\textwidth]{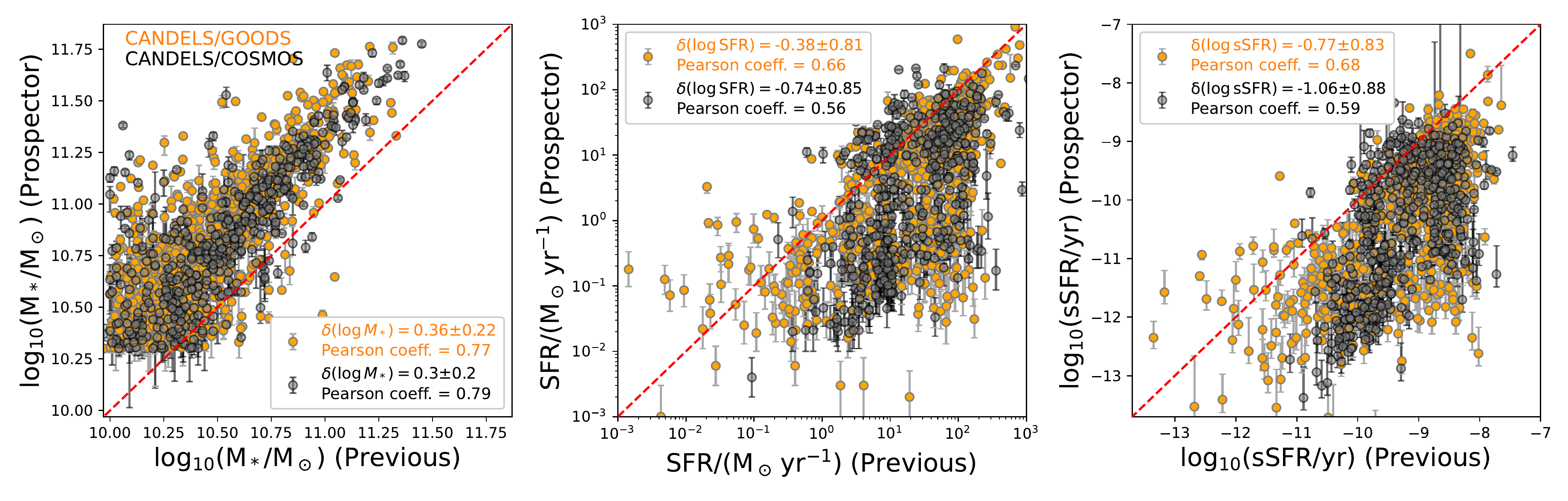}
    \caption{The comparisons between our \prospector fitting results (y-axis) and previous measurements (x-axis). Galaxies in the CANDELS/GOODS fields are shown in orange, while those in the CANDELS/COSMOS field are shown in black. In each panel, the red dashed line marks the one-to-one relation. In the legend we label the median and standard deviation of the differences between the two measurements, i.e. the \prospector one minus the previous one, and the coefficient of Pearson correlation test between the two measurements. Our measurements are in great qualitative agreement with previous ones, although systematic offsets are also clearly observed between the two measurements (see section \ref{sec:prosp_compare} for a detailed discussion).}
    \label{fig:compare_pre}
\end{figure*}

We first compare our measurements of $M_*$, SFR and sSFR with previous ones. Specifically, for the galaxies in the CANDELS/COSMOS field we compare our measures with those done by \citet{Nayyeri2017}, who assumed a parametric, exponentially declining SFH, i.e. $\rm{SFR(t)\propto e^{-t/\tau}}$. Their SFRs at \zobs were estimated by combining the UV and IR luminosities, i.e. $\rm{SFR_{UV}^{Obs} + SFR_{IR}}$,  which is the sum of the observed far-UV emission and the reprocessed thermal dust emission in the IR, after calibration onto a scale of SFR \citep{Kennicutt1998}. For the galaxies in the CANDELS/GOODS fields we compare our measures with those done by \citet{Lee2018}, who treated the functional form of the SFH as a free parameter, chosen from a set of five parametric analytical models. Their SFRs at $z_{obs}$ were also estimated using $\rm{SFR_{UV}^{Obs} + SFR_{IR}}$ when a galaxy is detected in and/or at wavelengths longer than the 24$\micron$ \textit{Spitzer}/MIPS images, or using the SED-derived SFRs otherwise. In Figure \ref{fig:compare_pre} we show the comparisons in the CANDELS/COSMOS and GOODS fields, respectively. We also run the Pearson correlation tests between our \prospector measurements and the previous ones for each parameter. In all cases we find a rather tight correlation between the two sets of measurements, with a Pearson correlation coefficient $\gtrsim0.6$. Yet, systematic deviations are also clearly seen. 

As Figure \ref{fig:compare_pre} shows, on average our \prospector fitting returns a $0.3-0.4$ dex larger $M_*$ than the previous measures. This systematically larger $M_*$ has been extensively discussed and now well understood  \citep{Leja2019,Leja2021}. Using nonparametric SFHs returns older stellar ages, and hence larger stellar masses, than using parametric ones, because they can easily accommodate a larger fraction of older stellar populations for a given SED \citep{Carnall2019,Leja2019}. Evidence that this systematically larger $M_*$ is actually more accurate than that found when using parametric forms for the SFH has also been found, including (1) the evolution of galaxy stellar-mass function inferred using nonparametric SFHs during SED fitting is more consistent with direct observations \citep{Leja2020}; (2) tests using synthetic observations generated from cosmological simulations show that $M_*$ derived from the SED fitting with nonparametric SFHs is unbiased, and in much better agreement with the intrinsic value than that derived with parametric SFHs \citep{Lower2020}. 

Regarding the SFR measures, our \prospector fitting procedure returns values that are $0.3-0.8$ dex smaller than the previous measures, in quantitative agreement with the findings of \citet{Leja2020} and \citet{Leja2021}. These authors have shown that this is primarily due to older, evolved stellar populations in massive galaxies. Specifically, the older stellar populations can contribute to the IR flux via dust heating, which consequentially leads to the overestimated $\rm{SFR_{IR}}$. This systematics is particularly important in low-sSFR galaxies \citep{Conroy2013}, and it becomes increasingly important at higher redshifts when the evolved stellar populations were more luminous just as they were younger. For QGs, their SFR and sSFR measures thus are arguably more accurate via panchromatic SED fitting than via $\rm{SFR_{UV}^{Obs} + SFR_{IR}}$, considering that SED fitting takes into account the full shape of SEDs and self-consistently models stellar populations with different ages. In fact, this point has been demonstrated using mock observations of galaxies from semi-analytic models (e.g., see Figure 6 of \citealt{Lee2018}).  This is also supported by the fact that we find a larger Pearson correlation coefficient, i.e. a stronger correlation, when comparing our new SFR and sSFR measures with those from \citet{Lee2018} than with those from \citet{Nayyeri2017}. As Figure \ref{fig:compare_pre} shows, the correlation improvement is mainly caused by galaxies with low sSFRs, because \citet{Lee2018} used SED-derived SFRs for  galaxies without MIR/FIR detections and most of the low-sSFR galaxies are not detected in the MIR/FIR bands. 

\begin{figure}
    \centering
    \includegraphics[width=0.47\textwidth]{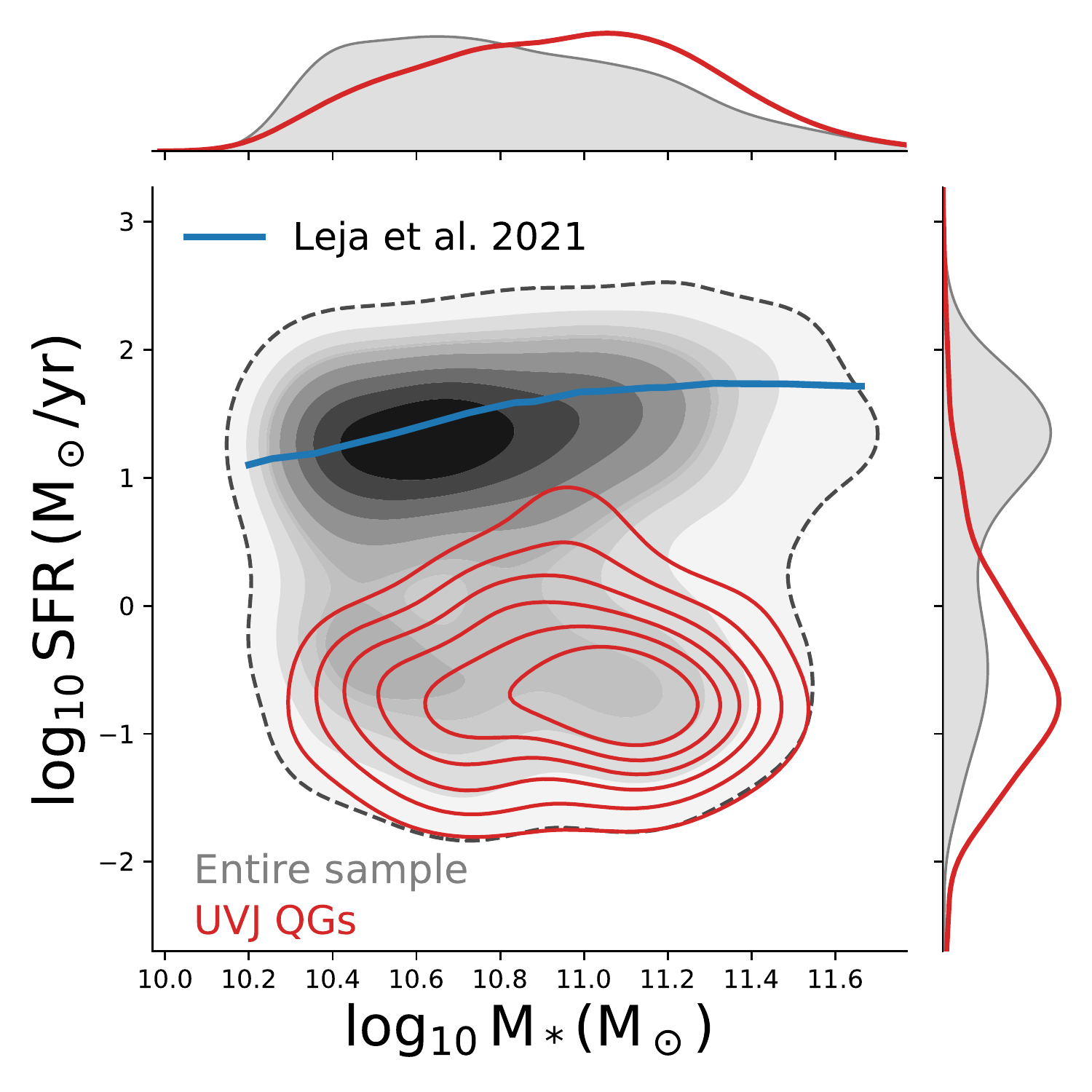}
    \caption{The plot of star-forming main sequence. Grey contours show the distribution of the entire sample in this study. Red contours show the distribution of a subsample that is classified as QGs using the UVJ technique. All distributions are estimated with a Gaussian kernel. The blue solid line marks the star-forming main sequence of \citet{Leja2021} who also used \prospector for their SED fittings.}
    \label{fig:sfms}
\end{figure}

We then proceed to compare the results from our fitting procedure with the star-forming main sequence measured by \citet{Leja2021}, who also used \prospector. Instead of separating galaxy populations according to their star-formation properties prior to measuring the star-forming main sequence, \citet{Leja2021} utilized the normalizing flow technique to model the density distribution of full galaxy populations, from which they measured the redshift evolution of the ridge (i.e. mode) and the mean of the density distribution in the $M_*$-SFR parameter space, i.e. the star-forming main sequence. In this way they demonstrated that systematic errors of the star-forming main sequence, which are introduced by the selection methods of SFGs, can be effectively mitigated. More importantly, \citet{Leja2021} showed that within this framework they were able to substantially mitigate, if not resolve, a long-standing systematic offset of the star-forming main sequence between the observations and the predictions from cosmological simulations (see their section 7.1). In Figure \ref{fig:sfms}, we compare the distribution of our sample with the density ridge\footnote{We have checked that using the mean density of \citet{Leja2021} as the main sequence does not change any of our conclusions of this work.} (their equation 9) of \citet{Leja2021}. Because the star-forming main sequence evolves with redshift, for a better comparison, instead of plotting \citet{Leja2021}'s star-forming main sequence at the median redshift of our entire sample, we first bin galaxies in our sample according to their $M_*$, and then within individual $M_*$ ranges we plot the star-forming main sequence at the median redshift value of that given $M_*$ bin (blue solid line in Figure \ref{fig:sfms}). Our measurement is in great agreement with that of \citet{Leja2021}. In Figure \ref{fig:sfms} we also plot the subsample of galaxies which are classified as QGs with the UVJ technique. They are well below the star-forming main sequence, again showing the consistency between the SED fitting and the UVJ selection that we already saw in Figure \ref{fig:uvj}.

\begin{figure}
    \centering
    \includegraphics[width=0.47\textwidth]{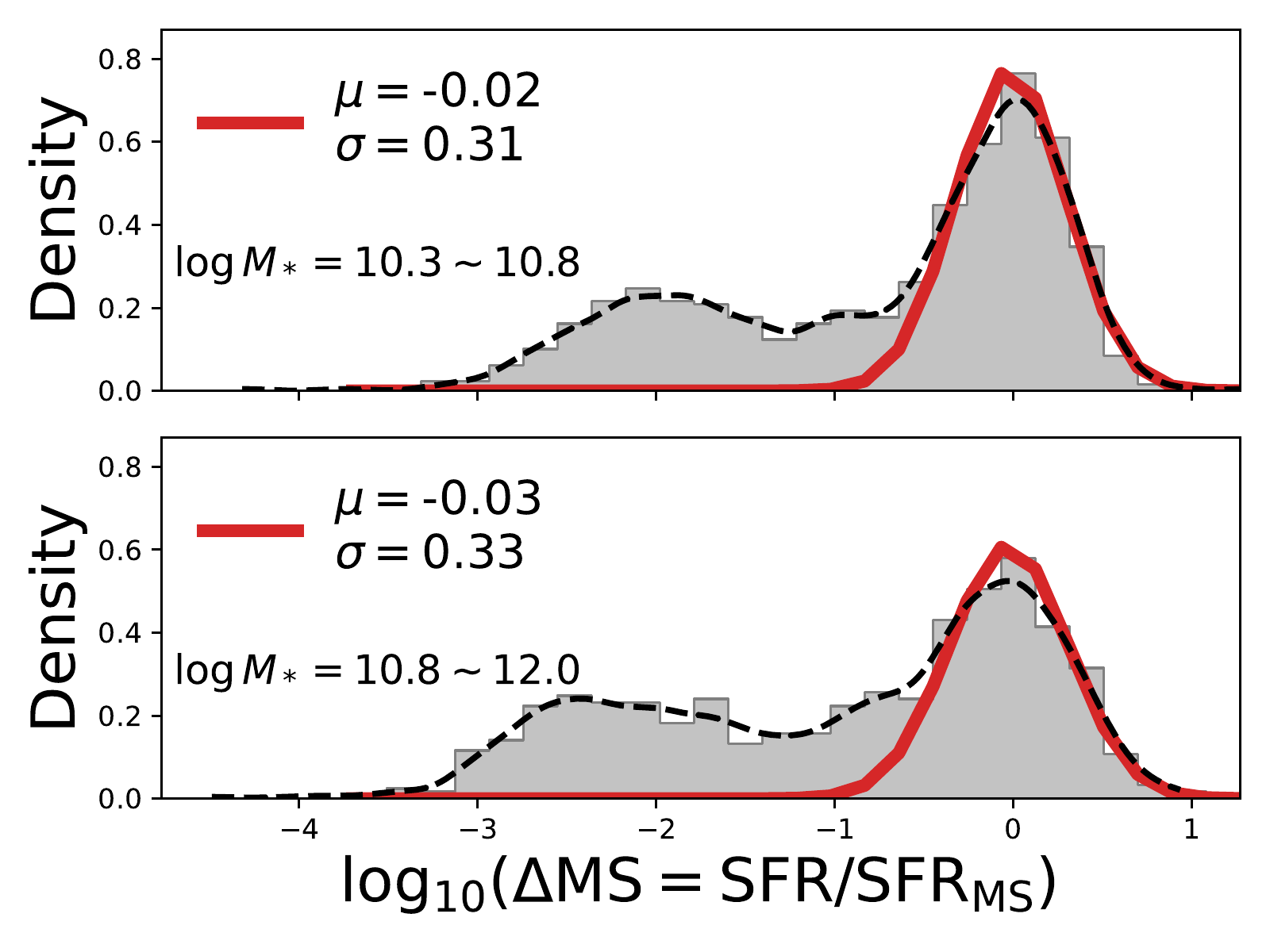}
    \caption{The distribution of \dms, i.e. the distance to the star-forming main sequence of \citet{Leja2021}. The entire sample is divided into two groups using the median stellar mass. The red solid line shows the Gaussian fit to the distribution of galaxies with \dms $\ge0.1$, the best-fit center ($\mu$) and width ($\sigma$) are shown in the plot. The $\mu($\dms) being consistent with one shows the consistency between our star-forming main sequence and \citet{Leja2021}'s. } 
    \label{fig:rsb_dist}
\end{figure}

The other way to compare our measurements with \citet{Leja2021}'s is through the distribution of \dms\ \citep{Elbaz2011}, i.e. the distance to the star-forming main sequence that is defined as
\begin{equation}
    \rm{\Delta MS = \frac{SFR}{SFR_{MS}}} \label{eq:dms}
\end{equation}
where $\rm{SFR_{MS}}$ is the star formation rate of a galaxy on the star-forming main sequence of \citet{Leja2021}. We show the comparison in Figure \ref{fig:rsb_dist} where we also fit a Gaussian distribution to the galaxies with \dms $\ge0.1$. While this study and that work both use \prospector, we note that our fiducial SED model is not the same as \citet{Leja2021}'s, because they used the Continuity prior for the nonparametric SFH reconstructions and assumed a different dust attenuation law. Yet, the Gaussian fit shows on average \dms $\approx 1$, showing no systematic shifts between our star-forming main sequence and \citet{Leja2021}'s. The Gaussian fit also shows that the width (i.e. dispersion) of the star-forming main sequence is $0.3-0.4$ dex, which is broadly consistent with \citet{Leja2021} and other recent studies \citep[e.g.][]{Whitaker2012,Speagle2014,Schreiber2015,Lee2018}.

\subsubsection{Examples of Results of the SED Fitting Procedure}

\begin{figure*}
    \centering
    \includegraphics[width=0.97\textwidth]{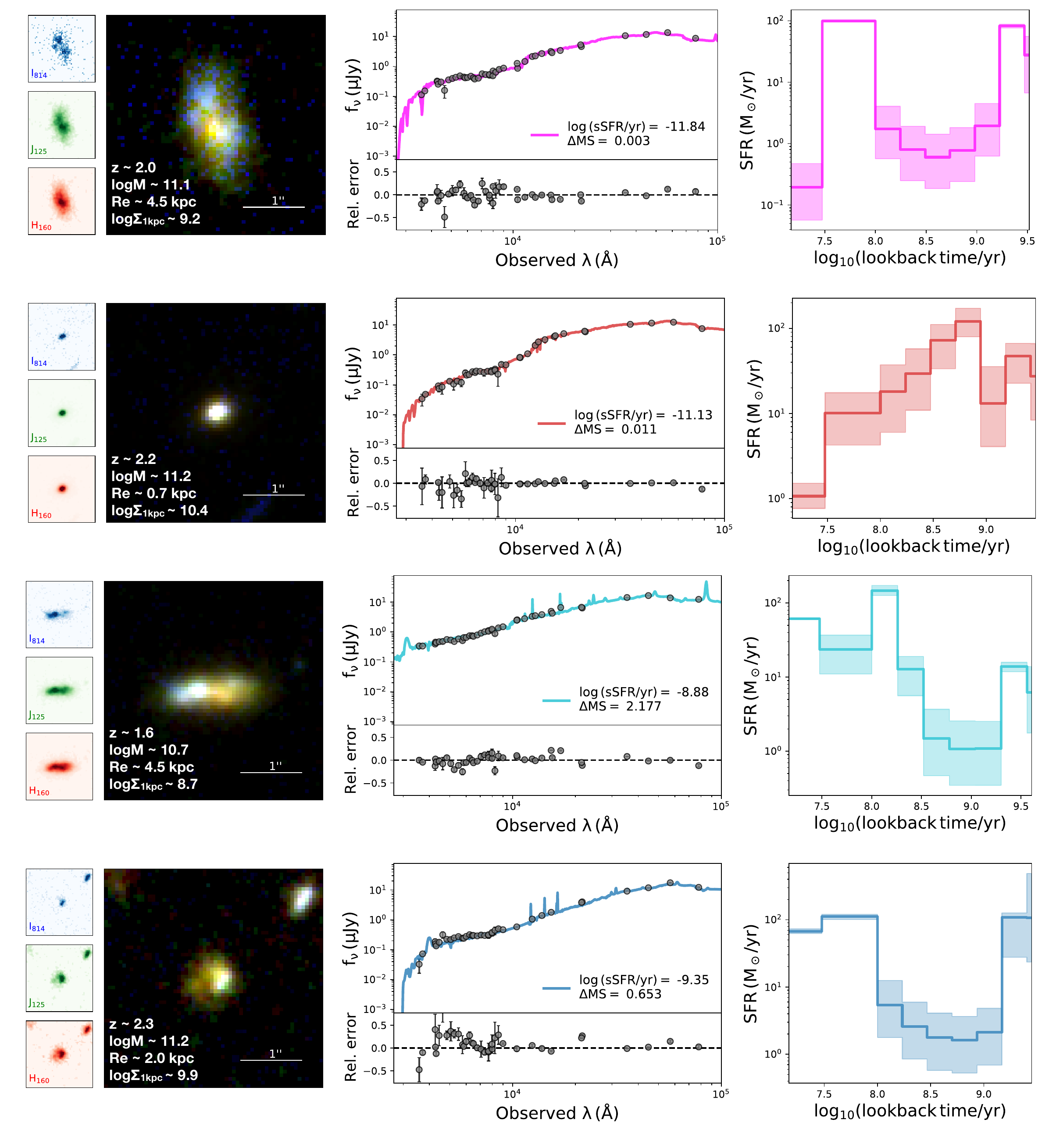}
    \caption{Examples of four different kinds of galaxies in this study, from top to bottom, namely an extended QG, a compact QG, an extended SFG and a compact SFG. The left column shows the RGB images made with multi-band \hst images. Basic information of individual examples is listed in each panel. The middle column shows the best-fit SED derived with \prospector, where in the bottom inset we show the \textit{relative} error between data and model-predicted photometry. The right column shows the reconstructed nonparametric SFH. }
    \label{fig:examples}
\end{figure*}

In Figure \ref{fig:examples} we show the SED fitting results, together with the \hst images, of four examples of massive galaxies at \zobs$\sim2$ with $\log_{10}(M_*/M_\sun)\sim11$ in our sample whose SFHs and morphological properties are illustrative of the variety that we have encountered in this study. The first row shows a QG with sSFR $<0.01$ Gyr$^{-1}$, and whose morphology is extended, with \re $\approx$ 4.5 kpc in the \H band. The SFH shows that the galaxy is a post-starburst one, because it experienced a recent burst of star formation $\approx 0.1$ Gyr prior to \zobs and it has a relatively low-level of ongoing SFR, which in this case is the SFR within 30 Myr, i.e the first lookback time bin of the nonparametric SFH. Interestingly, its \hst images in bluer bands, e.g. the \I band that probes rest-frame $\sim 2800$\AA, show features consistent with a merging event, fully in line with it being a post-starburst galaxy. The second row shows a very compact QG with \re $\approx$ 0.7 kpc, whose SFH shows a relatively gradual decline over the past 1 Gyr. The third and fourth rows show an extended SFG and a compact SFG, respectively. While in the \H band the size (\re) of the former is about two times larger than the latter, the images of  both galaxies show a very compact central component in the \I band, and their reconstructed SFHs suggest they are experiencing an on-going burst of star formation. These four galaxies are representative of the types of correlations between the morphological properties and SFHs at the core of this study, and we will discuss and quantify them for the entire sample in the remaining of this paper.

\subsection{Morphological Measurements} \label{sec:morph}

Similarly to \citetalias{Ji2022}, the morphological measures of the sample galaxies are taken from \citet{vanderWel2012}, where the light profile of galaxies was modelled with a 2-dimensional \sersic function using \galfit \citep{galfit}. We take the measurements in the \H band, because it better probes the stellar morphology for galaxies at $z\sim2$ as it is the reddest \hst imaging available in the three fields. To secure good quality, we follow the recommendation of \citet{vanderWel2012} to only use the measurements with \texttt{GALFITFlag = 0}. This means that the galaxies with unreliable single \sersic fitting results are excluded, including when (1) the \galfit-derived total flux significantly deviates from that derived with \sextractor \citep{sextractor}, (2) at least one parameter reaches either bounds of the range of allowed values set prior to the fitting (e.g. $0.2<n<8$) and (3) the \sersic fitting is unable to converge.  

Morphological parameters considered in this work include the 
\begin{itemize}
    \item \sersic index $n$
    \item circularized effective/half-light radius \re $=R_{e,maj}\times \sqrt{b/a}$, where $R_{e,maj}$ is the effective semi-major axis and $b/a$ is the axis ratio
    \item stellar-mass surface density within the effective radius \Se $=M_*/(2\pi R_e^2)$
    \item stellar-mass surface density within the central radius of 1kpc, \Sone \citep{Cheung2012}. As shown in \citet{Ji2022a}, because \Sone is derived as the combination of \re and $n$, this helps to reduce the uncertainty stemming from the strong covariance between the two parameters, making \Sone a more robust parameter to quantify galaxy compactness than $n$ or \re individually. 
    \item fractional mass within the central radius of 1 kpc, \Mone \citep{Ji2022a}. Similar to the stellar-mass surface density, \Mone also is a good metric of the compactness of galaxies. Because $\Sigma$ is strongly correlated with $M_*$, it is hard to interpret correlations of any property with $\Sigma$, because they could be driven by $M_*$ rather than the compactness. Because the dependence of \Mone on $M_*$ is very weak \citep{Ji2022a}, using it can provide a more direct view on the links between compactness and other physical properties.
\end{itemize}

\section{Results}

\subsection{The Progenitor Effect} \label{sec:progenitor}

\begin{figure*}
    \centering
    \includegraphics[width=1\textwidth]{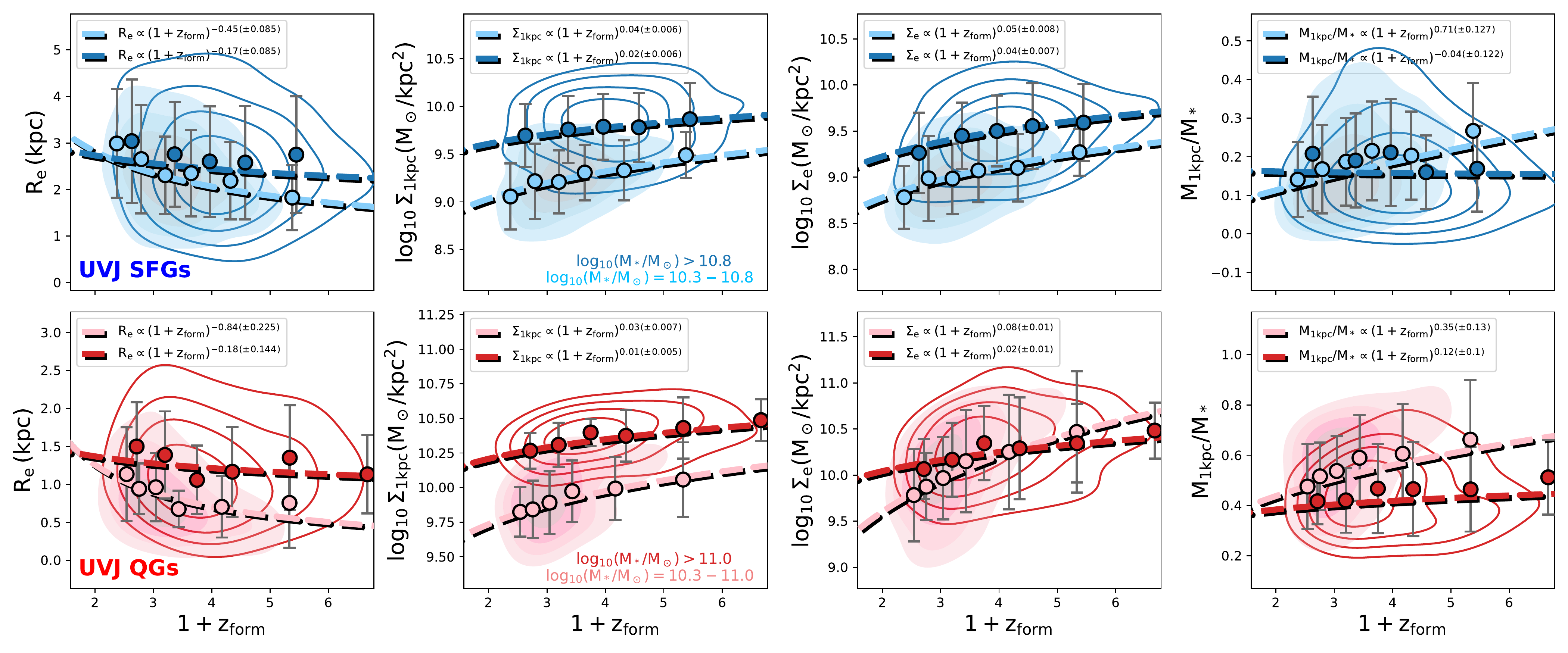}
    \caption{The relationships between the morphological properties of galaxies and their \zf. The top and bottom rows show the results of UVJ-selected SFGs and QGs, respectively. We remind that the y-axis' scale of the two rows are different. Each galaxy sample is further divided using its median stellar mass. In each panel, the contours show the density distributions estimated using a Gaussian kernel, the points with error bars show the medians and standard deviations of the corresponding morphological property of galaxies in the individual bins of \zf, and the dashed lines show the results of power-law fittings where the best-fit relations are labeled in the panel legend. The plots clearly show the dependence of galaxies' morphology on their formation time, i.e. the progenitor effect.}
    \label{fig:morph_zf}
\end{figure*}

To begin, we investigate the progenitor effect on the apparent structural evolution of galaxies by studying the correlations between structural properties and \zf. We first divide the entire sample into UVJ-selected SFG and QG subsamples, and then divide each subsample using its median $M_*$, motivated by \citetalias{Ji2022} where we found that among QGs the strength of the progenitor effect depends on $M_*$. In Figure \ref{fig:morph_zf}, we show the relationships of \zf with \re, \Sone, \Se and \Mone, respectively. For each relationship we also fit a power-law function, i.e. (1+\zf)$^{-\beta}$, which is plotted as a dashed line in Figure \ref{fig:morph_zf}. To estimate the uncertainty of best-fit power-law relations, similarly to what we did in \citetalias{Ji2022}, we bootstrap the sample 1000 times, during each time we resample the value of each measurement using a normal distribution with a width equal to the measurement uncertainty. The best-fit power-law relation and the corresponding uncertainty are labeled in the legend of each panel of Figure \ref{fig:morph_zf}. 

As Figure \ref{fig:morph_zf} shows, all considered structural properties depend on the formation epoch (\zf) of galaxies, which, in essence, is the progenitor effect. Regarding the relationship between \re and \zf, galaxies formed earlier, i.e. having larger \zf, tend to have smaller sizes. A clear dependence on $M_*$ is found for QGs where the relationship is steeper for lower-mass QGs than for higher-mass ones. This is most likely because of the increasing importance of repeated minor merging events that take place in more massive galaxies after quenching. The miner mergers drive the after-quenching size growth that flattens the \re-\zf relationship of more massive QGs, as we have already discussed in detail in \citetalias{Ji2022}. Compared to QGs, the \re of SFGs decreases with (1+\zf) seemingly at a slower rate. Quantitatively interpreting this relationship of SFGs requires the full knowledge\footnote{Because  rejuvenation of star formation apparently is rare, we consider the whole history of mass assembly of any galaxy as the period that starts from the Big Bang and ends at the time when a galaxy becomes quiescent.} of their mass assembly histories, which we do not have because (1) by the time of \zobs SFGs are still undergoing active star formation and have not yet completed assembling their masses, and (2) we cannot predict their SFHs after \zobs. Nevertheless, the decreasing trend of \re with \zf shows, even for SFGs,  that the progenitor effect also plays a role in their apparent size evolution. 

Because the measurements of \Sone, \Se and \Mone directly depend on \re (section \ref{sec:morph}), it is naturally expected that these metrics of compactness also depend on \zf, a relationship that we indeed observe and show in the left three columns of Figure \ref{fig:morph_zf}. Regardless of star-formation properties at \zobs, galaxies that formed earlier, i.e. having larger \zf, tend to have a more compact morphology, i.e. have larger \Sone, \Se and \Mone. 

Unlike using \zobs to select samples, which results in galaxies formed at different epochs of $z\ge$ \zobs being  grouped together, using \zf naturally separates galaxies formed at different epochs, which automatically allows one to investigate, and account for, the progenitor effect. Figure \ref{fig:morph_zf} shows that the progenitor effect at least in part contributes to the apparent evolutions of the size and of the compactness metrics considered in this work with \zobs, if galaxies are mixed together with no accounting for their formation epochs. As Figure \ref{fig:morph_zf} shows, variations up to 50\% can be observed in \re and up to 1 full dex in \Sone, depending on the stellar-mass range, can be accounted for solely because of the progenitor effect. This highlights the importance of accounting for its contribution before attempting any interpretation or modeling of the apparent structural evolution in terms of different physical mechanisms. 

\subsection{Correlations Between the Structural and Star-formation Properties of Galaxies Formed at a Similar Epoch} \label{sec:pistol}

\begin{figure*}
    \centering
    \includegraphics[width=1\textwidth]{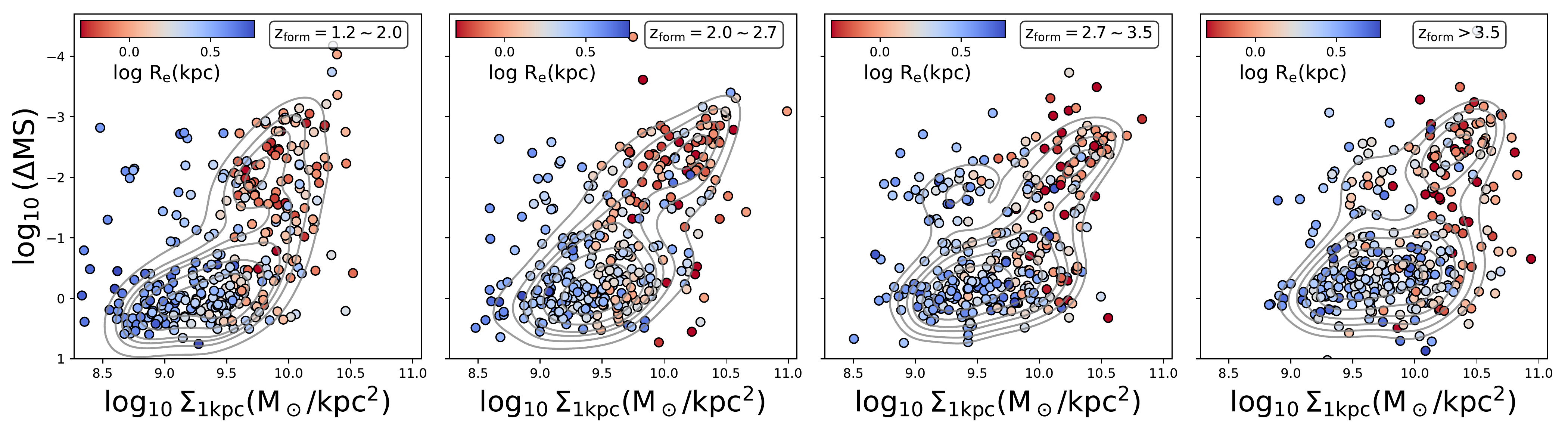}
    \includegraphics[width=1\textwidth]{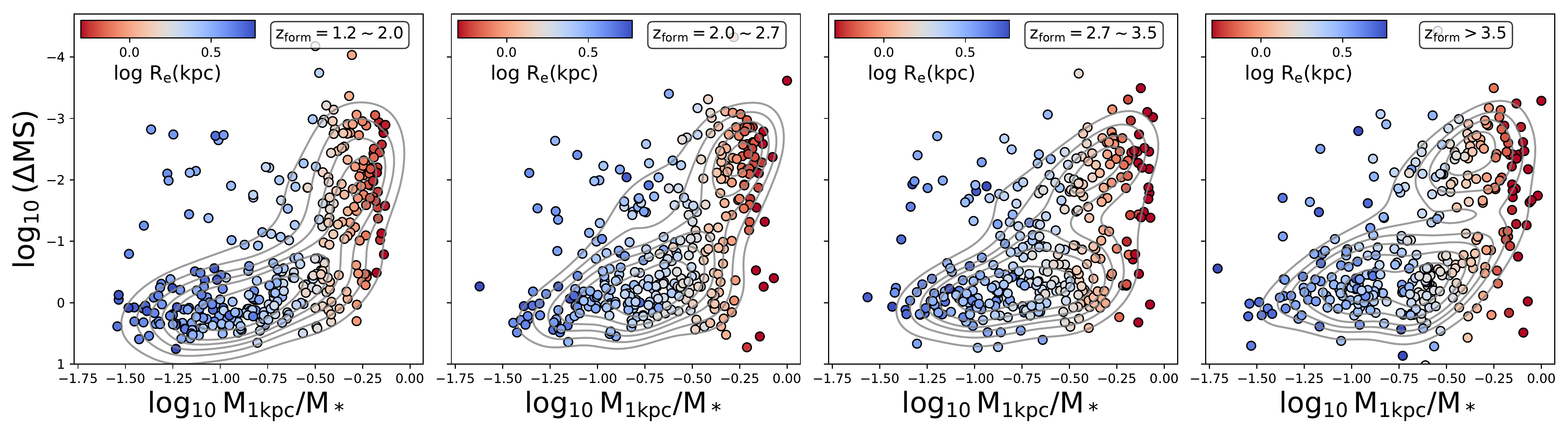}
    \includegraphics[width=1\textwidth]{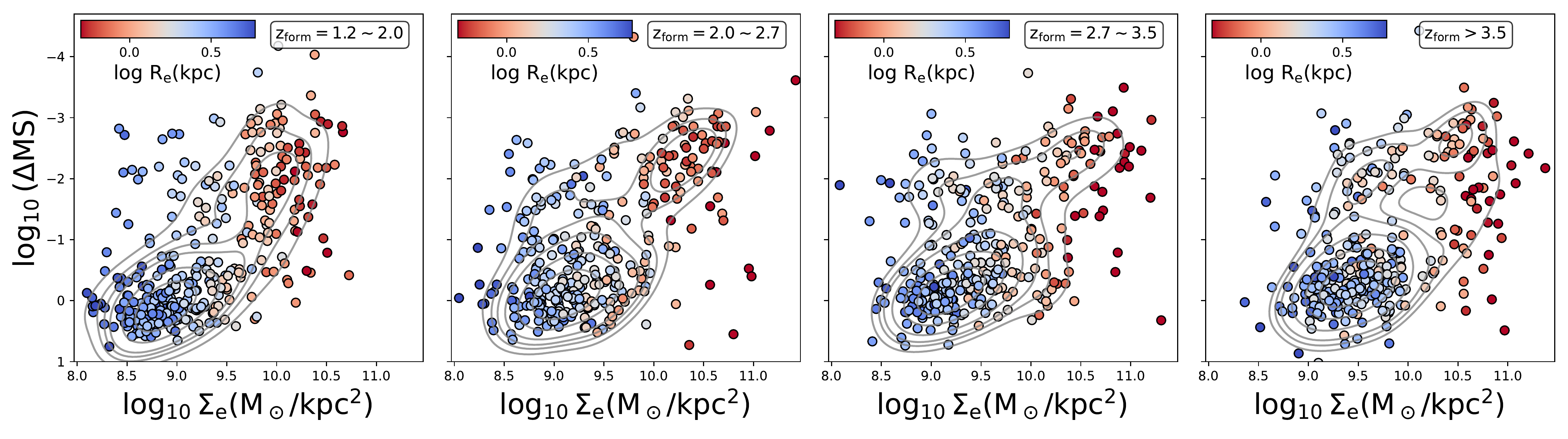}
    \caption{ The relationships between \dms (equation \ref{eq:dms}) and galaxy compactness metrics \Sone (the top row), \Mone (the middle row) and \Se (the bottom row). To mitigate the progenitor effect, we use the quartiles of the distribution of \zf to divide the entire sample into four subsamples, which are shown in individual columns, such that galaxies shown in the same panel formed at a similar epoch. We remind that the scale of y-axis is reversed, so galaxies being more quiescent, i.e. having smaller \dms, appear in upper regions of the plots. Each galaxy is color-coded according to its size \re. Background grey contours show the density distribution estimated using a Gaussian kernel. The dependence of the compactness metrics adopted here on \dms is still observed even after we mitigate the progenitor effect by binning the sample by \zf, implying a physical link between the compactness of galaxies and their star-formation properties.}
    \label{fig:pistol}
\end{figure*}

Taking advantage of the reconstructed SFHs, we can group galaxies according to their formation epochs. Once the formation epoch is fixed, the progenitor effect should then be essentially eliminated from the apparent evolution. If correlations between star-formation and structural properties are still observed, then these will provide strong empirical constraints on physical mechanisms of the observed structural evolution. In what follows, we use \zf as a proxy for the formation epoch of galaxies. Specifically, we divide the entire sample into four subsamples using quartiles of \zf, and study the relationships of the compactness metrics of galaxies with \dms (equation \ref{eq:dms}). The results are plotted in Figure \ref{fig:pistol}.

In all bins of \zf, the majority of galaxies are distributed following a ``pistol''-shape pattern, both in the diagram of \dms vs. \Sone (the top row of Figure \ref{fig:pistol}), and in the diagram of \dms vs. \Mone (the middle row of Figure \ref{fig:pistol}). At a fixed \zf, galaxies with smaller \dms, i.e. being more quiescent, have larger \Sone and \Mone, i.e. being more compact, while at the same time the distributions of \Sone and \Mone also become narrower. In the diagram of \dms vs. \Se (the bottom row of Figure \ref{fig:pistol}), a similar trend is also observed for the QGs, which in general have larger \Se whose distribution also is narrower than that of the SFGs, although the knee of the pistol pattern is not as obviously seen as that in the diagrams of \Sone and of \Mone. 

A similar pistol pattern between the star-formation properties and the central density of galaxies have been observed in recent literature \citep[e.g.][]{Barro2013,Barro2017,Lee2018}. However, in those studies the pistol pattern was produced \textit{without} regard to the formation epoch of galaxies, meaning that the quantitative details and overall significance of the observed patterns were at least in part due to the progenitor effect. In this work we refine the characterization of the evolutionary trends in the sense that we utilize the information extracted from the reconstructed SFHs to eliminate the contribution from the progenitor effect to the observed correlations. As Figure \ref{fig:pistol} shows, the pistol pattern, although different in shape, is still observed after the elimination of the progenitor effect, demonstrating the existence of a physical link between the effective radius, central mass density and compactness of galaxies and their star-formation properties. Overall as galaxies quench, their effective radii become smaller, central mass densities increase and thus their compactness also increases.  

\subsection{Automated Morphological Classification: the Dawn of Spheroids?}
While on average galaxies with smaller \dms are more compact, Figure \ref{fig:pistol} shows that there are also substantial numbers of galaxies with suppressed star formation, i.e. \dms$<0.1$, that have a extended morphology. Morphological metrics derived from the single \sersic fitting, such as the ones considered above, have been extensively adopted to characterize the overall shape of the light distribution of galaxies. However, those metrics entirely ignore substructures such as clumps and nonaxisymmetric features like bars and tidal features that contain key information about the evolutionary mechanism of galaxies' structures. In fact, those substructures are commonly observed not only in SFGs at high redshifts \citep[e.g.][]{Elmegreen2007,ForsterSchreiber2011, Guo2012} but also in some QGs as well (Giavalisco \& Ji in preparation). Although in the nearby universe visual classifications are effective in identifying them \citep[e.g.][]{Lintott2008}, at high redshifts identifying the substructures are very challenging because of (1) the cosmological dimming of surface brightness and (2) the limitations imposed by the available angular resolution that even with \hst, and now JWST, are such that the substructures are still hard to visually identify. Rapidly developing techniques based on machine learning and artificial intelligence hopefully can greatly help improve the morphological classification of galaxies in the early universe. For example, utilizing deep learning techniques, \citet{HuertasCompany2015} trained the Convolutional Neural Networks to classify the morphological types of CANDELS' galaxies brighter than 24.5 magnitude (AB) in the \H band. The quality of their machine-based classifications is very high, namely they have no bias (with a scatter of $\sim10\%$) compared to those done by human classifiers and the fraction of mis-classifications is better than $1\%$. 

\begin{figure*}
    \centering
    \includegraphics[width=1\textwidth]{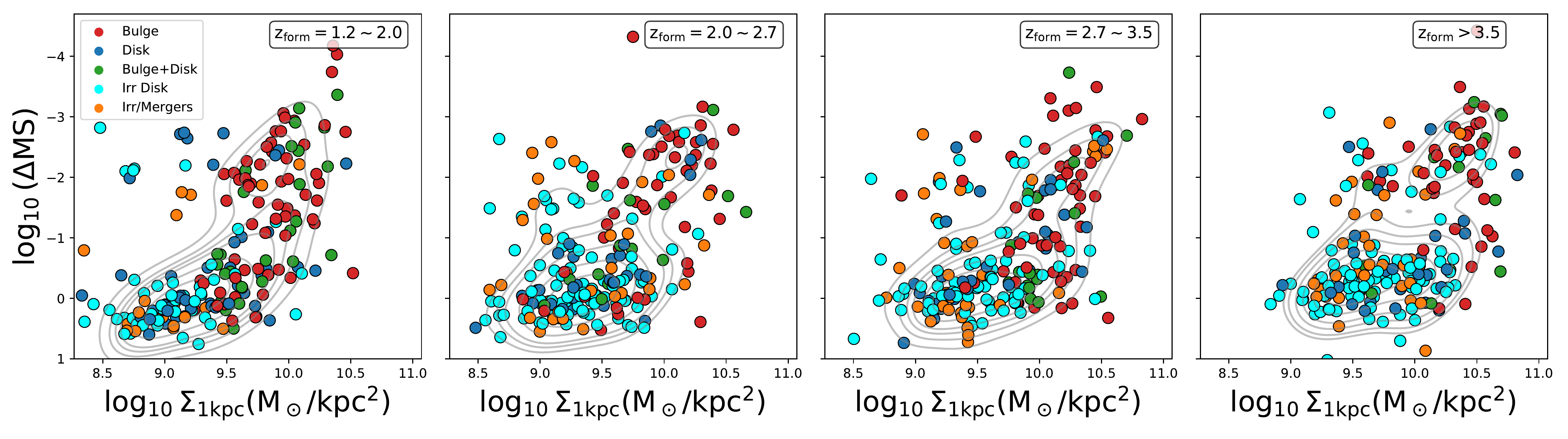}
    \includegraphics[width=1\textwidth]{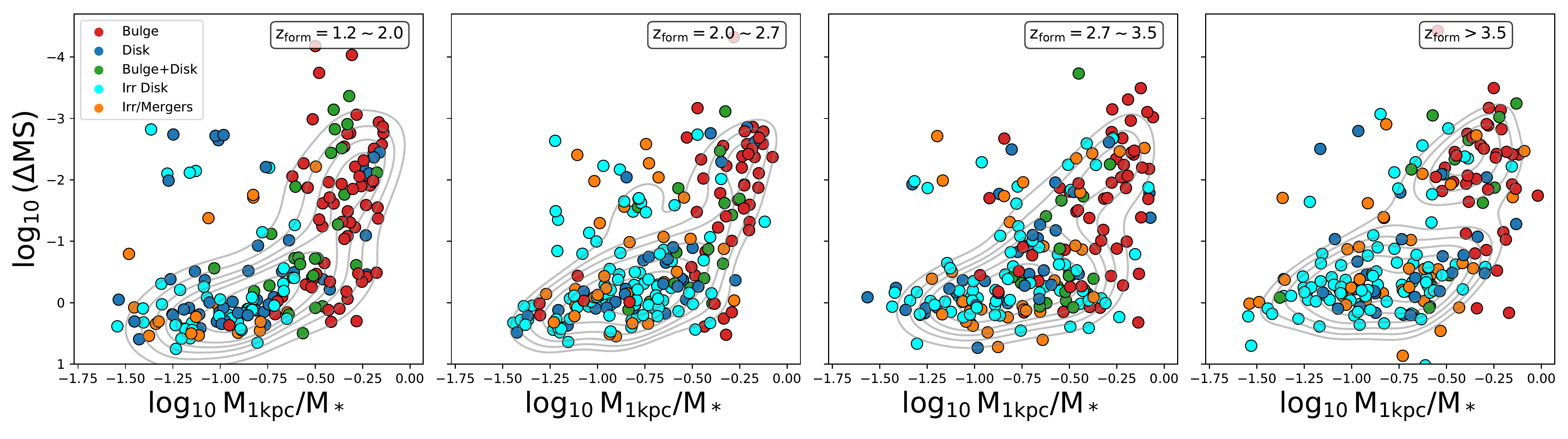}
    \includegraphics[width=1\textwidth]{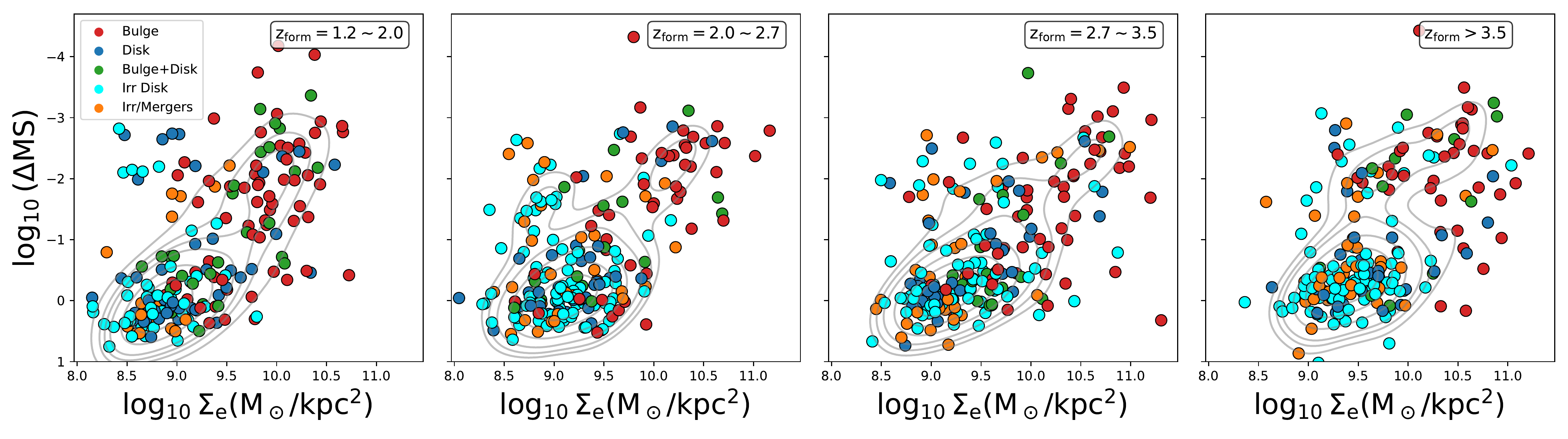}
    \caption{Similar to Figure \ref{fig:pistol}, but for the subsample of galaxies that are successfully matched to the catalog of \citet{HuertasCompany2015} where the Convolutional Neural Networks were used to classify the morphological type of galaxies, including a purely bulge- (red), a purely disk- (blue), a bulge+disk- (green), an irregular disk- (cyan) and a merger- (orange) type morphologies. The morphology of galaxies changes from disk-like to bulge-like as they become more compact and quench.}
    \label{fig:pistol_ML}
\end{figure*}

We cross match our sample with the catalog of \citeauthor{HuertasCompany2015}, and then plot the successfully-matched subsample in Figure \ref{fig:pistol_ML}. Overall, a clear trend can be observed such that the morphology of galaxies changes from disk-like to spheroidal/bulge-like as they become more compact and quench. With the progenitor effect eliminated, this apparent morphological transformation of galaxies is closer to the intrinsic structural transformations that galaxies undergo as they evolve and quench in time. Thus, morphologically spheroidal structures (no dynamical information is included in this discussion), whether bulges or elliptical galaxies, emerge as galaxies suppress their star formation and terminate the assembly of the bulk of their stellar masses.  We shall return later on this point from a different perspective.

Regarding galaxies with suppressed star formation in particular, the morphology of the extended ones is not only more disky, but also more disturbed compared to the compact ones. In particular, we use the median values of \Sone, \Se and \Mone of SFGs to divide the sample galaxies into compact and extended ones. The key result is that, regardless of the metrics adopted to classify galaxies' compactness, the probability of a QG to have a disturbed morphology is $\sim3-4$ times higher when it is extended relative to when it is compact. Specifically, for galaxies with \dms $<0.1$, the fraction of those having a disturbed morphology, namely being classified as an irregular disk or a merger according to \citeauthor{HuertasCompany2015}'s criteria (see their section 6 for details), is $68.6\pm12.9\%$ (48/70) for those with $\log_{10}$\Sone$<9.5$ compared to $18.2\pm0.3\%$ (42/231) for those with $\log_{10}$\Sone$>9.5$. If we use \Mone$=0.1$ as the dividing line between extended and compact galaxies, the fraction is $71.7\pm16.4\%$ (33/46) for the former compared to $22.4\pm0.3\%$ (57/255) for the latter. Similarly, if we use $\log_{10}\,$\Se$=9.2$ instead, the fraction is $64.3\pm12.3\%$ (45/70) for the extended ones compared to $19.5\pm0.3\%$ (45/231) for the compact ones. This distinct morphological difference \textit{within} the population of QGs already indicates the possibility that they reflect markedly different formation and evolutionary paths. We are going to elaborate this point more in the next section.

\subsection{The Relationship Between the Morphology and the Assembly History of Galaxies} \label{sec:morph_SFH}

So far, we have only considered the formation epoch (\zf) of galaxies as a key piece of information stemming from the knowledge of SFHs. In this section we take the full shape of reconstructed SFHs into account to further study the relationship between the morphological and star-formation properties. To better illustrate our findings, we divide the entire sample into four galaxy groups, namely (1) extended quiescent galaxies (eQGs); (2) compact quiescent galaxies (cQGs); (3) extended star-forming galaxies (eSFGs) and (4) compact star-forming galaxies (cSFGs). 

We have verified that the results presented below are not sensitive to the exact criteria adopted to select the four galaxy groups. Specifically, to separate SFGs and QGs, we have tested our results by both using the UVJ technique and using the distance from the star-forming main sequence provided by the dividing line of \dms $=0.1$. To separate compact and extended galaxies, we have tested our results by using both the fixed dividing line of $\log_{10}$\Sone$=9.5$ and also the varying dividing line of the  median \Sone of individual \zf bins. We have not observed any substantial changes in our results. Our results also appear to be insensitive to the exact choice of the compactness metrics adopted to separate compact and extended galaxies, as we demonstrate in Appendix \ref{app:stack_SFH_other} using \Mone and \Se. In what follows, we show the results of using \dms$=0.1$ to separate star-forming and quiescent galaxies, and using $\log_{10}$\Sone$=9.5$ to separate compact and extended galaxies.

We first measure the \textit{average} SFHs. Similarly to what we did in the previous sections, we first bin the galaxies according to their \zf. In each bin of \zf, instead of calculating the median/mean of best-fit SFHs of the individual galaxy groups, we stack the posteriors derived from the \prospector fittings, hence the full probability density distributions of the individual fittings are taken into account. Because in this work we only consider galaxies within a relatively narrow range of stellar mass, i.e. $\log_{10}(M_*/M_\sun)=10.3\sim11.7$, to emphasize the shape of SFHs, before stacking we normalize each SFH by $M_*$ observed at \zobs so that
\begin{equation}
    \int_{0}^{t_H}\widetilde{\rm{SFR}}(t)\,dt = 1,
\end{equation}
where $\widetilde{\rm{SFR}}(t)=\rm{SFR}(t)/M_*$. We then calculate the median and standard deviation (1$\sigma$) from the stacked posteriors, and finally obtain the average SFHs.

In Figure \ref{fig:stack_SFH_S1}, we show the average SFHs of eQGs (magenta), cQGs (red), eSFGs (cyan) and cSFGs (blue), respectively. Clear differences in the shape of the SFH of each subgroup are clearly observed. While SFGs have either an overall flat or rising SFH, QGs show a clear decline of recent SFR, as expected. The novel aspect is the dependence of SFHs on morphology: while cQGs underwent an early phase of intense star formation followed by a gradual decline and eventual quiescence, eQGs followed the opposite trend, namely, a gradual rise of star formation toward a relatively recent peak followed by a rapid decrease toward quiescence. Among SFGs, the differences in the average SFH as a function of morphology seem to be more subtle, but we observe that the overall shape of their average SFH is similar to that of the eQGs minus the decrease toward quiescence. We also observe that the rapid rise toward the very early peak seems to be absent among SFGs, suggesting that this type of SFH is typical of the early type galaxies but becomes rare or absent in galaxies that are still forming stars later on in the cosmic evolution, even among cSFGs. In the following, we will study in detail the relationships between the SFH and the compactness of QGs (section \ref{sec:sfh_morp_qg}), and of SFGs (section \ref{sec:sfh_morp_sfg}), respectively.

\subsubsection{The Case of QGs} \label{sec:sfh_morp_qg}

\begin{figure*}
    \centering
    \includegraphics[width=1\textwidth]{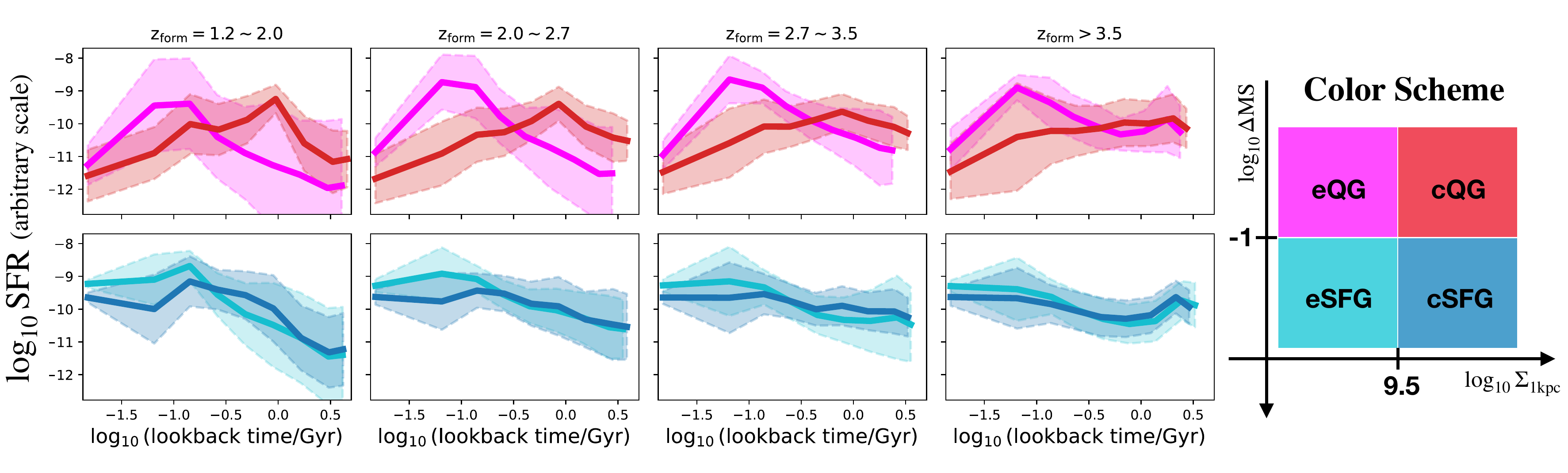}
    \caption{The stacked SFHs of QGs (the top row) and SFGs (the bottom row), where the solid line and shaded region correspond to the median and 1$\sigma$ range derived from the stacked posteriors. Because each SFH has been normalized before stacking, the y-axis has an arbitrary scale (see section \ref{sec:morph_SFH} for details). Different columns show galaxies formed at different epochs. Galaxies are also divided according to their compactness. These end up with four groups of galaxies, namely extended/compact QGs (e/c QGs) and extended/compact SFGs (e/c SFGs). The selection  criteria of the individual groups, as well as the color scheme, are illustrated in the right-most panel. The differences in stacked SFHs between eQGs and cQGs are significant, while they are subtle between eSFGs and cSFGs.}
    \label{fig:stack_SFH_S1}
\end{figure*}

Clear trends are observed between the assembly history and the morphology of QGs. As Figure \ref{fig:stack_SFH_S1} shows, the stacked SFH of eQGs peaks at a much later time, i.e. smaller lookback time, of $\lesssim 0.5$ Gyr compared to the cQGs whose stacked SFH peaks at $\gtrsim1$ Gyr ago. Thus, while the stacked SFH of cQGs is very similar to that of typical QGs, which have assembled most mass early on followed by a gradual decline of their SFRs, the stacked SFH of eQGs is very similar to that of post-starburst galaxies, which shows a recent (a short lookback time from \zobs) burst of star formation followed by a rapid decline of SFR. 

\begin{figure*}
    \centering
    \includegraphics[width=1\textwidth]{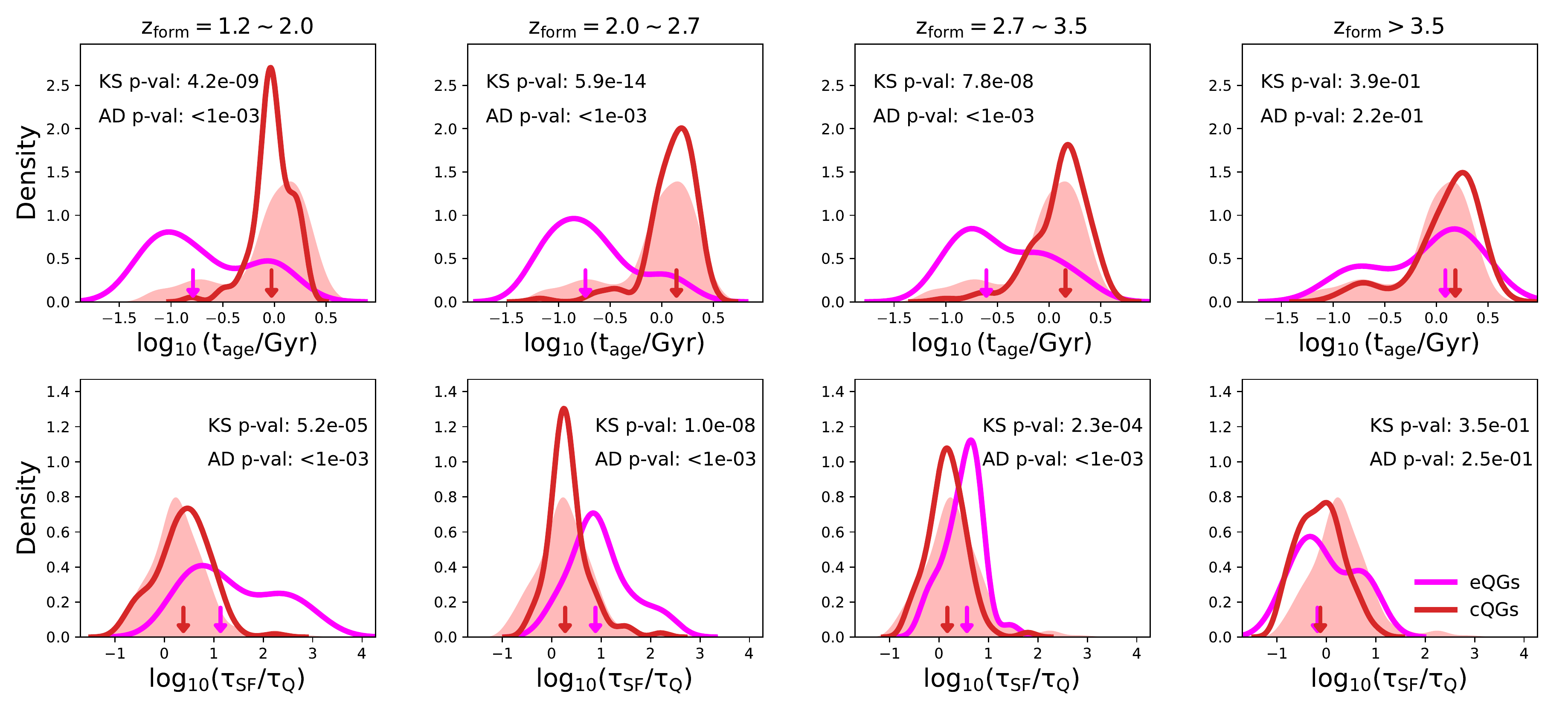}
    \caption{The distributions of the mass-weighted age (\tage, the top row) and the Asymmetry of SFHs (\tausf/\tauq, the bottom row) of QGs. Similar to Figure \ref{fig:stack_SFH_S1}, here we refer to galaxies with \dms $<0.1$ as the QGs. Different columns show galaxies formed at different epochs. The magenta and red solid lines show the distributions of eQGs and cQGs, respectively. The median of each distribution is labeled using a downward arrow with the corresponding color. For reference, in every panel the distribution of the entire QG sample is shown as the filled area in light red. Also labeled are the $p$-values of the Kolmogorov–Smirnov (KS) and the Anderson-Darling (AD) tests on the null hypothesis that the distributions of eQGs and cQGs are identical. The histograms again show the significant differences between the stacked SFHs of eQGs and of cQGs.}
    \label{fig:QG_sfh_shape_dist}
\end{figure*}

In Figure \ref{fig:QG_sfh_shape_dist} we compare the distributions of \tage and of \tausf/\tauq. The eQGs have smaller \tage (i.e. younger) and larger \tausf/\tauq compared to the cQGs, which are consistent with the shape of the stacked SFHs. We also run both the Kolmogorov–Smirnov and the Anderson-Darling tests on the null hypothesis that the distributions of the two QG groups are the same. We can reject the null hypothesis with a $\gtrsim3\sigma$ confidence level, except in the largest \zf bin where we can reject it with an $\sim2.5\sigma$ confidence level. Thus, compared to the eQGs, in general the cQGs have both a larger mass-weighted age and a shorter star-formation timescale relative to the quenching timescale. 

Taken together with the higher frequency of finding eQGs with a disturbed morphology, as we have already observed in Figure \ref{fig:pistol_ML} and discussed in section \ref{sec:pistol}, the finding of eQGs' average SFH being consistent with that of post-starburst galaxies supports  a scenario that some eQGs might be merging transients, namely, they are merger remnants observed at a stage when the SFR rapidly declines after a merger-induced recent starburst. The stacked SFH of eQGs suggests that the decline of SFR happened $\lesssim 0.5$ Gyr ago, which is in broad agreement with studies of galaxy mergers based on hydrodynamical simulations \citep[e.g.][]{Springel2005,Hopkins2008}. A complete merging event very likely includes multiple episodes of starburst, and the whole process can help galaxies to form dense central cores because strong gravitational torques induced by the merging galaxy can effectively drive gas to the center and then trigger central star formation \citep[e.g.][]{Sanders1988,Hopkins2010}. Because the eQGs do not seem to have built up their central cores by the time of observation\footnote{It is possible that some of the eQGs might have highly dust-obscured dense cores. Unfortunately, the sensitivity of existing MIR/FIR observations in the three fields is relatively low. This possibility can be tested with future observations such as the upcoming \jwst surveys.}, if they indeed come from mergers, then they likely have just finished one earlier episode of merger-induced starburst, meaning that their star formation can possibly rejuvenate. Interestingly, $23\pm3 \%$ QGs in our sample are the eQGs, and this fraction is quantitatively consistent with the rejuvenation rate measured in recent spectroscopic studies of QGs at lower redshift $z=0.5\sim1$ \citep[e.g.][]{Belli2017,Chauke2019}, although we note that using different criteria for the rejuvenation event can result in quantitatively different rates \citep[e.g., section 6.5 of][]{Tacchella2022}. 

\subsubsection{The Case of SFGs} \label{sec:sfh_morp_sfg}

\begin{figure}
    \centering
    \includegraphics[width=0.447\textwidth]{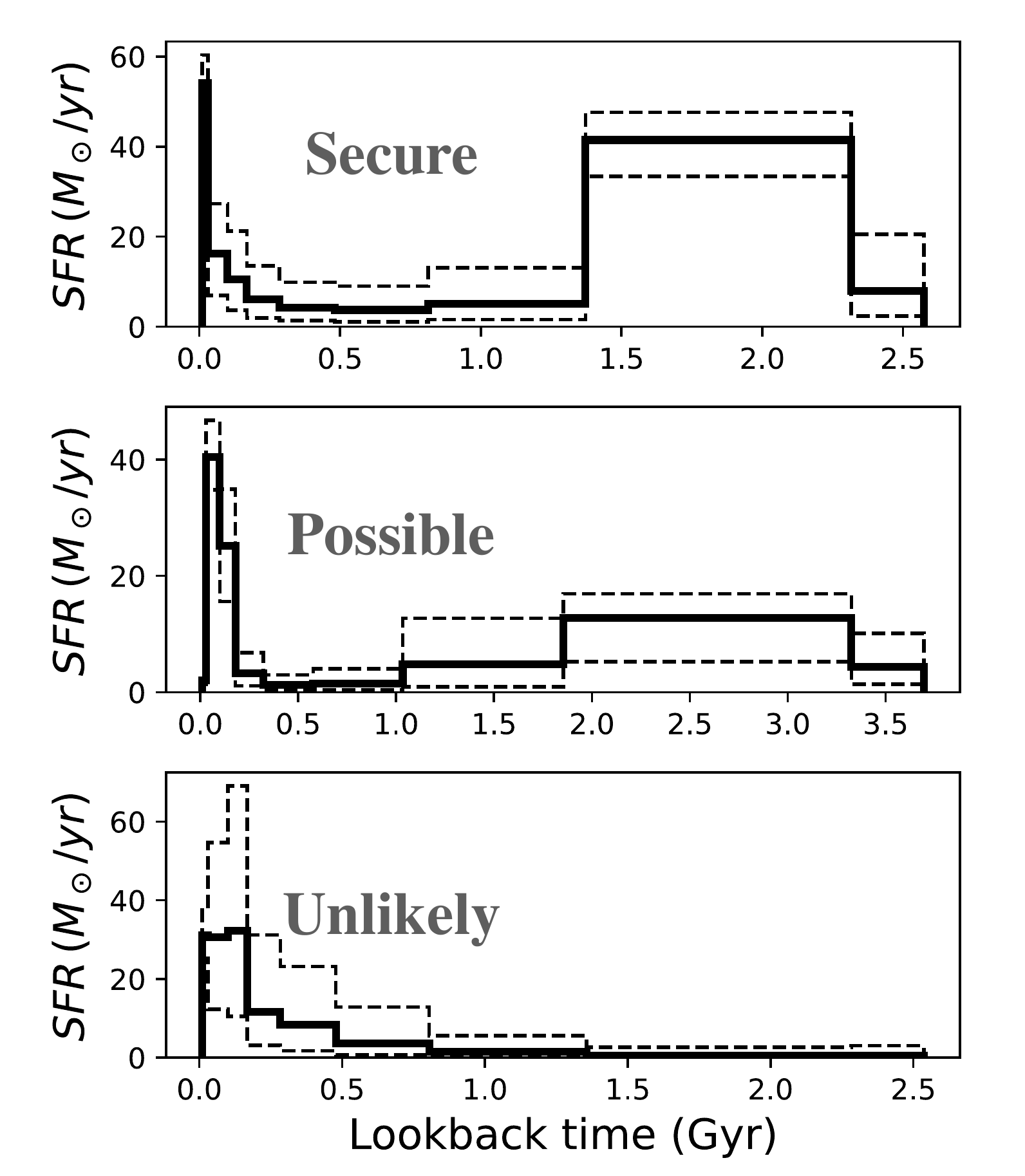}
    \caption{Examples of individual reconstructed SFHs to illustrate our visual selection of SFGs with more than one prominent episode of star formation. The black solid line shows the best-fit reconstructed SFH, and the dashed line shows the 1$\sigma$ uncertainty. To highlight the older age bins, unlike previous figures about SFHs, we remind that we use the linear scale along the x-axis here. The upper panel shows an ideal case where two episodes of star formation are clearly seen. The middle panel shows a possible case where the uncertainty of the reconstructed SFH makes it hard to judge the robustness of the $\sim2.5$ Gyr-old star-formation episode, although the best-fit SFH indeed suggests a two-epoch formation. The bottom panel shows an unlikely case where the SFR monotonically rises since the Big Bang.  }
    \label{fig:visual_sfh}
\end{figure}

Although \textit{on average} the SFHs of eSFGs and of cSFGs are very similar (Figure \ref{fig:stack_SFH_S1}), a correlation indeed is found between the central stellar-mass surface density of SFGs and a more detailed SFH classification that we describe in what follows. Observations of nearby elliptical galaxies\footnote{Given the stellar mass range the majority of the SFGs considered in this study have likely become massive elliptical galaxies at $z=0$.} suggest that their masses were assembled via multiple episodes that are responsible for the formation of different structures \citep[e.g.][]{Huang2013}. Motivated by this, we visually inspect individual SFHs to identify the SFGs with multiple, prominent episodes of star formation, i.e. two or more peaks that are clearly shown in a reconstructed SFH. As Figure \ref{fig:visual_sfh} illustrates, both the best-fit SFH and the corresponding uncertainty are taken into account during the process of visual classifications. We then study the relationship between \Sone and the fraction of SFGs with multiple star-formation episodes, i.e.
\begin{equation}
    \mathcal{F}_{\rm{multi-SF}}=\frac{N_{\rm{multi-SF}}}{N_{\rm{tot}}},
\end{equation}
where $N_{\rm{tot}}$ is the total number of SFGs. 

Before proceeding to discuss the results, we clarify in more detail our visual classification procedure, and also we caution about possible systematics affecting  \fmulsf. First, given the limitation imposed by the data, the SFH reconstruction in this work is done with the \textit{integrated} photometry and those refer to the galaxy as a whole. Also, the time resolution of the SFH reconstruction remains relatively low compared to the timescale of starbursts, i.e. Gyr vs. tens or hundreds of Myr. The combination of these two effects means that every star-formation feature, such as a peak or a burst, identified in a reconstructed SFH might in fact consist of more than one independent star-formation events that happen at approximately the same epoch but are not physically associated, e.g., two or more episodes of star formation in different, physically uncorrelated HII regions. As a result, for each galaxy we are unable to identify all independent star-formation events, which however is not our goal. Our goal simply is to find SFGs that by the time of \zobs have had more than one enhanced and distinct epoch of mass assembly (as opposed to independent star-formation events). Because we only have nine bins of look-back time (section \ref{sec:prosp}), for nearly all cases of our SFGs the ``multiple star-formation episodes'' really means two well-separated episodes, in the form of peaks of SFR enhancement, a younger one and an older one with a typical dividing line of 1 Gyr. Because all SFGs, by definition, include the younger, on-going star-formation episode in their SFHs, what really is measured by \fmulsf can be considered as the fraction of SFGs with the clear presence of older stellar populations created in previous episodes of star formation. Second, ambiguous cases certainly exist such as the ``possible'' example illustrated in Figure \ref{fig:visual_sfh} where the uncertainty of reconstructed SFHs makes it hard to tell the robustness of individual star-formation episodes. While the measures presented below only include the ``secure'' cases, we have checked that, qualitatively, our findings do not change if the ``possible'' cases are also included. Finally, although for each galaxy we have $\approx40$ bands photometry, most of these cover the spectral range bluer than the rest-frame V-band for galaxies at $z\sim2$. In this wavelength range young, bright stars outshine older stars, meaning that either a large mass fraction of older stellar populations is present or a comparatively high S/N of the photometric data (especially in the rest-frame optical and NIR) is required to robustly recover the older stellar populations in the reconstructed SFHs \citep[e.g.][]{Papovich2001}. Therefore, even for the ``unlikely'' cases illustrated in Figure \ref{fig:visual_sfh} it is still possible that the galaxies in fact have older stellar populations which, however, cannot be recovered by our \prospector fitting because of wavelength coverage and data quality at the moment. This can lead to under-estimated \fmulsf. 

\begin{figure*}
    \centering
    \includegraphics[width=0.7\textwidth]{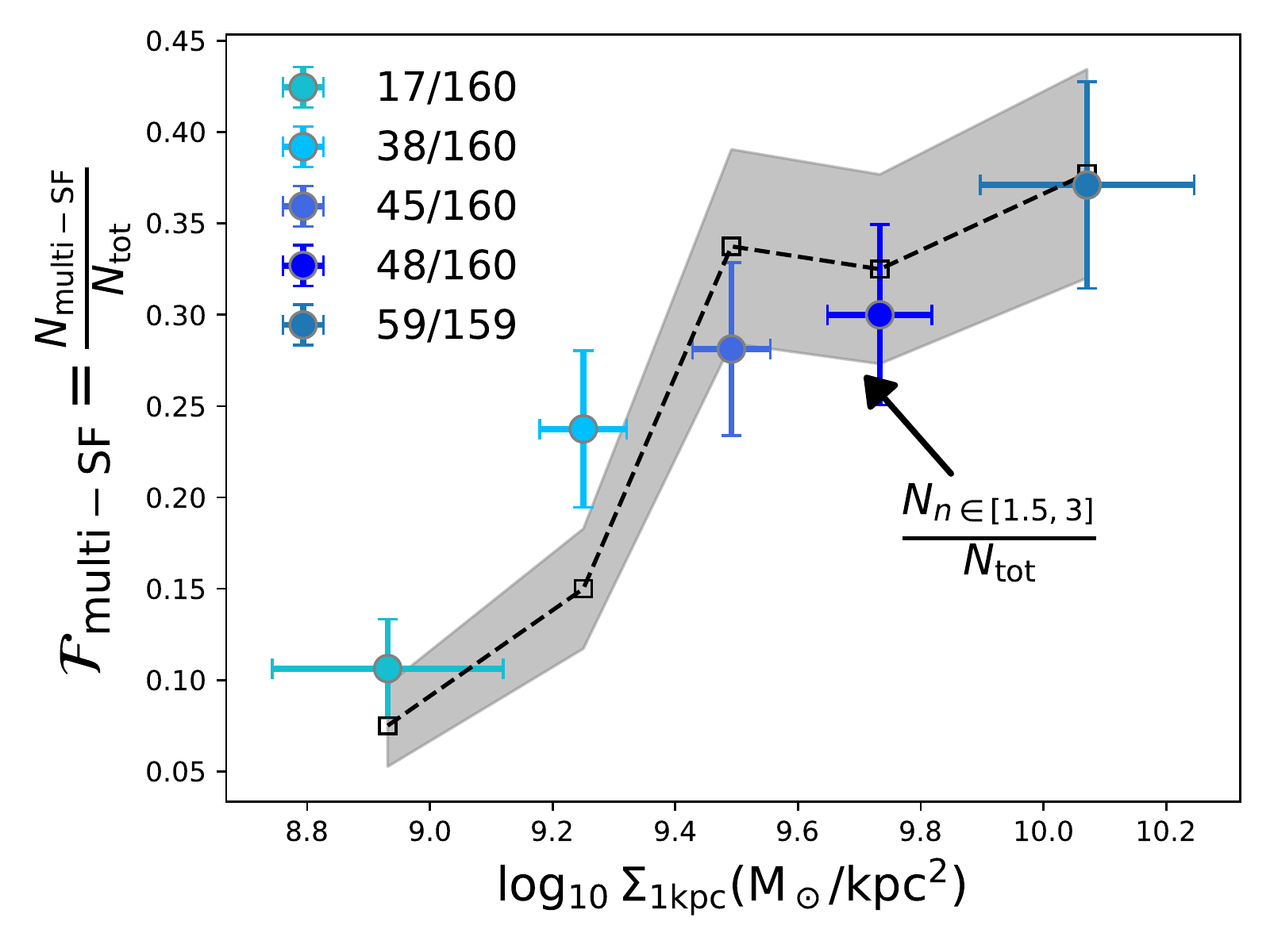}
    \caption{The relationship between \Sone and \fmulsf, i.e. the fraction of SFGs (i.e. \dms $\ge0.1$) showing multiple, prominent star-formation episodes in their reconstructed SFHs. We remind that in this plot only the ``secure'' cases (see Figure \ref{fig:visual_sfh} for illustration) are included, but the overall increasing trend does not change if the ``possible'' cases are also included. In the top-left corner we list in detail the number of galaxies in each bin of \Sone. Also plotted is the fraction of SFGs having \sersic index $n=1.5\sim3$, which is shown as the black dashed line and the grey shaded region that marks the 1$\sigma$ range. All uncertainties in the plot are estimated using the Poisson errors. This plot shows that the probability of finding SFGs that have a sizeable stellar mass in older stellar populations increases as the central stellar-mass surface density grows (see section \ref{sec:sfh_morp_sfg} for a detailed discussion).}
    \label{fig:F_mulSF}
\end{figure*}

The key result of this investigation is shown in Figure \ref{fig:F_mulSF}. We find that \fmulsf generally increases with \Sone, implying that the probability of finding SFGs that have a sizeable stellar mass in older stellar populations increases as their central stellar mass densities grow. The grey line in Figure \ref{fig:F_mulSF} illustrates the same point in a different way by plotting the probability of finding SFGs that have a \sersic index \n between 1.5 and 3, i.e. whose light profiles in the \H band indicate their morphology likely is a combination of a dense central core plus an extended outskirt, as a function of \Sone. We have checked that this trend does not change if we use a slightly different range of \n, e.g., between 1.5 and 2.5. We stress that the two trends of \Sone that we have just presented, one with \fmulsf and the other with the shape of light profiles rely on entirely independent observables, the former based on the SFH and the latter based on morphology. Yet we find that the two trends depict a consistent picture, which on the one hand points to the  substantial robustness of the SFH reconstructions, and on the other to the power of incorporating the SFHs to study the structural evolution of galaxies. 

\begin{figure}
    \centering
    \includegraphics[width=0.47\textwidth]{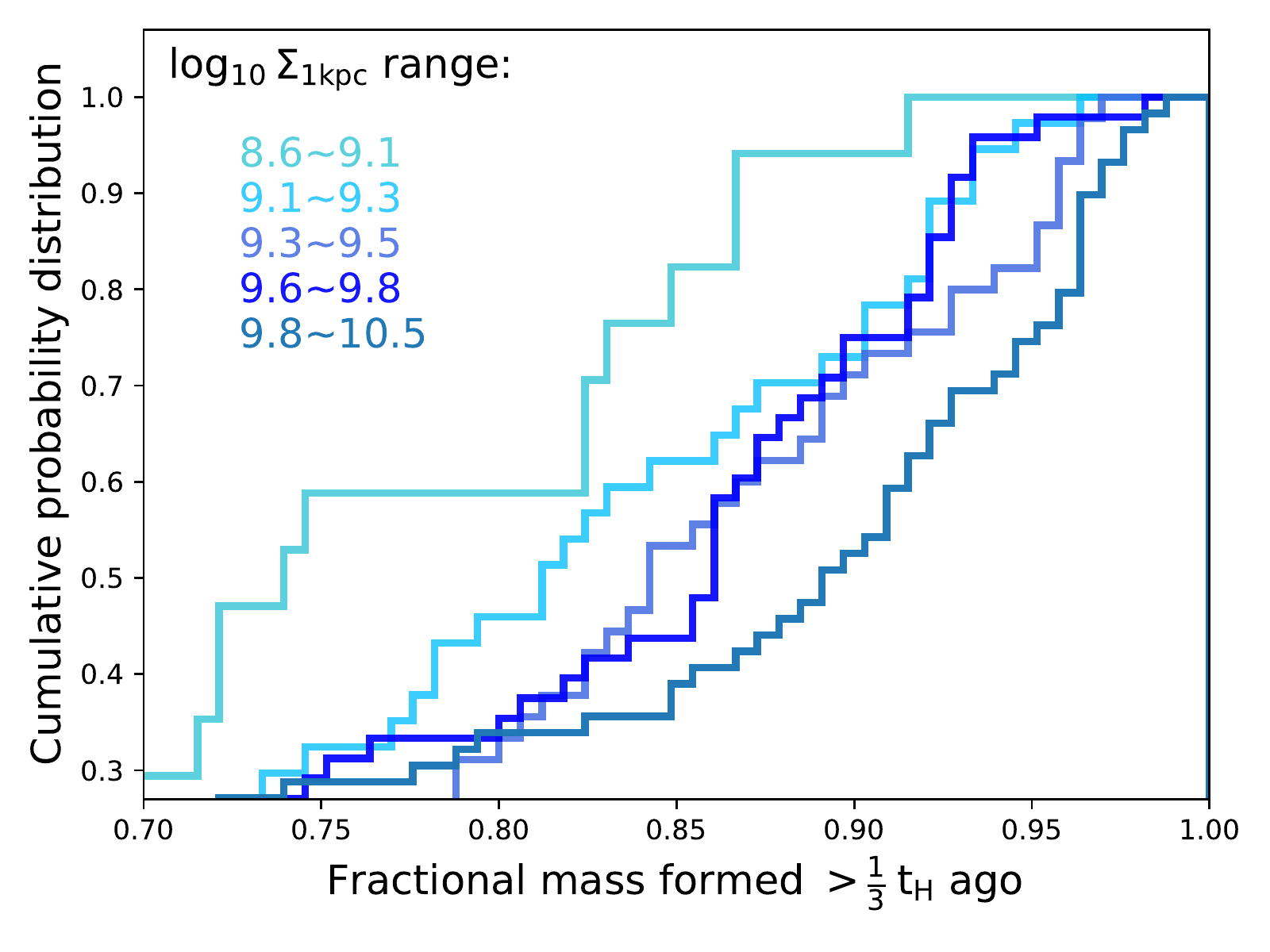}
    \caption{The cumulative distribution of fractional mass of the older stellar populations of the SFGs with multiple, prominent star-formation episodes. Here the older stellar populations are referred to those with age larger than one third of the Hubble Time at \zobs, i.e. $\frac{1}{3}\rm{t_H}$. For reference, $\frac{1}{3}\rm{t_H}$ corresponds to $\approx1$ Gyr at $z=2$. Galaxies with different \Sone are colored-coded using the same color scheme as Figure \ref{fig:F_mulSF} which is also labeled in the upper-left.}
    \label{fig:old_mass_frac}
\end{figure}

For the SFGs with multiple episodes of star formation, we have further studied the fractional mass of stellar populations assembled earlier, namely the mass of the stellar populations older than $\rm{\frac{1}{3}t_H}$ divided by the total stellar mass observed at \zobs. The choice of $\rm{\frac{1}{3}t_H}$ is based on results from  hydrodynamical simulations, which suggest that the characteristic timescale of galaxy compaction via dissipative processes at the redshift considered here is roughly a constant fraction ($0.3\sim0.4\times$) of the Hubble Time \citep[see section 5.2 of][]{Zolotov2015}. We have also checked our results by using the fixed 1 Gyr, and found no substantial changes in our results. Figure \ref{fig:old_mass_frac} shows the cumulative distribution of the fractional mass of older stellar populations in bins of \Sone. The fractional mass increases with \Sone. Taken together with the strong, increasing trend of \fmulsf with \Sone, this provides strong evidence that massive galaxies at cosmic noon assembled first their central regions which kept growing both in mass and density as the outer regions developed. These make the galaxies appear more compact and nucleated, and with a shrinking effective radius. This picture is broadly consistent with  conclusions reached by using independent observables such as the radial profile of SFR obtained from spatially-resolved maps of H$\alpha$ emission \citep[e.g.][]{Nelson2012,Nelson2016}. 

\section{Discussion}

Galaxies form and evolve accompanied by changes of their structural properties which, if their evolutionary tracks can be reconstructed, contain key empirical information on the physical drivers of structural evolution. Unlike cosmological simulations where the evolutionary path of galaxies and the contributions from different physical processes are known at any time, the morphology and other properties of real galaxies, however, can \textit{only} be known at the time of observation, \zobs. Here we propose the following methodology to statistically reconstruct the structural evolution of individual galaxies using their reconstructed SFHs. Because our method relies on good morphological measures which are only available at $z\lesssim3$ at the moment, only galaxies with \zf $\le 3$ are included to the following analysis. 

\subsection{The Method} \label{sec:diss_method}

The method is straightforward, and it is built upon two simple basic assumptions. Thanks to the \hst's high angular resolution, over the last two decades significant progress has been made in measuring the morphology of massive galaxies of all spectral types up to $z\sim3$ \citep[][just to name a few]{vanderWel2014,Shibuya2015,Mowla2019}. A common result from these studies is that, at least for massive galaxies, structural properties are strongly correlated with $M_*$ (e.g. mass-size relation, \citealt{Shen2003, vanderWel2014}), star-formation activity (e.g. SFR, \citealt{Salim2007,Elbaz2007, Whitaker2012}) and redshift. At least to first order, it is thus reasonable for us to assume that the structure of a galaxy is determined once its redshift, $M_*$ and SFR are known. But we also have to keep in mind substantial, non-negligible intrinsic scatters in such relationships exist. For example the intrinsic scatter of galaxy's mass-size relation has been found to be $\sim0.2$ dex at $0<z<3$ \citep{vanderWel2014}. This suggests the possibility that additional parameters also determine the structure of galaxies. Nonetheless, because little information on how additional parameters drive the scatter is available, here we ignore this issue entirely. The second assumption of the method is that existing galaxy samples observed at any given redshift are representative of the population of massive ($>10^{10}M_\sun$) galaxies at that redshift. We do not see any strong evidence that this assumption is grossly violated at the redshifts discussed here. We remind, however, that given the relatively small area covered by the three legacy fields considered in this study, the results could still be biased if significant cosmic variance exists, which remains to be tested with future larger-area surveys.

\begin{figure}
    \centering
    \includegraphics[width=0.47\textwidth]{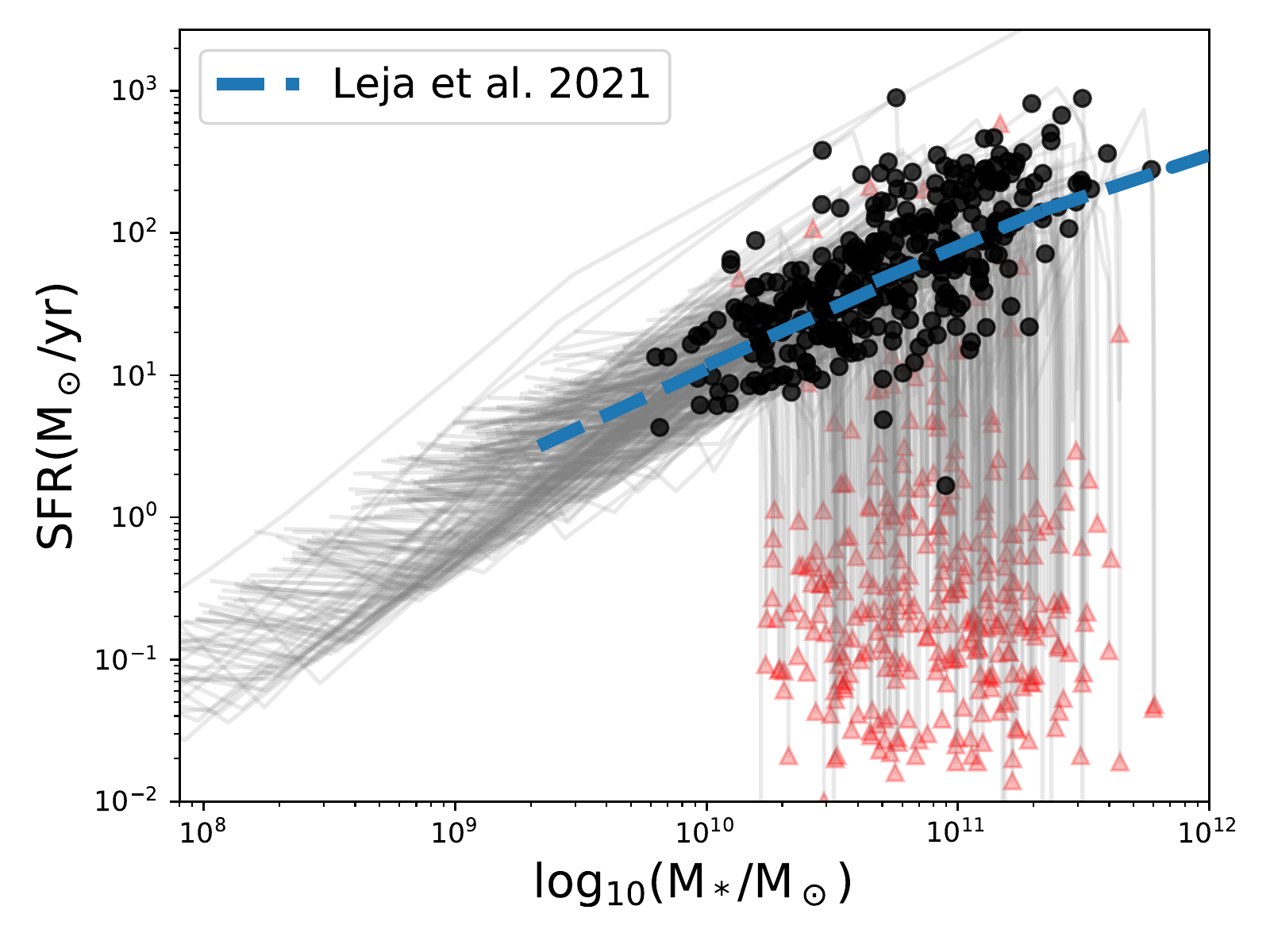}
    \caption{Predicting the star-forming main sequence using the reconstructed SFHs of the UVJ-selected QGs. The individual evolutionary tracks are plotted as the light grey lines. The red triangles show the $M_*$ and SFR of the galaxies at \zobs. The black circles show the $M_*$ and SFR of the galaxies at \zf as inferred by their reconstructed SFHs. The blue dashed line marks the star-forming main sequence of \citet{Leja2021} measured at \zf. Great agreement is seen between the measured and the predicted star-forming main sequence.}
    \label{fig:predict_ms}
\end{figure}

The method includes three steps. First, for any given galaxy  observed at \zobs we reconstruct its SFH and we use it to measure \zf and from this we obtain the galaxy's stellar mass and star-formation rate at \zf, i.e. $M_*^{\rm{z_{form}}}$ and SFR$\rm{^{z_{form}}}$. This first step requires accurate multi-band photometry (section \ref{sec:sample}) which is critical for the robust reconstruction of the SFH and thus of the evolutionary track of individual galaxies in the plane of SFR vs. $M_*$. About this point, we remind that the use of \prospector with nonparametric SFHs \citep{Leja2020} has essentially eliminated the long-standing tension between the observed cosmic star-formation rate density and the cosmic stellar-mass density, where the time integral of the former used to over-predict the latter by $\approx60\%$ \citep{Madau2014}. In Figure \ref{fig:predict_ms} we demonstrate this point in a more direct way using our SFH reconstructions. In particular, we use the SFHs of the UVJ-selected QGs in our sample to predict their distribution in the SFR-$M_*$ diagram at \zf, and then compare the distribution with the star-forming main sequence \textit{measured} at that redshift. Similarly to what we did in section \ref{sec:prosp_compare}, the star-forming main sequence used for the comparison is the one from \citet{Leja2021}. As Figure \ref{fig:predict_ms} shows, there is excellent agreement between the prediction from the reconstructed SFHs and the observed star-forming main sequence, demonstrating that \prospector is able to robustly reconstruct the evolutionary tracks of galaxies in the $M_*$ vs. SFR plane.

The second step is to select from the existing samples those galaxies whose (\zobs, $M_*$, SFR) are similar to the values of (\zf, $M_*^{\rm{z_{form}}}$, SFR$\rm{^{z_{form}}}$). We utilize the posteriors from individual \prospector fittings, and select all galaxies whose (\zobs, $M_*$, SFR) are within the 2$\sigma$ posterior contours of (\zf, $M_*^{\rm{z_{form}}}$, SFR$\rm{^{z_{form}}}$). We have also tested our results using 1$\sigma$ contours, and found no substantial changes. 

We remind that using different SED fitting procedures generally results in systematic shifts in the measures of physical properties of galaxies (section \ref{sec:prosp_compare}). Here we use the entire sample of this work as the default, because all physical properties are measured consistently, namely using the same SED fitting procedure and under the same SED assumptions. However, our selection criteria are very strict (section \ref{sec:sample}) because of our emphasis on the robustness of SFH reconstructions, which requires high-quality, densely sampled photometric data. Because all we need for this step are (\zobs, $M_*$, SFR), which compared to the measurement of SFHs are less sensitive to the data quality, it is possible to use an enlarged sample to increase the statistics at the potential cost of introducing systematics stemming from blending measures derived from different SED fitting procedures. Specifically, we have tested the robustness of our results using the enlarged samples of the CANDELS/COSMOS and GOODS fields where \textit{all} galaxies with reliable \galfit fitting results have been included. Because running \prospector is computationally quite expensive and most galaxies in the enlarged sample do not have available \prospector fitting results, we decide to empirically correct the existing measurements of $M_*$ and SFR from the enlarged CANDELS catalogs using the median systematic shifts found between previous measurements and the ones we obtained with \prospector (section \ref{sec:prosp_compare} and Figure \ref{fig:compare_pre}). To ensure a uniform correction, for all galaxies in the enlarged sample we use the SFRs estimated using the UV+IR ladder of \citet{Barro2019} and then correct them by -0.7 dex (the middle panel of Figure \ref{fig:compare_pre}), and the median stellar mass of the CANDELS catalogs using the Hodges–Lehmann method \citep{Santini2015} and then correct them by +0.3 dex (the left panel of Figure \ref{fig:compare_pre}). As one can see in the inset of Figure \ref{fig:time_evo_morph}, using the enlarged sample does not substantially change our results. In the following, we therefore choose the results from the sample of our own selection as the fiducial ones, although we do also report and compare the results from the enlarged sample in the following discussions.

The final step of our method is to compute the weighted average of the structural properties of the galaxies selected from the step two, where we use the corresponding posterior as the weights, and use them as the inferred structural properties of a galaxy at its formation epoch \zf. Accordingly, the uncertainties of the inferred structural properties are computed as the weighted standard deviation of the selected galaxies. By comparing the inferred properties at \zf with the observed ones at \zobs, we can get the evolutionary track of structural properties. 

Finally, before proceeding to discuss the results, we highlight the important differences of our method from earlier studies. Previous works have attempted to reconstruct the evolutionary trajectories of galaxies in planes defined by observables that include morphological (structural) properties, $M_*$ and star-formation properties \citep[e.g.][]{Barro2013,Barro2017}, with the ultimate goal of identifying the underlying physics that drives specific evolutionary phases. The method described above of reconstructing the evolutionary tracks provide another such attempt. What is different here, and provides a crucial advantage, is that we are able to statistically estimate the true tracks of the individual galaxies from the time of their formation, estimated by \zf, to that of their observation, i.e. \zobs, thanks to the reconstruction of SFHs without mixing together galaxies born at different times. This difference from previous works is key and an important step forward in the sense that we can avoid bundling together galaxies that formed at different epochs and experienced different assembly histories. This means that we can mitigate, in fact eliminate, any form of the progenitor effect solely based on empirical quantities. And the population averages, which define the general trends, are naturally obtained from the individual galaxies as they evolve, and are not predictions from models.

\subsection{The Structural Evolution of Galaxies Since \zf} \label{sec:diss_change}

\begin{figure*}
    \centering
    \includegraphics[width=1\textwidth]{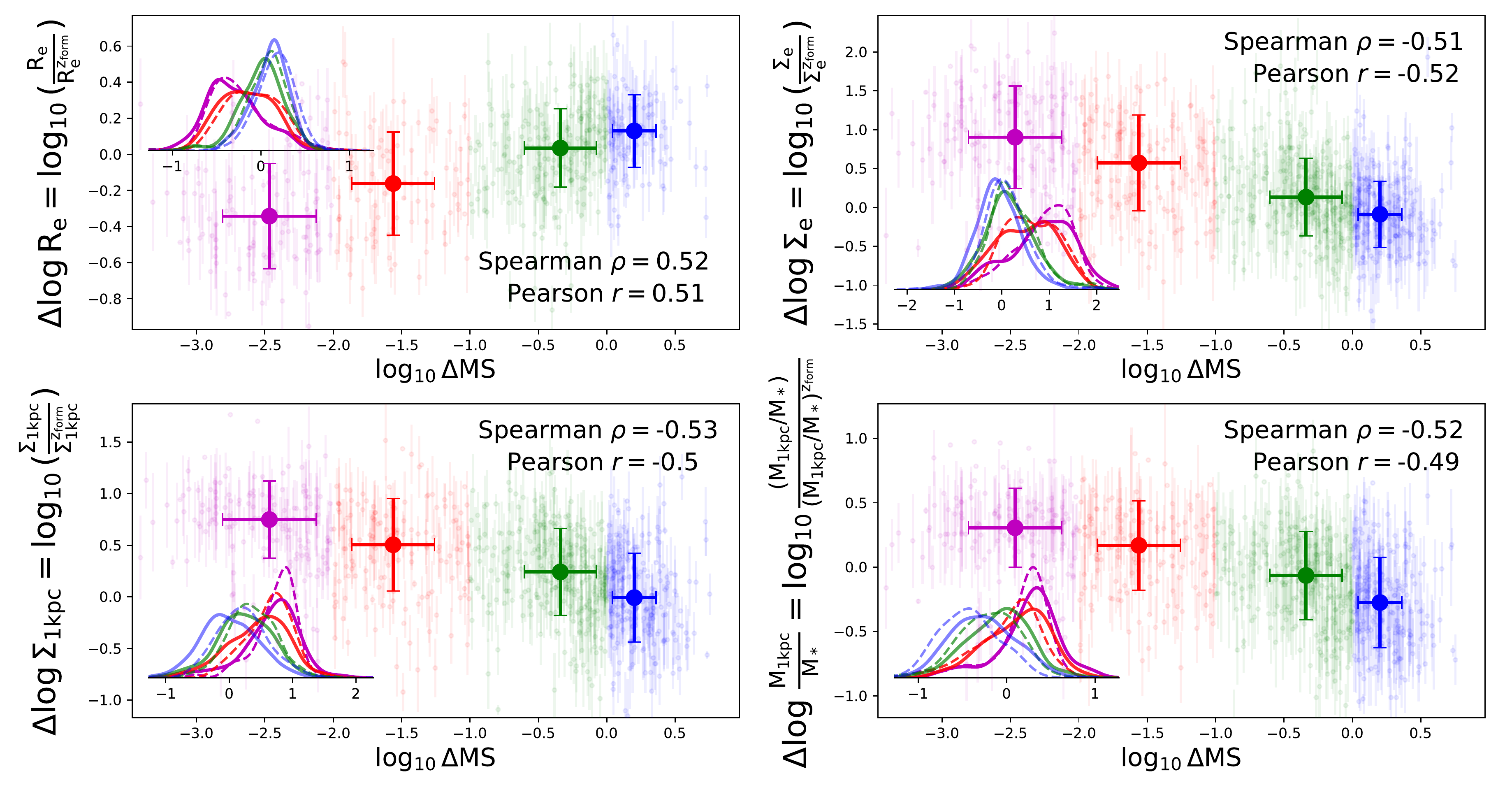}
    \caption{The inferred structural changes since \zf of \textit{individual} galaxies using the evolution of \re (upper-left), \Se (upper-right), \Sone (bottom-left) and \Mone (bottom-right) that we reconstruct with the SFHs (see section \ref{sec:diss_method} for detail). Galaxies are color coded according to \dms. Individual galaxies are plotted as the light-colored, small dots. The large dots with thick error bars show the medians and 1$\sigma$ ranges of galaxies within the individual bins of \dms. In the inset of each panel, we plot the probability density distribution of the corresponding morphological property, where the solid line shows our fiducial results and the dashed line shows the results of using an enlarged sample for our reconstruction of evolution of the structural property (section \ref{sec:diss_method}). Also labeled in each panel are the Spearman coefficient $\rho$ and the Pearson coefficient $r$ of the correlation tests between the corresponding structural change and the \dms. The inferred structural changes clearly depend on star-formation properties of galaxies.} 
    \label{fig:time_evo_morph}
\end{figure*}

Figure \ref{fig:time_evo_morph} shows the change of structural parameters, namely the observed structural properties at \zobs divided by the inferred ones at \zf, as a function of the star-formation property of individual galaxies. Specifically, the structural changes considered are: $\rm{\Delta log R_e}$, $\rm{\Delta log \Sigma_e}$, $\rm{\Delta log \Sigma_{1kpc}}$ and $\rm{\Delta log (M_{1kpc}/M_*)}$. We list the inferred median changes of these structural properties as a function of \dms in Table \ref{tab:struc_change}, where we also include the results from the enlarged sample as well. Also listed in Table \ref{tab:struc_change} are the uncertainties of the median values that we estimate by (1) bootstrapping the sample 1000 times and (2) during each bootstrapping using a normal distribution to resample the value of inferred structural properties at \zf with the estimated uncertainties that we described in the final step in section \ref{sec:diss_method}. 

\begin{table*}[]
    \centering
    \begin{tabular}{|c|c|c|c|c|}
    \hline
     \dms & $\rm{\Delta log\, R_e}$ & $\rm{\Delta log\, \Sigma_e}$ & $\rm{\Delta log\, \Sigma_{1kpc}}$ & $\rm{\Delta log \,(M_{1kpc}/M_*)}$ \\
    \hline
    $>1$ & $0.11\pm0.03$ ($0.17\pm0.03$) & $-0.07\pm0.04$ ($0.08\pm0.05$) & $-0.01\pm0.04$ ($0.21\pm0.03$) &  $-0.27\pm0.04$ ($-0.40\pm0.03$) \\
    $0.1-1$ & $0.02\pm 0.02$ ($0.07\pm 0.02$) & $0.15\pm 0.05$ ($0.22\pm 0.05$) & $0.24\pm 0.04$ ($0.36\pm0.03$) & $-0.08\pm 0.03$ ($-0.19\pm 0.03$) \\
    $0.01-0.1$ & $-0.19\pm 0.05$ ($-0.11\pm 0.04$) & $0.55\pm 0.09$ ($0.65\pm 0.07$) & $0.48\pm0.05$ ($0.65\pm 0.05$) & $0.14\pm 0.05$ ($0.05\pm 0.04$) \\
    $<0.01$ & $-0.34\pm 0.04$ ($-0.28\pm 0.04$) & $0.91\pm 0.08$ ($0.93\pm 0.07$) & $0.72\pm 0.05$ ($0.76\pm 0.04$) & $0.29\pm 0.04$ ($0.21\pm 0.04$) \\
    \hline
    \end{tabular}
    \caption{The median inferred structural changes since \zf for galaxies with different \dms (also see Figure \ref{fig:time_evo_morph}). The values in the parentheses are the results of using the enlarged sample for the reconstruction of structural evolution (section \ref{sec:diss_method}). The errors represent the uncertainties of the median.}
    \label{tab:struc_change}
\end{table*}

A clear trend, with an absolute correlation coefficient $\sim0.5$ of both the Spearman's rank and the Pearson tests, of the structural evolution since \zf with \dms is observed for all parameters considered. The $\rm{\Delta log R_e}$ increases with the \dms. Galaxies observed to have on-going star formation (i.e. large \dms) grew in size since \zf. For galaxies with suppressed star formation we see that \re decreases and the decrease continuously becomes more prominent as \dms becomes smaller, i.e. the galaxies become more quiescent: galaxies do shrink as they approach quiescence, at least when the size is measured by \re.

\begin{figure}
    \centering
    \includegraphics[width=0.47\textwidth]{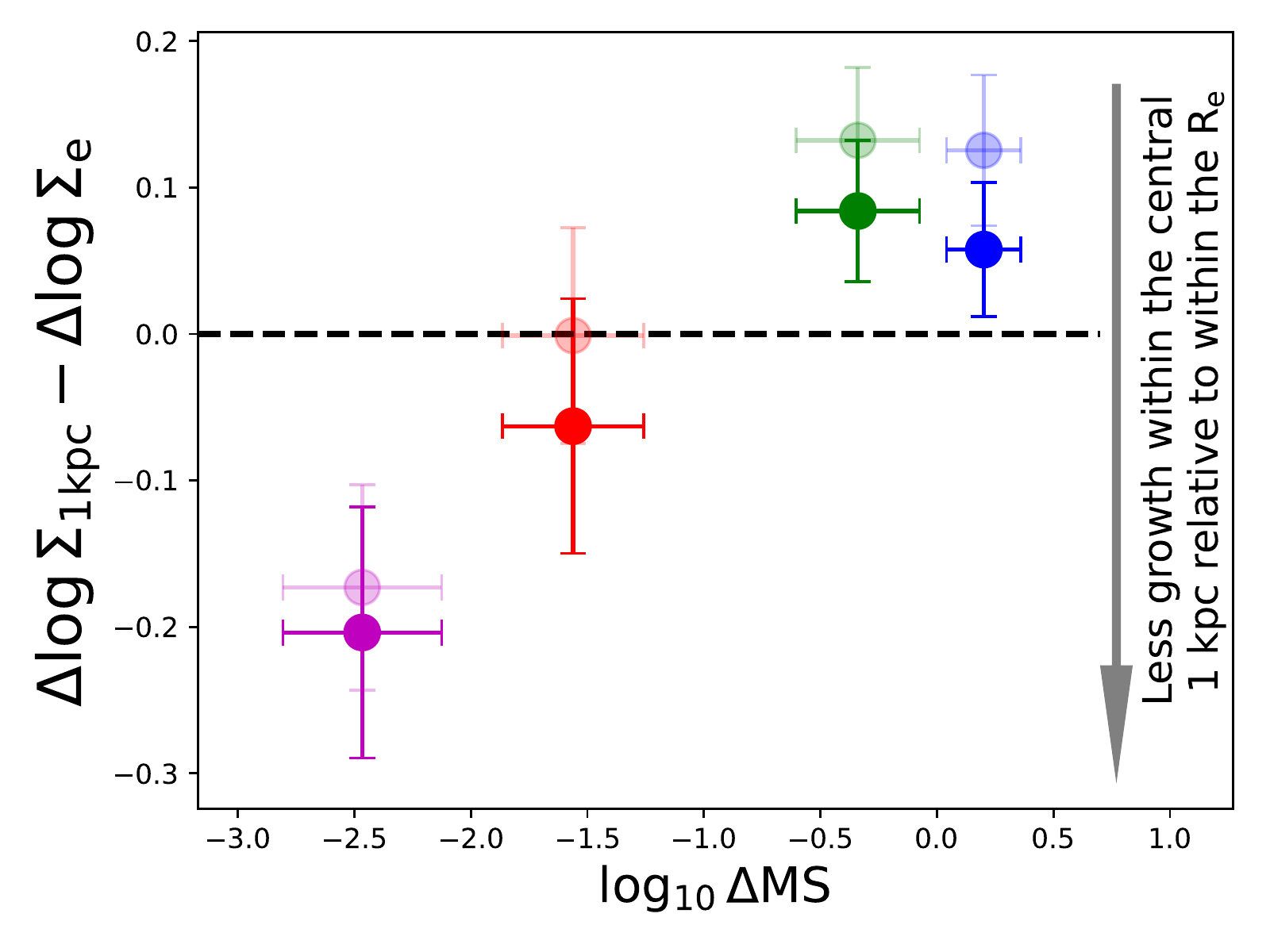}
    \caption{The difference between the inferred changes of \Sone and of \Se, i.e. $\rm{\Delta log \Sigma_{1kpc} - \Delta log \Sigma_{e}}$. If it equals to zero (the black horizontal dashed line), this means the changes of \Sone and of \Se are the same. The more negative the value is, the less growth took place within the central radius of 1 kpc compared to that within the \re since \zf. The dark-colored circles show the result of using the sample of our own selection for the reconstruction of the structural evolution, while the light-colored circles show the results of using the enlarged sample. The plot shows that, relative to the regions within \re, galaxies preferentially grow in mass density in the central regions of 1 kpc when they are still in star-forming phase, while the growth slows down as they become quiescent.}
    \label{fig:delta1kpc_deltare}
\end{figure}

In general, $\rm{\Delta log \Sigma_e}$, $\rm{\Delta log \Sigma_{1kpc}}$ and $\rm{\Delta log (M_{1kpc}/M_*)}$ decrease with increasing \dms. One the one hand, for galaxies above the star-forming main sequence, i.e. \dms$>1$, \Se does not significantly evolve since \zf. Depending on the exact sample selection (the step two described in section \ref{sec:diss_method}), \Sone either remains constant using the reference sample adopted in this work, or it exhibits a very mild increase of $\approx0.2$ dex if using the enlarged sample. On the other hand, since \zf, for galaxies with very little or no on-going star formation by the time of \zobs, \Se and \Sone significantly increase by $\approx0.9$ dex and $\approx0.7$ dex, respectively, regardless of the adopted  sample. More importantly, as Figure \ref{fig:delta1kpc_deltare} shows, we find that the inferred evolution of \Sone is larger than that of \Se for galaxies with \dms$>0.1$, while for those with \dms$<0.1$ the opposite is found, namely the $\rm{\Delta log \Sigma_e}$ is larger than the $\rm{\Delta log \Sigma_{1kpc}}$ and the difference, $\rm{\Delta log \Sigma_{1kpc}-\Delta log \Sigma_{e}}$, becomes more negative as the galaxies become more quiescent. 

The findings above have two important implications. First, galaxies grow in mass preferentially in the central regions as they are still in the star-forming phase, which explains $\rm{\Delta log \Sigma_{1kpc}\gtrsim\Delta log \Sigma_{e}}$ that we find for \dms$>0.1$ galaxies. This is consistent with what we found earlier in section \ref{sec:sfh_morp_sfg}, namely that compact SFGs are not only more likely to have a sizeable presence of older stellar populations (Figure \ref{fig:F_mulSF}) but also their fractional mass is larger (Figure \ref{fig:old_mass_frac}) than extended SFGs. We notice this conclusion is also consistent with that reached by \citet{Barro2017} (see their section 4) which, however, is based upon totally different assumptions. \citeauthor{Barro2017} showed that the scatter of the relationship between $\Sigma$ and $M_*$ is smaller than the expected growth in mass within the redshift intervals covered in their analysis. Because of this, they argued that it was reasonable to assume the observed $\Sigma$-$M_*$ relationship as a proxy for the evolutionary track of SFGs. Because the observed slope of the \Sone-$M_*$ relationship is steeper than that of the \Se-$M_*$ relationship, \citeauthor{Barro2017} concluded that high-redshift galaxies preferentially build up their centeral regions within 1 kpc. Our approach also remains entirely empirical, and includes the substantial improvement of getting rid of the key assumption of \citet{Barro2017}. Our approach uses the SFHs to statistically reconstruct the structural evolution of individual galaxies, including their central densities, from the data alone. It also avoids the progenitor effect stemming by avoiding to bundle together galaxies with different formation histories, which adds a spurious contribution to the intrinsic structural evolution, as we already discussed in section \ref{sec:progenitor}. 

The second important implication is that the growth in central stellar-mass surface density after galaxies go through their last major episode of star formation, i.e. after they go through quenching, is slower within the central radius of 1 kpc than within \re. We interpret this as a consequence of two effects. First, as they are still in the star-forming phase galaxies continuously grow their central stellar mass densities. When the quenching starts, the central density, e.g. \Sone, is about to reach the maximum value (i.e. the knee of the pistol pattern seen in Figure \ref{fig:pistol}), and its growth rate slows down. Meanwhile, after star formation quenches in the center, galaxies can still keep growing their masses and sizes in part through circum-nuclear, in-situ star formation, and in part through galaxy mergers \citep[e.g.][]{Bezanson2009,vanDokkum2010,Oser2012,Ji2022}. However, unlike the growth of the central regions which is likely fed by dissipative accretion of gas (see next section), this after-quenching growth also includes, and very likely is dominated by less dissipative processes such as minor mergers that are more likely to affect the outer regions. These two effects combined can lead to the observed $\rm{\Delta log \Sigma_{1kpc}\lesssim\Delta log \Sigma_{e}}$ that we find for the galaxies with small \dms. 

\subsection{On the Co-happening of Structural Transformations and Quenching} \label{sec:diss_imp}

The physics behind the empirically well-established correlations between galaxy structural and star-formation properties has long been eluded us. Critical, but yet open questions are whether a causal link exists between structural transformations and quenching, or whether they both are the results of a third, unidentified process, and what the relative timing of the two phenomena is \citep[e.g.][]{Khochfar2006,Wellons2015,Zolotov2015,Tacchella2016,Lilly2016}. 
To empirically disentangle the timing sequence and also to investigate the possibility of a direct causal link between quenching and structural transformations from other possibilities, the first key step is to eliminate any contribution from the progenitor effect \citep{Lilly2016, Ji2022a}, as we already explained in section \ref{sec:intro}, namely to select galaxies with similar \zf and assembly histories.

So far, we have used the SFHs to separate galaxies formed at different epochs and having different assembly histories. We have shown that, while the progenitor effect clearly exists and is observed (section \ref{sec:progenitor}), it alone is not enough to explain the observed correlations among \dms, \Sone, \Se, \Mone and \re (section \ref{sec:pistol} and \ref{sec:diss_change}), pointing to the possibility of a physical link between quenching and structural transformations. Our findings are in broad agreement with the theoretical study of \citet{Wellons2015}, where they found that both the physical compaction of galaxies and the progenitor effect are responsible for the formation of compact QGs in the Illustris cosmological simulations.

Generally speaking, galaxies at cosmic noon have much larger gas fractions than local ones, around 50\% and reaching up to $\approx 80$\% \citep{Tacconi2020}, and the rate of gas inflow is also higher \citep[e.g.][]{Dekel2013,Scoville2017}. As the result of direct accretion, dynamical instabilities or wet mergers, inflowing gas would sink to the center of gravitational potential, trigger central starbursts and build up dense central regions of galaxies, a process often referred to as wet compaction. Recent ALMA observations of galaxies at this epoch indeed found that the distribution of cold gas is more compact than that of stars \citep[e.g.][]{Barro2016,Spilker2016,Tadaki2017,Kaasinen2020,GomezGuijarro2022}, supporting the notion that high-redshift galaxies build up their centers through dissipative processes associated with cold gas, and hence adding empirical support to the scenario of wet compaction. 

\begin{figure*}
    \centering
    \includegraphics[width=1\textwidth]{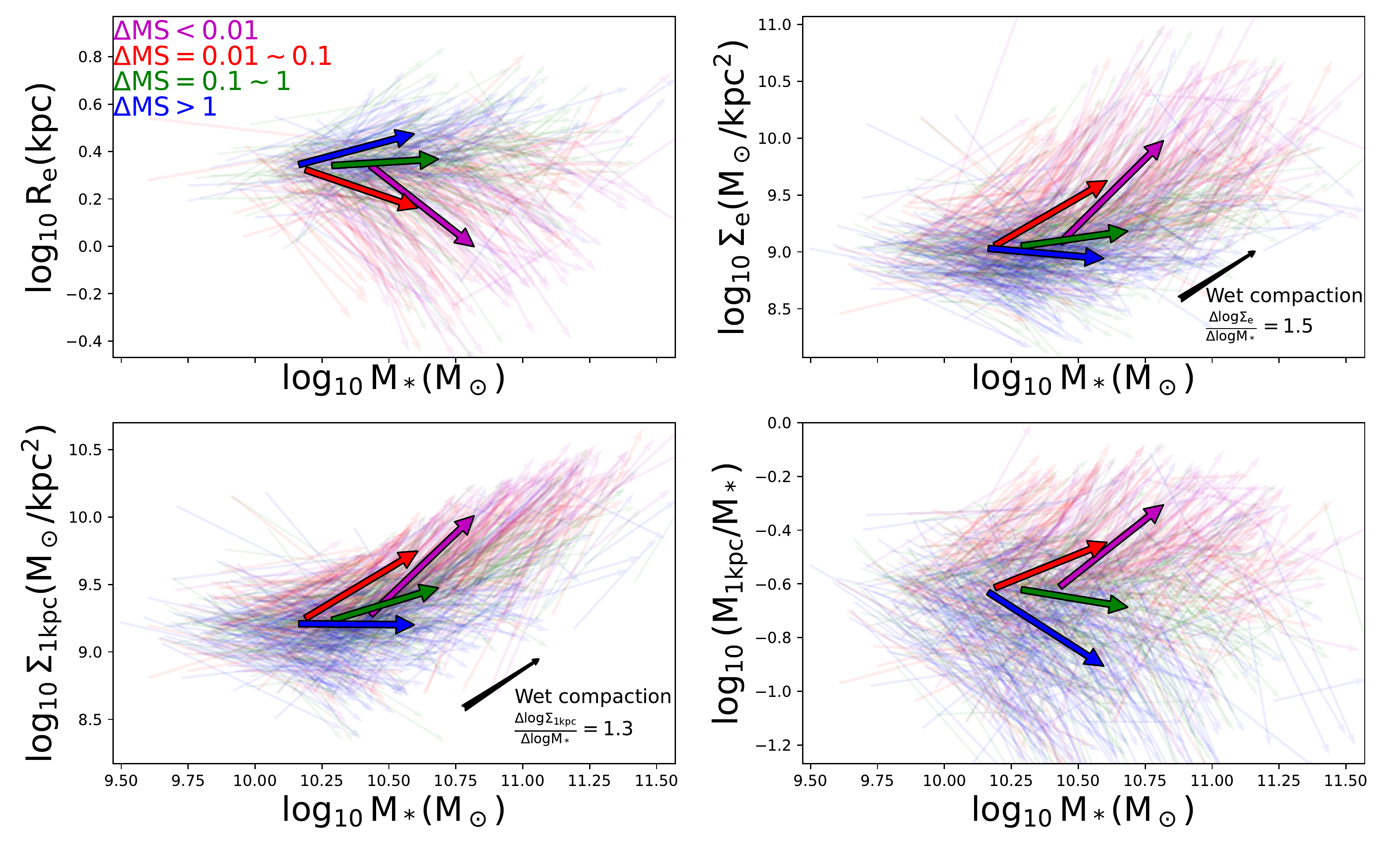}
    \caption{The inferred evolutionary vectors of the structure of \textit{individual} galaxies from \zf to \zobs. We use the same color scheme as Figure \ref{fig:time_evo_morph} to divide the galaxies into four groups according to their \dms. The evolutionary path of each galaxy is plotted as a light-colored arrow that starts from its inferred structural properties at \zf and ends at its observed properties at \zobs. The large, thick arrows show the median structural evolution of the groups of galaxies with different \dms. In the panels of \Se and \Sone, we also use the black arrow to show the inferred structural evolution due to the highly dissipative process, i.e. wet compaction, from \citet{Zolotov2015} and \citet{Barro2017}. We see great agreement between our purely empirical constraints of the structural transformation of galaxies and the predictions of wet compaction made by hydrodynamical simulations.}
    \label{fig:time_evo_diagrams}
\end{figure*}

\begin{figure}
    \centering
    \includegraphics[width=0.47\textwidth]{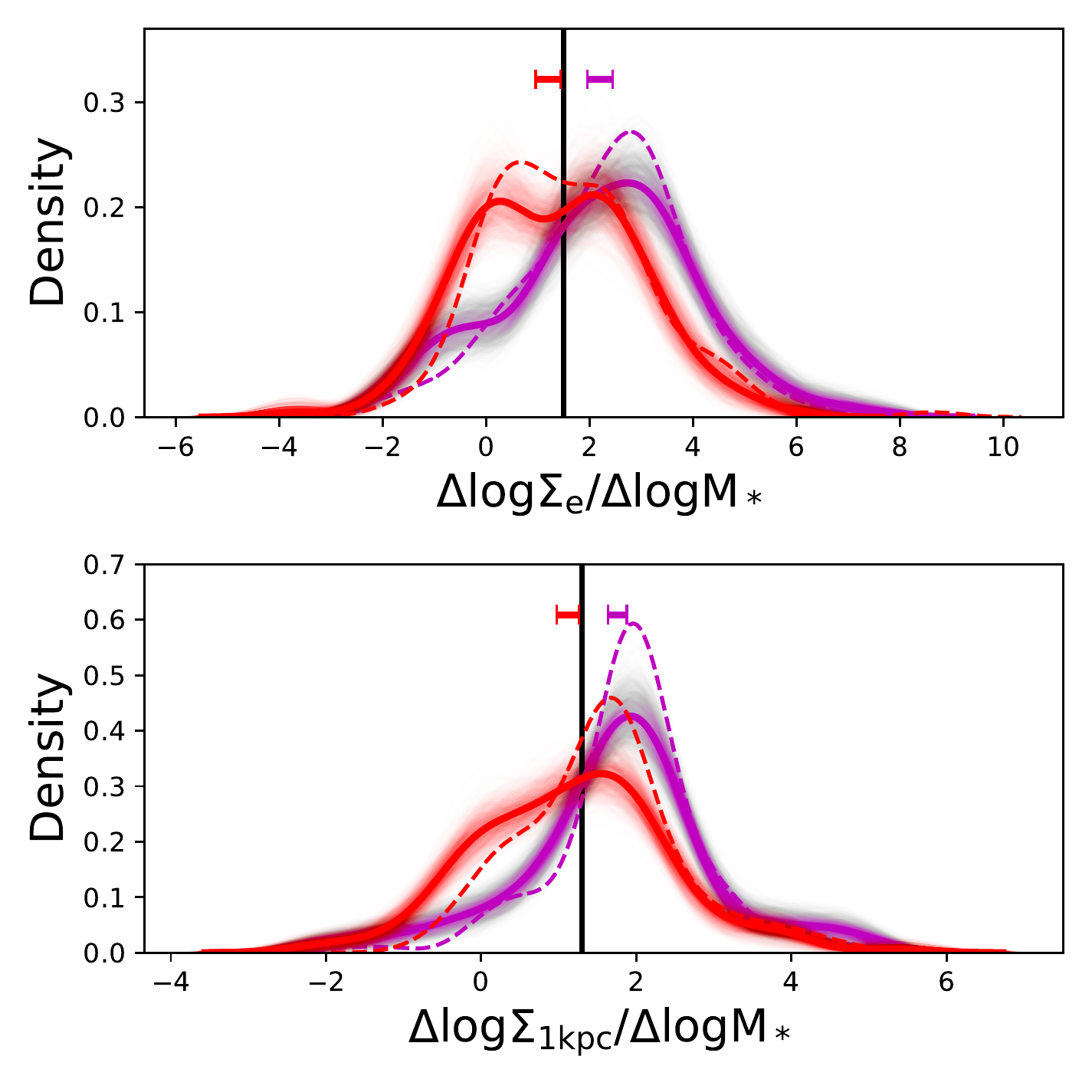}
    \caption{The distributions of $\rm{\Delta log\Sigma_e/\Delta logM_*}$ (upper) and $\rm{\Delta log\Sigma_{1kpc}/\Delta logM_*}$ (bottom), i.e. the slope of the evolutionary vector of Figure \ref{fig:time_evo_diagrams}, for the galaxies with \dms$<0.01$ (magenta) and \dms$=0.01-0.1$ (red). The dashed curves show the results of using the enlarged sample for the reconstruction of the structural evolution, and the solid curves show the results of using the sample of our own selection. The medians and their 1$\sigma$ uncertainties of the 1000-time realizations are marked as the horizontal bars (see section \ref{sec:diss_imp} for details). The vertical, black lines show the slopes predicted by the simulations of wet compaction, i.e. the same as those in Figure \ref{fig:time_evo_diagrams}. The plots show quantitative agreement between our empirical measurements and the predictions of hydrodynamical simulations of wet compaction.}
    \label{fig:slope_dist}
\end{figure}

Thanks to the reconstruction of the structural evolution of individual galaxies that we have described in section \ref{sec:diss_method} and \ref{sec:diss_change}, we are able to empirically derive their trajectories, from \zf to \zobs, in the planes of $M_*$ vs. structural property proxies, which are shown in Figure \ref{fig:time_evo_diagrams}. While there is a relatively large dispersion among the individual evolutionary vectors, the median trajectories gradually change as a function of \dms toward the direction expected from wet compaction processes and they are consistent with what we found in section \ref{sec:diss_change} and Figure \ref{fig:time_evo_morph}. The slope of the evolutionary vectors, which shows the relationship between the growth in stellar mass since \zf, i.e. $\rm{\Delta logM_*}$, and the change of structural properties, can be used to quantitatively compare with the predictions of theories. Intriguingly, Figure \ref{fig:time_evo_diagrams} shows that for galaxies with \dms$<0.1$ our reconstructed median evolutionary trajectory of the central density, both \Se (top-right panel) and \Sone (bottom-left panel), closely mirrors the one predicted by the simulations of wet compaction \citep{Zolotov2015}. We show this in a more quantitative way in Figure \ref{fig:slope_dist}, where we plot the distributions of the slope of the inferred evolutionary vectors, i.e. $\rm{\Delta log\Sigma_e/\Delta logM_*}$ and $\rm{\Delta log\Sigma_{1kpc}/\Delta logM_*}$, of galaxies with \dms$<0.1$. To estimate the uncertainties of the distributions, similarly to what we did in section \ref{sec:diss_change}, we bootstrap the sample 1000 times and during each bootstrapping we also resample the individual values using the measurement uncertainties. As Figure \ref{fig:slope_dist} shows, the median evolutionary slope measured by our method is in quantitative agreement with that predicted by the hydrodynamical simulations of wet compaction.

Although the consistency between our empirical constraints and the simulations shows wet compaction is a viable mechanism, however, we stress that we do not know what exact process(es) is responsible for the compaction. Physically,  wet compaction can be triggered by several physical processes, including galaxy mergers \citep{Hopkins2006}, counter-rotating streams \citep{Danovich2015}, recycled gas with low angular momentum \citep{Dekel2014} and violent disc instabilities \citep{Dekel2009}. A common feature of all these mechanisms, however, is that once the inflow of gas ends, it is the central regions that first become gas-depleted, halting further stellar mass growth (e.g. as traced by \Sone) leading to inside-out quenching \citep{Ceverino2015,Zolotov2015,Tacchella2016}. This scenario provides a broad explanation for the correlations observed among $M_*$, $\Sigma$ and SFR \citep{Barro2013,Barro2016,Barro2017}. It also is in good quantitative agreement with what we found in Figure \ref{fig:delta1kpc_deltare}. 

In conclusion, with the progenitor effect accounted for, the findings described here suggest a direct connection, at least in the form of simultaneous happening, of the process of quenching and the structural transformations. This temporal coincidence does not necessarily imply a causal link between these two phenomena. We suggest that this link is provided by the mechanisms that first triggers gas accretion into the central regions, with subsequent star formation and then shuts it down, eventually resulting in quenching. The accreted mass is large, and and it becomes compact (likely due to dissipative gaseous accretion), changing the dynamical state of the central region (Ji \& Giavalisco 2022, in preparation), and it also causes adiabatic contraction (Giavalisco \& Ji 2022, in preparation). As we have seen, the magnitude of the structural transformations that appear to take place is quantitatively consistent with the predictions from the wet compaction models, where this terminology is generically representative of a family of processes of growth of the central mass of galaxies through dissipative gaseous accretion. Regardless of the exact physical mechanism (or mechanisms)  behind compaction, one important conclusion of our study is that quenching takes place at the same time as the the formation of a compact, dense central core, which reaches its maximum density as star formation terminates. 

\section{Caveats }

We now explicitly address a few caveats of this study. First, the current nonparametric SFH reconstruction requires the assumption of a prior about how the SFR changes as a function of time on small timescales to enable the SED fitting to converge. Using different priors can lead to systematic shifts of the measurement of physical parameters. We have run a number of tests to constrain the magnitude of this uncertainty. As we already mentioned in section \ref{sec:prosp} and in \citetalias{Ji2022}, however, unfortunately at the moment we do not know what the optimal prior is, a situation that can hopefully be improved with future large, spectroscopic surveys at high redshifts. Luckily, as one can see in Appendix \ref{app:cont_diri}, tight correlations are observed between the parameters of interest to this study obtained from different priors. Therefore, there is evidence that the results of this work should not be overly sensitive to the assumed nonparametric SFH priors. Second, throughout the paper we used \zf as the proxy for the formation epoch of galaxies. This is purely from an empirical stand point, but there could be other better characteristic timescales (redshifts) to use. For example, in \citetalias{Ji2022} we argued that the best characteristic redshift to mitigate the progenitor effect should be the one that minimizes the scatter of the distribution of \Sone (see section 3.3 of \citetalias{Ji2022}). Perhaps due to the relatively small sample size and large scatter (both intrinsic and introduced by the current measurements), the best characteristic redshift to use remains inconclusive, and we plan to investigate this aspect further in the future. Third, in this work SED modeling was done with the integrated photometry, and we ignored possible color gradients inside the galaxies. Such gradients can make the light distribution different from the mass distribution, hence introduce systematics in our interpretations on the evolution of morphological properties. We plan to expand this study to include spatially resolved measurements using upcoming data from JWST. Finally, we remind that we ignore entirely the environmental effect on galaxy evolution in our interpretation of the observations, because in this work we only focus on massive galaxies whose evolution should be primarily driven by internal processes more than the external environment based on current observations, as  mentioned in section \ref{sec:intro}. That said, however, a detailed understanding of when and how the environmental effect on galaxy evolution becomes important remains incomplete. A few recent studies found evidence that the dense environment might even be able to affect the evolution of very massive galaxies ($\log_{10}(M_*/M_\sun)\ge10.5$) at cosmic noon by altering their molecular gas content \citep[e.g.,][]{Wang2018,Zavala2019}. However, these studies were based on a very limited number of known galaxy (proto-)clusters at $z>2$, statistical samples are still required to confirm these findings. Therefore, we cannot exclude that the environmental effect might still contribute to some part of the apparent structural evolution described here, and this will need further investigations in the future.

\section{Summary}

In this work we studied the relationships among the structural and star-formation properties, and mass assembly history of a carefully characterized sample of galaxies, with stellar mass $\log_{10}(M_*/M_\sun)\ge10.3$, at cosmic noon epoch \zobs$\sim2$. We have reconstructed high-fidelity, nonparametric SFHs of these galaxies using the fully Bayesian SED fitting code \prospector. We found strong correlations between the structural properties of galaxies and their assembly history. Specifically,
\begin{itemize}
    \item for QGs (section \ref{sec:sfh_morp_qg}), we found that the SFH of compact QGs, i.e. those having larger \Sone, \Se and \Mone, is similar to that of the average QG who has assembled most of its mass through an old burst of star formation and then its SFR gradually declined toward the quenching phase. The SFH of extended QGs, however, is very similar to the SFH of post-starburst galaxies, namely galaxies that have just experienced a recent major, massive burst of star formation $<1$ Gyr ago. Their morphology is also more disturbed compared to the compact ones.
    \item for SFGs (section \ref{sec:sfh_morp_sfg}), we found that as they become more compact, their SFHs are more likely to show multiple, prominent star-forming episodes. In addition, among those with multiple star-formation episodes, we found the fractional mass formed during the older episode, typically with an age $>1$ Gyr, also increases as the SFGs become more compact. 
\end{itemize}

With the availability of SFHs, we were able to separate galaxies formed at different epochs and with different assembly histories, hence to mitigate any form of the progenitor effect (e.g. the dependence of the size and central mass density on the epoch of formation) which otherwise biases the observed correlations between galaxies' structures and star-formation activities. We showed that
\begin{itemize}
    \item the progenitor effect is clearly observed in the sample galaxies discussed here (section \ref{sec:progenitor}). We showed that galaxies that formed earlier, i.e. with larger \zf, tend to have smaller sizes and larger densities.
    \item once the formation epoch (\zf) of the galaxies is controlled and accounted for, the distributions of galaxies in the diagrams of \dms, i.e. the distance from the star-forming main sequence, vs. various compactness metrics, including \Sone, \Se and \Mone, still exhibit a well-known pistol pattern (section \ref{sec:pistol}). Namely, as galaxies become quiescent they also become more compact, and the distributions of the compactness metrics also become narrower. These suggest the progenitor effect alone is not enough to explain the apparent correlations between the structural and star-formation properties of galaxies at cosmic noon.
\end{itemize}

Finally, we introduce a novel, purely empirical approach, which exploits the SFHs to reconstruct the evolution of structural properties of individual galaxies (section \ref{sec:diss_method}). Differently from earlier studies, our method naturally takes the assembly history of galaxies into account, and the population averages, which define the general trends, are obtained from the individual galaxies as they evolve, and are not predictions from models. We showed that since \zf
\begin{itemize}
    \item the change of structural properties are strongly correlated with \dms (section \ref{sec:diss_change}). In particular, for SFGs, i.e. galaxies with large \dms, we found that they grew in size, and their compactness (\Sone, \Se and \Mone) is consistent either with remaining constant or with a slight increase. QGs, i.e. those with small \dms, exhibit a significant decrease in their \re and increase in their compactness. Importantly, on the one hand, we found the change of \Se is smaller than that of \Sone for SFGs, suggesting galaxies preferentially grow the central regions as they are still in the star-forming phase.  On the other hand, an opposite trend, namely the change of \Se is larger than that of \Sone, was found for QGs, suggesting that galaxies, after they go through quenching, grow more slowly in their central radius of 1 kpc than they do within \re.
    \item the evolutionary vectors in the diagrams of stellar mass vs. structural properties enable us to compare with the predictions from cosmological simulations (section \ref{sec:diss_imp}). Our reconstructed evolutions of \Se and of \Sone for QGs are in quantitative agreement with the predictions of wet compaction made by simulations.
\end{itemize}

Combining all these results together, we converge to a consistent picture where the quenching of star formation in massive galaxies and their structural transformations in the form of decreasing \re and increasing central stellar-mass density take place roughly at the same time. To see this, first the progenitor effect must be accounted for when interpreting any correlation between the structural and star-formation properties. Second, the progenitor effect can only be partly responsible for the observed evolution, implying the existence of some physical link between galaxies' structural transformations and quenching. Our reconstructed structural evolution suggests that the wet compaction is one viable such link that helps build up the central dense regions of galaxies via highly dissipative gaseous accretion at cosmic noon epoch. After galaxies go through their final major episode of star formation and eventually become quiescent, the growth of their central regions (i.e. $<1$ kpc) slows down, while galaxy mergers can continue growing their outskirts. 

\section{ACKNOWLEDGMENT}

We thank the anonymous referee for their useful comments. This work was completed in part with resources provided by the University of Massachusetts' Green High Performance Computing Cluster (GHPCC).  

\software{GALFIT \citep{Peng2002,galfit}, Prospector \citep{Leja2017,Johnson2021}, FSPS \citep{Conroy2009,Conroy2010}, MIST \citep{Choi2016,Dotter2016}, MILES \citep{Falcon-Barroso2011}} 

\appendix

\section{Comparison of the physical parameters derived using the Dirichlet prior and the Continuity prior} \label{app:cont_diri}

\begin{figure*}
    \centering
    \includegraphics[width=0.97\textwidth]{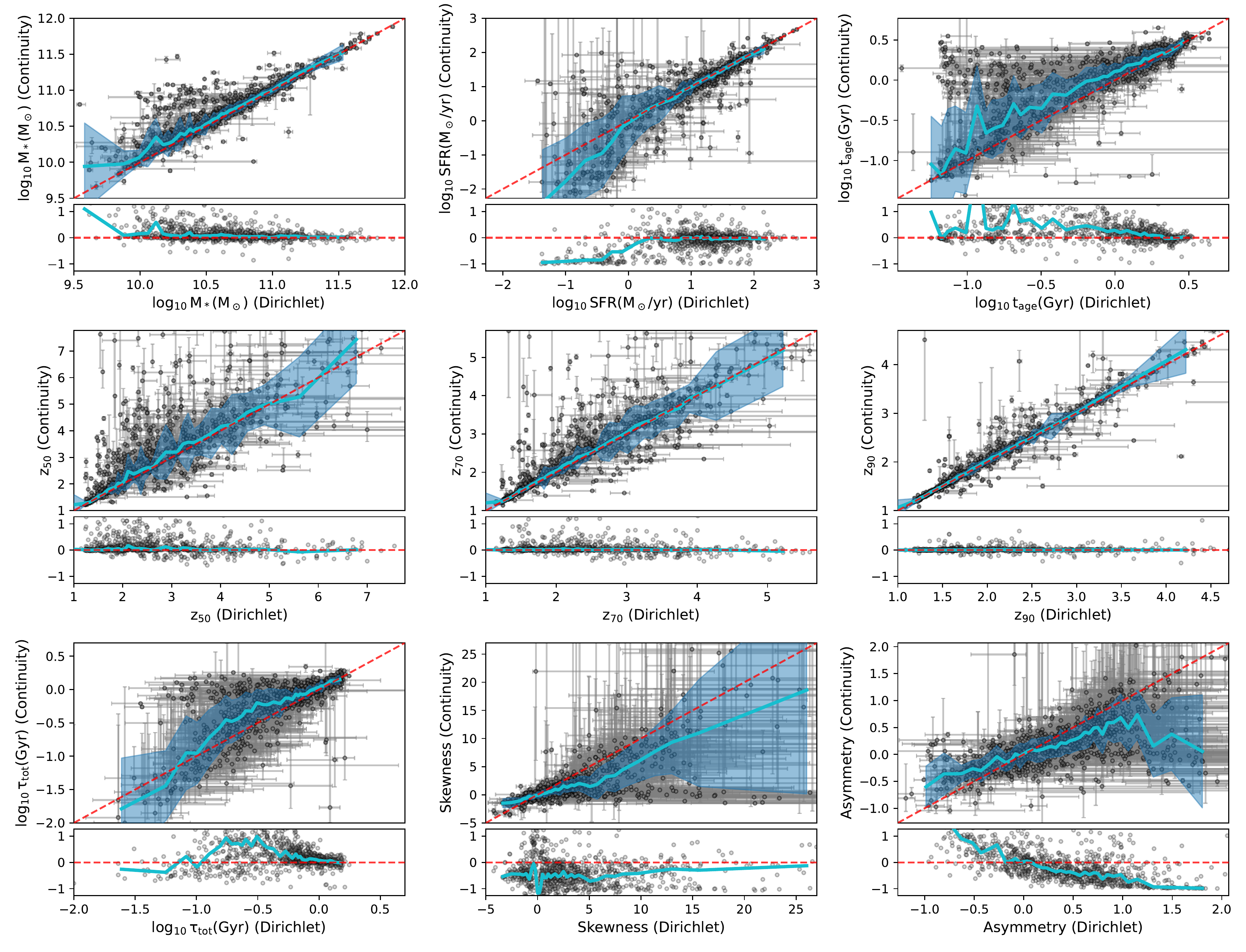}
    \caption{The comparisons of parameters derived using the Dirichlet (horizontal axis) and the Continuity (vertical axis) priors. Black dots with error bars show the individual measurements. In each panel, the cyan solid line with shaded region mark the median relationship and 1$\sigma$ range. The red dashed line shows the one-to-one relationship. The subpanel at the bottom shows the relative differences between the two measurements, i.e. (Continuity - Dirichlet) / Dirichlet.  }
    \label{fig:D_v_C}
\end{figure*}

In Figure \ref{fig:D_v_C} we compare the physical properties derived using the Dirichlet prior (the fiducial one in the main text) with the ones derived using the Continuity prior which unlike the former is strongly against the sharp change of SFR in adjacent lookback time bins. The comparison is not only for the basic properties such as $M_*$ and SFR, but also the parameters that we introduced in \citetalias{Ji2022} to quantify the shape of reconstructed SFHs. Systematic offsets and scatter notwithstanding, for all parameters we see strong correlations between the two measurements, suggesting that our results of this work should qualitatively not be sensitive to the prior assumed.

\section{The stacked SFH of individual galaxy groups divided using \Se and \Mone} \label{app:stack_SFH_other}

\begin{figure*}
    \centering
    \includegraphics[width=1\textwidth]{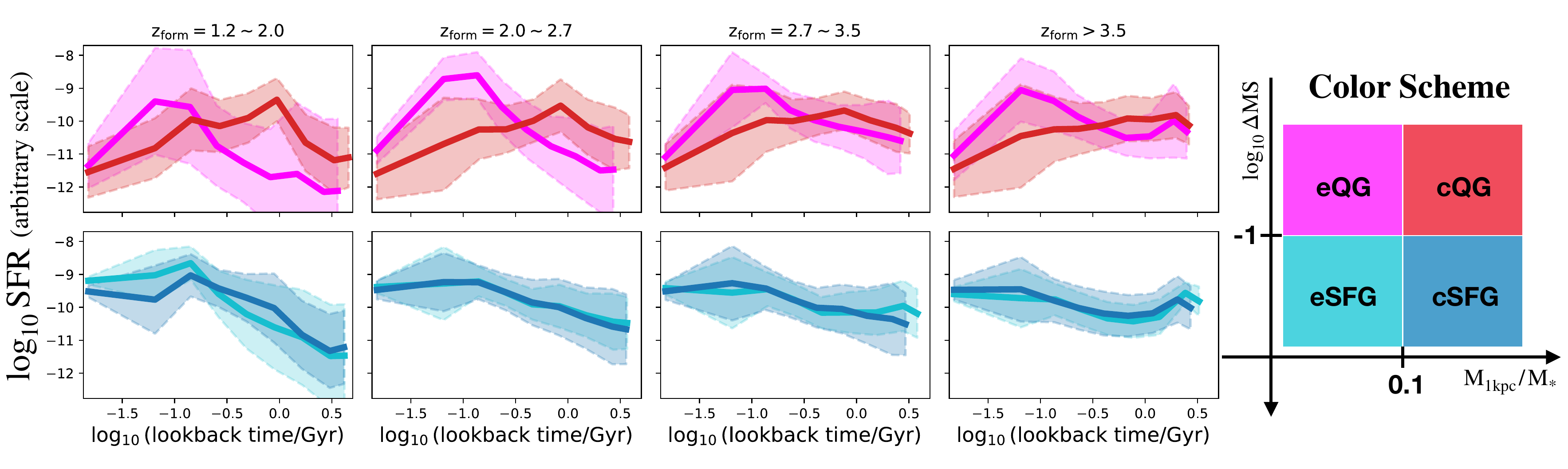}
    \includegraphics[width=1\textwidth]{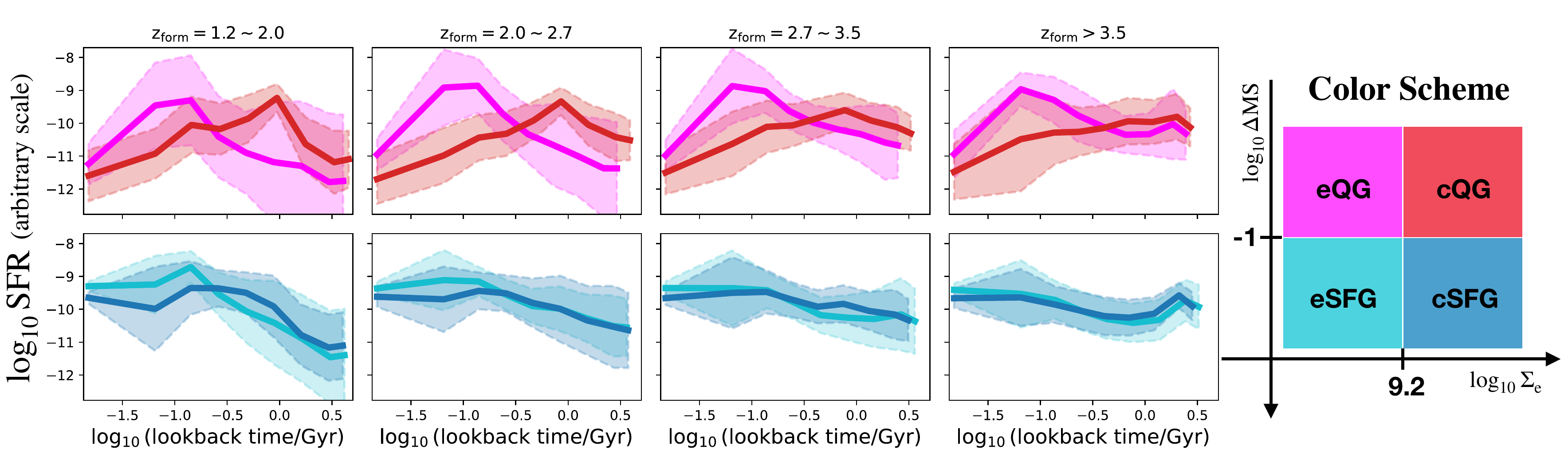}
    \caption{Similar to Figure \ref{fig:stack_SFH_S1}, but the extended and compact galaxies are classified according to \Mone (the top row) and \Se (the bottom row), respectively.}
    \label{fig:stack_SFH_other}
\end{figure*}

In section \ref{sec:morph_SFH} of the main text, we use \Sone to separate compact and extended galaxies. Here we use \Se and \Mone to do the separation, and then stack the SFHs of the individual groups of galaxies. The results are shown in Figure \ref{fig:stack_SFH_other} where no substantial change is found to our results.  

\bibliography{ji_2022_compaction_sfh}{}

\begin{thebibliography}{}
\expandafter\ifx\csname natexlab\endcsname\relax\def\natexlab#1{#1}\fi
\providecommand{\url}[1]{\href{#1}{#1}}
\providecommand{\dodoi}[1]{doi:~\href{http://doi.org/#1}{\nolinkurl{#1}}}
\providecommand{\doeprint}[1]{\href{http://ascl.net/#1}{\nolinkurl{http://ascl.net/#1}}}
\providecommand{\doarXiv}[1]{\href{https://arxiv.org/abs/#1}{\nolinkurl{https://arxiv.org/abs/#1}}}

\bibitem[{{Baldry} {et~al.}(2004){Baldry}, {Glazebrook}, {Brinkmann},
  {Ivezi{\'c}}, {Lupton}, {Nichol}, \& {Szalay}}]{Baldry2004}
{Baldry}, I.~K., {Glazebrook}, K., {Brinkmann}, J., {et~al.} 2004, \apj, 600,
  681, \dodoi{10.1086/380092}

\bibitem[{{Barro} {et~al.}(2013){Barro}, {Faber}, {P{\'e}rez-Gonz{\'a}lez},
  {Koo}, {Williams}, {Kocevski}, {Trump}, {Mozena}, {McGrath}, {van der Wel},
  {Wuyts}, {Bell}, {Croton}, {Ceverino}, {Dekel}, {Ashby}, {Cheung},
  {Ferguson}, {Fontana}, {Fang}, {Giavalisco}, {Grogin}, {Guo}, {Hathi},
  {Hopkins}, {Huang}, {Koekemoer}, {Kartaltepe}, {Lee}, {Newman}, {Porter},
  {Primack}, {Ryan}, {Rosario}, {Somerville}, {Salvato}, \& {Hsu}}]{Barro2013}
{Barro}, G., {Faber}, S.~M., {P{\'e}rez-Gonz{\'a}lez}, P.~G., {et~al.} 2013,
  \apj, 765, 104, \dodoi{10.1088/0004-637X/765/2/104}

\bibitem[{{Barro} {et~al.}(2016){Barro}, {Kriek}, {P{\'e}rez-Gonz{\'a}lez},
  {Trump}, {Koo}, {Faber}, {Dekel}, {Primack}, {Guo}, {Kocevski},
  {Mu{\~n}oz-Mateos}, {Rujopakarn}, \& {Seth}}]{Barro2016}
{Barro}, G., {Kriek}, M., {P{\'e}rez-Gonz{\'a}lez}, P.~G., {et~al.} 2016,
  \apjl, 827, L32, \dodoi{10.3847/2041-8205/827/2/L32}

\bibitem[{{Barro} {et~al.}(2017){Barro}, {Faber}, {Koo}, {Dekel}, {Fang},
  {Trump}, {P{\'e}rez-Gonz{\'a}lez}, {Pacifici}, {Primack}, {Somerville},
  {Yan}, {Guo}, {Liu}, {Ceverino}, {Kocevski}, \& {McGrath}}]{Barro2017}
{Barro}, G., {Faber}, S.~M., {Koo}, D.~C., {et~al.} 2017, \apj, 840, 47,
  \dodoi{10.3847/1538-4357/aa6b05}

\bibitem[{{Barro} {et~al.}(2019){Barro}, {P{\'e}rez-Gonz{\'a}lez}, {Cava},
  {Brammer}, {Pandya}, {Eliche Moral}, {Esquej}, {Dom{\'\i}nguez-S{\'a}nchez},
  {Alcalde Pampliega}, {Guo}, {Koekemoer}, {Trump}, {Ashby}, {Cardiel},
  {Castellano}, {Conselice}, {Dickinson}, {Dolch}, {Donley}, {Espino Briones},
  {Faber}, {Fazio}, {Ferguson}, {Finkelstein}, {Fontana}, {Galametz},
  {Gardner}, {Gawiser}, {Giavalisco}, {Grazian}, {Grogin}, {Hathi}, {Hemmati},
  {Hern{\'a}n-Caballero}, {Kocevski}, {Koo}, {Kodra}, {Lee}, {Lin}, {Lucas},
  {Mobasher}, {McGrath}, {Nandra}, {Nayyeri}, {Newman}, {Pforr}, {Peth},
  {Rafelski}, {Rodr{\'\i}guez-Munoz}, {Salvato}, {Stefanon}, {van der Wel},
  {Willner}, {Wiklind}, \& {Wuyts}}]{Barro2019}
{Barro}, G., {P{\'e}rez-Gonz{\'a}lez}, P.~G., {Cava}, A., {et~al.} 2019, \apjs,
  243, 22, \dodoi{10.3847/1538-4365/ab23f2}

\bibitem[{{Bell} {et~al.}(2004){Bell}, {Wolf}, {Meisenheimer}, {Rix}, {Borch},
  {Dye}, {Kleinheinrich}, {Wisotzki}, \& {McIntosh}}]{Bell2004}
{Bell}, E.~F., {Wolf}, C., {Meisenheimer}, K., {et~al.} 2004, \apj, 608, 752,
  \dodoi{10.1086/420778}

\bibitem[{{Belli} {et~al.}(2019){Belli}, {Newman}, \& {Ellis}}]{Belli2019}
{Belli}, S., {Newman}, A.~B., \& {Ellis}, R.~S. 2019, \apj, 874, 17,
  \dodoi{10.3847/1538-4357/ab07af}

\bibitem[{{Belli} {et~al.}(2017){Belli}, {Genzel}, {F{\"o}rster Schreiber},
  {Wisnioski}, {Wilman}, {Wuyts}, {Mendel}, {Beifiori}, {Bender}, {Brammer},
  {Burkert}, {Chan}, {Davies}, {Davies}, {Fabricius}, {Fossati}, {Galametz},
  {Lang}, {Lutz}, {Momcheva}, {Nelson}, {Saglia}, {Tacconi}, {Tadaki},
  {{\"U}bler}, \& {van Dokkum}}]{Belli2017}
{Belli}, S., {Genzel}, R., {F{\"o}rster Schreiber}, N.~M., {et~al.} 2017,
  \apjl, 841, L6, \dodoi{10.3847/2041-8213/aa70e5}

\bibitem[{{Bertin} \& {Arnouts}(1996)}]{sextractor}
{Bertin}, E., \& {Arnouts}, S. 1996, \aaps, 117, 393,
  \dodoi{10.1051/aas:1996164}

\bibitem[{{Bezanson} {et~al.}(2009){Bezanson}, {van Dokkum}, {Tal},
  {Marchesini}, {Kriek}, {Franx}, \& {Coppi}}]{Bezanson2009}
{Bezanson}, R., {van Dokkum}, P.~G., {Tal}, T., {et~al.} 2009, \apj, 697, 1290,
  \dodoi{10.1088/0004-637X/697/2/1290}

\bibitem[{{Brammer} {et~al.}(2009){Brammer}, {Whitaker}, {van Dokkum},
  {Marchesini}, {Labb{\'e}}, {Franx}, {Kriek}, {Quadri}, {Illingworth}, {Lee},
  {Muzzin}, \& {Rudnick}}]{Brammer2009}
{Brammer}, G.~B., {Whitaker}, K.~E., {van Dokkum}, P.~G., {et~al.} 2009, \apjl,
  706, L173, \dodoi{10.1088/0004-637X/706/1/L173}

\bibitem[{{Bruce} {et~al.}(2012){Bruce}, {Dunlop}, {Cirasuolo}, {McLure},
  {Targett}, {Bell}, {Croton}, {Dekel}, {Faber}, {Ferguson}, {Grogin},
  {Kocevski}, {Koekemoer}, {Koo}, {Lai}, {Lotz}, {McGrath}, {Newman}, \& {van
  der Wel}}]{Bruce2012}
{Bruce}, V.~A., {Dunlop}, J.~S., {Cirasuolo}, M., {et~al.} 2012, \mnras, 427,
  1666, \dodoi{10.1111/j.1365-2966.2012.22087.x}

\bibitem[{{Bundy} {et~al.}(2010){Bundy}, {Scarlata}, {Carollo}, {Ellis},
  {Drory}, {Hopkins}, {Salvato}, {Leauthaud}, {Koekemoer}, {Murray}, {Ilbert},
  {Oesch}, {Ma}, {Capak}, {Pozzetti}, \& {Scoville}}]{Bundy2010}
{Bundy}, K., {Scarlata}, C., {Carollo}, C.~M., {et~al.} 2010, \apj, 719, 1969,
  \dodoi{10.1088/0004-637X/719/2/1969}

\bibitem[{{Byler} {et~al.}(2017){Byler}, {Dalcanton}, {Conroy}, \&
  {Johnson}}]{Byler2017}
{Byler}, N., {Dalcanton}, J.~J., {Conroy}, C., \& {Johnson}, B.~D. 2017, \apj,
  840, 44, \dodoi{10.3847/1538-4357/aa6c66}

\bibitem[{{Calzetti} {et~al.}(2000){Calzetti}, {Armus}, {Bohlin}, {Kinney},
  {Koornneef}, \& {Storchi-Bergmann}}]{Calzetti2000}
{Calzetti}, D., {Armus}, L., {Bohlin}, R.~C., {et~al.} 2000, \apj, 533, 682,
  \dodoi{10.1086/308692}

\bibitem[{{Cappellari}(2013)}]{Cappellari2013}
{Cappellari}, M. 2013, \apjl, 778, L2, \dodoi{10.1088/2041-8205/778/1/L2}

\bibitem[{{Cardamone} {et~al.}(2010){Cardamone}, {van Dokkum}, {Urry},
  {Taniguchi}, {Gawiser}, {Brammer}, {Taylor}, {Damen}, {Treister}, {Cobb},
  {Bond}, {Schawinski}, {Lira}, {Murayama}, {Saito}, \&
  {Sumikawa}}]{Cardamone2010}
{Cardamone}, C.~N., {van Dokkum}, P.~G., {Urry}, C.~M., {et~al.} 2010, \apjs,
  189, 270, \dodoi{10.1088/0067-0049/189/2/270}

\bibitem[{{Carnall} {et~al.}(2019){Carnall}, {Leja}, {Johnson}, {McLure},
  {Dunlop}, \& {Conroy}}]{Carnall2019}
{Carnall}, A.~C., {Leja}, J., {Johnson}, B.~D., {et~al.} 2019, \apj, 873, 44,
  \dodoi{10.3847/1538-4357/ab04a2}

\bibitem[{{Carnall} {et~al.}(2018){Carnall}, {McLure}, {Dunlop}, \&
  {Dav{\'e}}}]{Carnall2018}
{Carnall}, A.~C., {McLure}, R.~J., {Dunlop}, J.~S., \& {Dav{\'e}}, R. 2018,
  \mnras, 480, 4379, \dodoi{10.1093/mnras/sty2169}

\bibitem[{{Ceverino} {et~al.}(2010){Ceverino}, {Dekel}, \&
  {Bournaud}}]{Ceverino2010}
{Ceverino}, D., {Dekel}, A., \& {Bournaud}, F. 2010, \mnras, 404, 2151,
  \dodoi{10.1111/j.1365-2966.2010.16433.x}

\bibitem[{{Ceverino} {et~al.}(2015){Ceverino}, {Dekel}, {Tweed}, \&
  {Primack}}]{Ceverino2015}
{Ceverino}, D., {Dekel}, A., {Tweed}, D., \& {Primack}, J. 2015, \mnras, 447,
  3291, \dodoi{10.1093/mnras/stu2694}

\bibitem[{{Chang} {et~al.}(2013){Chang}, {van der Wel}, {Rix}, {Holden},
  {Bell}, {McGrath}, {Wuyts}, {H{\"a}ussler}, {Barden}, {Faber}, {Mozena},
  {Ferguson}, {Guo}, {Galametz}, {Grogin}, {Kocevski}, {Koekemoer}, {Dekel},
  {Huang}, {Hathi}, \& {Donley}}]{Chang2013}
{Chang}, Y.-Y., {van der Wel}, A., {Rix}, H.-W., {et~al.} 2013, \apj, 773, 149,
  \dodoi{10.1088/0004-637X/773/2/149}

\bibitem[{{Chauke} {et~al.}(2019){Chauke}, {van der Wel}, {Pacifici},
  {Bezanson}, {Wu}, {Gallazzi}, {Straatman}, {Franx}, {Bari{\v{s}}i{\'c}},
  {Bell}, {van Houdt}, {Maseda}, {Muzzin}, {Sobral}, \& {Spilker}}]{Chauke2019}
{Chauke}, P., {van der Wel}, A., {Pacifici}, C., {et~al.} 2019, \apj, 877, 48,
  \dodoi{10.3847/1538-4357/ab164d}

\bibitem[{{Cheung} {et~al.}(2012){Cheung}, {Faber}, {Koo}, {Dutton}, {Simard},
  {McGrath}, {Huang}, {Bell}, {Dekel}, {Fang}, {Salim}, {Barro}, {Bundy},
  {Coil}, {Cooper}, {Conselice}, {Davis}, {Dom{\'\i}nguez}, {Kassin},
  {Kocevski}, {Koekemoer}, {Lin}, {Lotz}, {Newman}, {Phillips}, {Rosario},
  {Weiner}, \& {Willmer}}]{Cheung2012}
{Cheung}, E., {Faber}, S.~M., {Koo}, D.~C., {et~al.} 2012, \apj, 760, 131,
  \dodoi{10.1088/0004-637X/760/2/131}

\bibitem[{{Choi} {et~al.}(2016){Choi}, {Dotter}, {Conroy}, {Cantiello},
  {Paxton}, \& {Johnson}}]{Choi2016}
{Choi}, J., {Dotter}, A., {Conroy}, C., {et~al.} 2016, \apj, 823, 102,
  \dodoi{10.3847/0004-637X/823/2/102}

\bibitem[{{Ciesla} {et~al.}(2016){Ciesla}, {Boselli}, {Elbaz}, {Boissier},
  {Buat}, {Charmandaris}, {Schreiber}, {B{\'e}thermin}, {Baes}, {Boquien}, {De
  Looze}, {Fern{\'a}ndez-Ontiveros}, {Pappalardo}, {Spinoglio}, \&
  {Viaene}}]{Ciesla2016}
{Ciesla}, L., {Boselli}, A., {Elbaz}, D., {et~al.} 2016, \aap, 585, A43,
  \dodoi{10.1051/0004-6361/201527107}

\bibitem[{{Conroy}(2013)}]{Conroy2013}
{Conroy}, C. 2013, \araa, 51, 393, \dodoi{10.1146/annurev-astro-082812-141017}

\bibitem[{{Conroy} \& {Gunn}(2010)}]{Conroy2010}
{Conroy}, C., \& {Gunn}, J.~E. 2010, \apj, 712, 833,
  \dodoi{10.1088/0004-637X/712/2/833}

\bibitem[{{Conroy} {et~al.}(2009){Conroy}, {Gunn}, \& {White}}]{Conroy2009}
{Conroy}, C., {Gunn}, J.~E., \& {White}, M. 2009, \apj, 699, 486,
  \dodoi{10.1088/0004-637X/699/1/486}

\bibitem[{{Danovich} {et~al.}(2015){Danovich}, {Dekel}, {Hahn}, {Ceverino}, \&
  {Primack}}]{Danovich2015}
{Danovich}, M., {Dekel}, A., {Hahn}, O., {Ceverino}, D., \& {Primack}, J. 2015,
  \mnras, 449, 2087, \dodoi{10.1093/mnras/stv270}

\bibitem[{{Dekel} \& {Burkert}(2014)}]{Dekel2014}
{Dekel}, A., \& {Burkert}, A. 2014, \mnras, 438, 1870,
  \dodoi{10.1093/mnras/stt2331}

\bibitem[{{Dekel} {et~al.}(2009){Dekel}, {Sari}, \& {Ceverino}}]{Dekel2009}
{Dekel}, A., {Sari}, R., \& {Ceverino}, D. 2009, \apj, 703, 785,
  \dodoi{10.1088/0004-637X/703/1/785}

\bibitem[{{Dekel} {et~al.}(2013){Dekel}, {Zolotov}, {Tweed}, {Cacciato},
  {Ceverino}, \& {Primack}}]{Dekel2013}
{Dekel}, A., {Zolotov}, A., {Tweed}, D., {et~al.} 2013, \mnras, 435, 999,
  \dodoi{10.1093/mnras/stt1338}

\bibitem[{{Dotter}(2016)}]{Dotter2016}
{Dotter}, A. 2016, \apjs, 222, 8, \dodoi{10.3847/0067-0049/222/1/8}

\bibitem[{{Elbaz} {et~al.}(2007){Elbaz}, {Daddi}, {Le Borgne}, {Dickinson},
  {Alexander}, {Chary}, {Starck}, {Brandt}, {Kitzbichler}, {MacDonald},
  {Nonino}, {Popesso}, {Stern}, \& {Vanzella}}]{Elbaz2007}
{Elbaz}, D., {Daddi}, E., {Le Borgne}, D., {et~al.} 2007, \aap, 468, 33,
  \dodoi{10.1051/0004-6361:20077525}

\bibitem[{{Elbaz} {et~al.}(2011){Elbaz}, {Dickinson}, {Hwang},
  {D{\'\i}az-Santos}, {Magdis}, {Magnelli}, {Le Borgne}, {Galliano},
  {Pannella}, {Chanial}, {Armus}, {Charmandaris}, {Daddi}, {Aussel}, {Popesso},
  {Kartaltepe}, {Altieri}, {Valtchanov}, {Coia}, {Dannerbauer}, {Dasyra},
  {Leiton}, {Mazzarella}, {Alexander}, {Buat}, {Burgarella}, {Chary}, {Gilli},
  {Ivison}, {Juneau}, {Le Floc'h}, {Lutz}, {Morrison}, {Mullaney}, {Murphy},
  {Pope}, {Scott}, {Brodwin}, {Calzetti}, {Cesarsky}, {Charlot}, {Dole},
  {Eisenhardt}, {Ferguson}, {F{\"o}rster Schreiber}, {Frayer}, {Giavalisco},
  {Huynh}, {Koekemoer}, {Papovich}, {Reddy}, {Surace}, {Teplitz}, {Yun}, \&
  {Wilson}}]{Elbaz2011}
{Elbaz}, D., {Dickinson}, M., {Hwang}, H.~S., {et~al.} 2011, \aap, 533, A119,
  \dodoi{10.1051/0004-6361/201117239}

\bibitem[{{Elmegreen} {et~al.}(2007){Elmegreen}, {Elmegreen}, {Ravindranath},
  \& {Coe}}]{Elmegreen2007}
{Elmegreen}, D.~M., {Elmegreen}, B.~G., {Ravindranath}, S., \& {Coe}, D.~A.
  2007, \apj, 658, 763, \dodoi{10.1086/511667}

\bibitem[{{Estrada-Carpenter} {et~al.}(2020){Estrada-Carpenter}, {Papovich},
  {Momcheva}, {Brammer}, {Simons}, {Bridge}, {Cleri}, {Ferguson},
  {Finkelstein}, {Giavalisco}, {Jung}, {Matharu}, {Trump}, \&
  {Weiner}}]{EstradaCarpenter2020}
{Estrada-Carpenter}, V., {Papovich}, C., {Momcheva}, I., {et~al.} 2020, \apj,
  898, 171, \dodoi{10.3847/1538-4357/aba004}

\bibitem[{{Falc{\'o}n-Barroso} {et~al.}(2011){Falc{\'o}n-Barroso},
  {S{\'a}nchez-Bl{\'a}zquez}, {Vazdekis}, {Ricciardelli}, {Cardiel}, {Cenarro},
  {Gorgas}, \& {Peletier}}]{Falcon-Barroso2011}
{Falc{\'o}n-Barroso}, J., {S{\'a}nchez-Bl{\'a}zquez}, P., {Vazdekis}, A.,
  {et~al.} 2011, \aap, 532, A95, \dodoi{10.1051/0004-6361/201116842}

\bibitem[{{Fang} {et~al.}(2013){Fang}, {Faber}, {Koo}, \& {Dekel}}]{Fang2013}
{Fang}, J.~J., {Faber}, S.~M., {Koo}, D.~C., \& {Dekel}, A. 2013, \apj, 776,
  63, \dodoi{10.1088/0004-637X/776/1/63}

\bibitem[{{F{\"o}rster Schreiber} {et~al.}(2011){F{\"o}rster Schreiber},
  {Shapley}, {Genzel}, {Bouch{\'e}}, {Cresci}, {Davies}, {Erb}, {Genel},
  {Lutz}, {Newman}, {Shapiro}, {Steidel}, {Sternberg}, \&
  {Tacconi}}]{ForsterSchreiber2011}
{F{\"o}rster Schreiber}, N.~M., {Shapley}, A.~E., {Genzel}, R., {et~al.} 2011,
  \apj, 739, 45, \dodoi{10.1088/0004-637X/739/1/45}

\bibitem[{{Franx} {et~al.}(2008){Franx}, {van Dokkum}, {F{\"o}rster Schreiber},
  {Wuyts}, {Labb{\'e}}, \& {Toft}}]{Franx2008}
{Franx}, M., {van Dokkum}, P.~G., {F{\"o}rster Schreiber}, N.~M., {et~al.}
  2008, \apj, 688, 770, \dodoi{10.1086/592431}

\bibitem[{{Genzel} {et~al.}(2008){Genzel}, {Burkert}, {Bouch{\'e}}, {Cresci},
  {F{\"o}rster Schreiber}, {Shapley}, {Shapiro}, {Tacconi}, {Buschkamp},
  {Cimatti}, {Daddi}, {Davies}, {Eisenhauer}, {Erb}, {Genel}, {Gerhard},
  {Hicks}, {Lutz}, {Naab}, {Ott}, {Rabien}, {Renzini}, {Steidel}, {Sternberg},
  \& {Lilly}}]{Genzel2008}
{Genzel}, R., {Burkert}, A., {Bouch{\'e}}, N., {et~al.} 2008, \apj, 687, 59,
  \dodoi{10.1086/591840}

\bibitem[{{G{\'o}mez-Guijarro} {et~al.}(2022){G{\'o}mez-Guijarro}, {Elbaz},
  {Xiao}, {B{\'e}thermin}, {Franco}, {Magnelli}, {Daddi}, {Dickinson},
  {Demarco}, {Inami}, {Rujopakarn}, {Magdis}, {Shu}, {Chary}, {Zhou},
  {Alexander}, {Bournaud}, {Ciesla}, {Ferguson}, {Finkelstein}, {Giavalisco},
  {Iono}, {Juneau}, {Kartaltepe}, {Lagache}, {Le Floc'h}, {Leiton}, {Lin},
  {Motohara}, {Mullaney}, {Okumura}, {Pannella}, {Papovich}, {Pope}, {Sargent},
  {Silverman}, {Treister}, \& {Wang}}]{GomezGuijarro2022}
{G{\'o}mez-Guijarro}, C., {Elbaz}, D., {Xiao}, M., {et~al.} 2022, \aap, 658,
  A43, \dodoi{10.1051/0004-6361/202141615}

\bibitem[{{Grogin} {et~al.}(2011){Grogin}, {Kocevski}, {Faber}, {Ferguson},
  {Koekemoer}, {Riess}, {Acquaviva}, {Alexander}, {Almaini}, {Ashby}, {Barden},
  {Bell}, {Bournaud}, {Brown}, {Caputi}, {Casertano}, {Cassata}, {Castellano},
  {Challis}, {Chary}, {Cheung}, {Cirasuolo}, {Conselice}, {Roshan Cooray},
  {Croton}, {Daddi}, {Dahlen}, {Dav{\'e}}, {de Mello}, {Dekel}, {Dickinson},
  {Dolch}, {Donley}, {Dunlop}, {Dutton}, {Elbaz}, {Fazio}, {Filippenko},
  {Finkelstein}, {Fontana}, {Gardner}, {Garnavich}, {Gawiser}, {Giavalisco},
  {Grazian}, {Guo}, {Hathi}, {H{\"a}ussler}, {Hopkins}, {Huang}, {Huang},
  {Jha}, {Kartaltepe}, {Kirshner}, {Koo}, {Lai}, {Lee}, {Li}, {Lotz}, {Lucas},
  {Madau}, {McCarthy}, {McGrath}, {McIntosh}, {McLure}, {Mobasher},
  {Moustakas}, {Mozena}, {Nandra}, {Newman}, {Niemi}, {Noeske}, {Papovich},
  {Pentericci}, {Pope}, {Primack}, {Rajan}, {Ravindranath}, {Reddy}, {Renzini},
  {Rix}, {Robaina}, {Rodney}, {Rosario}, {Rosati}, {Salimbeni}, {Scarlata},
  {Siana}, {Simard}, {Smidt}, {Somerville}, {Spinrad}, {Straughn}, {Strolger},
  {Telford}, {Teplitz}, {Trump}, {van der Wel}, {Villforth}, {Wechsler},
  {Weiner}, {Wiklind}, {Wild}, {Wilson}, {Wuyts}, {Yan}, \& {Yun}}]{Grogin2011}
{Grogin}, N.~A., {Kocevski}, D.~D., {Faber}, S.~M., {et~al.} 2011, \apjs, 197,
  35, \dodoi{10.1088/0067-0049/197/2/35}

\bibitem[{{Guo} {et~al.}(2012){Guo}, {Giavalisco}, {Ferguson}, {Cassata}, \&
  {Koekemoer}}]{Guo2012}
{Guo}, Y., {Giavalisco}, M., {Ferguson}, H.~C., {Cassata}, P., \& {Koekemoer},
  A.~M. 2012, \apj, 757, 120, \dodoi{10.1088/0004-637X/757/2/120}

\bibitem[{{Guo} {et~al.}(2013){Guo}, {Ferguson}, {Giavalisco}, {Barro},
  {Willner}, {Ashby}, {Dahlen}, {Donley}, {Faber}, {Fontana}, {Galametz},
  {Grazian}, {Huang}, {Kocevski}, {Koekemoer}, {Koo}, {McGrath}, {Peth},
  {Salvato}, {Wuyts}, {Castellano}, {Cooray}, {Dickinson}, {Dunlop}, {Fazio},
  {Gardner}, {Gawiser}, {Grogin}, {Hathi}, {Hsu}, {Lee}, {Lucas}, {Mobasher},
  {Nandra}, {Newman}, \& {van der Wel}}]{Guo2013}
{Guo}, Y., {Ferguson}, H.~C., {Giavalisco}, M., {et~al.} 2013, \apjs, 207, 24,
  \dodoi{10.1088/0067-0049/207/2/24}

\bibitem[{{Guo} {et~al.}(2015){Guo}, {Ferguson}, {Bell}, {Koo}, {Conselice},
  {Giavalisco}, {Kassin}, {Lu}, {Lucas}, {Mandelker}, {McIntosh}, {Primack},
  {Ravindranath}, {Barro}, {Ceverino}, {Dekel}, {Faber}, {Fang}, {Koekemoer},
  {Noeske}, {Rafelski}, \& {Straughn}}]{Guo2015}
{Guo}, Y., {Ferguson}, H.~C., {Bell}, E.~F., {et~al.} 2015, \apj, 800, 39,
  \dodoi{10.1088/0004-637X/800/1/39}

\bibitem[{{Guo} {et~al.}(2017){Guo}, {Bell}, {Lu}, {Koo}, {Faber}, {Koekemoer},
  {Kurczynski}, {Lee}, {Papovich}, {Chen}, {Dekel}, {Ferguson}, {Fontana},
  {Giavalisco}, {Kocevski}, {Nayyeri}, {P{\'e}rez-Gonz{\'a}lez}, {Pforr},
  {Rodr{\'\i}guez-Puebla}, \& {Santini}}]{Guo2017}
{Guo}, Y., {Bell}, E.~F., {Lu}, Y., {et~al.} 2017, \apjl, 841, L22,
  \dodoi{10.3847/2041-8213/aa70e9}

\bibitem[{{Hopkins} {et~al.}(2008){Hopkins}, {Cox}, {Kere{\v{s}}}, \&
  {Hernquist}}]{Hopkins2008}
{Hopkins}, P.~F., {Cox}, T.~J., {Kere{\v{s}}}, D., \& {Hernquist}, L. 2008,
  \apjs, 175, 390, \dodoi{10.1086/524363}

\bibitem[{{Hopkins} {et~al.}(2006){Hopkins}, {Hernquist}, {Cox}, {Di Matteo},
  {Robertson}, \& {Springel}}]{Hopkins2006}
{Hopkins}, P.~F., {Hernquist}, L., {Cox}, T.~J., {et~al.} 2006, \apjs, 163, 1,
  \dodoi{10.1086/499298}

\bibitem[{{Hopkins} {et~al.}(2010){Hopkins}, {Bundy}, {Croton}, {Hernquist},
  {Keres}, {Khochfar}, {Stewart}, {Wetzel}, \& {Younger}}]{Hopkins2010}
{Hopkins}, P.~F., {Bundy}, K., {Croton}, D., {et~al.} 2010, \apj, 715, 202,
  \dodoi{10.1088/0004-637X/715/1/202}

\bibitem[{{Huang} {et~al.}(2013){Huang}, {Ho}, {Peng}, {Li}, \&
  {Barth}}]{Huang2013}
{Huang}, S., {Ho}, L.~C., {Peng}, C.~Y., {Li}, Z.-Y., \& {Barth}, A.~J. 2013,
  \apjl, 768, L28, \dodoi{10.1088/2041-8205/768/2/L28}

\bibitem[{{Huertas-Company} {et~al.}(2013){Huertas-Company}, {Mei}, {Shankar},
  {Delaye}, {Raichoor}, {Covone}, {Finoguenov}, {Kneib}, {Le}, \&
  {Povic}}]{HuertasCompany2013}
{Huertas-Company}, M., {Mei}, S., {Shankar}, F., {et~al.} 2013, \mnras, 428,
  1715, \dodoi{10.1093/mnras/sts150}

\bibitem[{{Huertas-Company} {et~al.}(2015){Huertas-Company}, {Gravet},
  {Cabrera-Vives}, {P{\'e}rez-Gonz{\'a}lez}, {Kartaltepe}, {Barro}, {Bernardi},
  {Mei}, {Shankar}, {Dimauro}, {Bell}, {Kocevski}, {Koo}, {Faber}, \&
  {Mcintosh}}]{HuertasCompany2015}
{Huertas-Company}, M., {Gravet}, R., {Cabrera-Vives}, G., {et~al.} 2015, \apjs,
  221, 8, \dodoi{10.1088/0067-0049/221/1/8}

\bibitem[{{Inoue} {et~al.}(2016){Inoue}, {Dekel}, {Mandelker}, {Ceverino},
  {Bournaud}, \& {Primack}}]{Inoue2016}
{Inoue}, S., {Dekel}, A., {Mandelker}, N., {et~al.} 2016, \mnras, 456, 2052,
  \dodoi{10.1093/mnras/stv2793}

\bibitem[{{Iyer} {et~al.}(2019){Iyer}, {Gawiser}, {Faber}, {Ferguson},
  {Kartaltepe}, {Koekemoer}, {Pacifici}, \& {Somerville}}]{Iyer2019}
{Iyer}, K.~G., {Gawiser}, E., {Faber}, S.~M., {et~al.} 2019, \apj, 879, 116,
  \dodoi{10.3847/1538-4357/ab2052}

\bibitem[{{Ji} \& {Giavalisco}(2022)}]{Ji2022}
{Ji}, Z., \& {Giavalisco}, M. 2022, arXiv e-prints, arXiv:2204.02414.
\newblock \doarXiv{2204.02414}

\bibitem[{{Ji} {et~al.}(2022){Ji}, {Giavalisco}, {Kirkpatrick}, {Kocevski},
  {Daddi}, {Delvecchio}, \& {Hatcher}}]{Ji2022a}
{Ji}, Z., {Giavalisco}, M., {Kirkpatrick}, A., {et~al.} 2022, \apj, 925, 74,
  \dodoi{10.3847/1538-4357/ac3837}

\bibitem[{{Ji} {et~al.}(2018){Ji}, {Giavalisco}, {Williams}, {Faber},
  {Ferguson}, {Guo}, {Liu}, \& {Lee}}]{Ji2018}
{Ji}, Z., {Giavalisco}, M., {Williams}, C.~C., {et~al.} 2018, \apj, 862, 135,
  \dodoi{10.3847/1538-4357/aacc2c}

\bibitem[{{Johnson} {et~al.}(2021){Johnson}, {Leja}, {Conroy}, \&
  {Speagle}}]{Johnson2021}
{Johnson}, B.~D., {Leja}, J., {Conroy}, C., \& {Speagle}, J.~S. 2021, \apjs,
  254, 22, \dodoi{10.3847/1538-4365/abef67}

\bibitem[{{Kaasinen} {et~al.}(2020){Kaasinen}, {Walter}, {Novak}, {Neeleman},
  {Smail}, {Boogaard}, {Cunha}, {Weiss}, {Liu}, {Decarli}, {Popping},
  {Diaz-Santos}, {Cort{\'e}s}, {Aravena}, {Werf}, {Riechers}, {Inami}, {Hodge},
  {Rix}, \& {Cox}}]{Kaasinen2020}
{Kaasinen}, M., {Walter}, F., {Novak}, M., {et~al.} 2020, \apj, 899, 37,
  \dodoi{10.3847/1538-4357/aba438}

\bibitem[{{Kauffmann} {et~al.}(2003){Kauffmann}, {Heckman}, {White}, {Charlot},
  {Tremonti}, {Peng}, {Seibert}, {Brinkmann}, {Nichol}, {SubbaRao}, \&
  {York}}]{Kauffmann2003}
{Kauffmann}, G., {Heckman}, T.~M., {White}, S. D.~M., {et~al.} 2003, \mnras,
  341, 54, \dodoi{10.1046/j.1365-8711.2003.06292.x}

\bibitem[{{Kawinwanichakij} {et~al.}(2017){Kawinwanichakij}, {Papovich},
  {Quadri}, {Glazebrook}, {Kacprzak}, {Allen}, {Bell}, {Croton}, {Dekel},
  {Ferguson}, {Forrest}, {Grogin}, {Guo}, {Kocevski}, {Koekemoer}, {Labb{\'e}},
  {Lucas}, {Nanayakkara}, {Spitler}, {Straatman}, {Tran}, {Tomczak}, \& {van
  Dokkum}}]{Kawinwanichakij2017}
{Kawinwanichakij}, L., {Papovich}, C., {Quadri}, R.~F., {et~al.} 2017, \apj,
  847, 134, \dodoi{10.3847/1538-4357/aa8b75}

\bibitem[{{Kennicutt}(1998)}]{Kennicutt1998}
{Kennicutt}, Robert~C., J. 1998, \araa, 36, 189,
  \dodoi{10.1146/annurev.astro.36.1.189}

\bibitem[{{Khochfar} \& {Silk}(2006)}]{Khochfar2006}
{Khochfar}, S., \& {Silk}, J. 2006, \apjl, 648, L21, \dodoi{10.1086/507768}

\bibitem[{{Koekemoer} {et~al.}(2011){Koekemoer}, {Faber}, {Ferguson}, {Grogin},
  {Kocevski}, {Koo}, {Lai}, {Lotz}, {Lucas}, {McGrath}, {Ogaz}, {Rajan},
  {Riess}, {Rodney}, {Strolger}, {Casertano}, {Castellano}, {Dahlen},
  {Dickinson}, {Dolch}, {Fontana}, {Giavalisco}, {Grazian}, {Guo}, {Hathi},
  {Huang}, {van der Wel}, {Yan}, {Acquaviva}, {Alexander}, {Almaini}, {Ashby},
  {Barden}, {Bell}, {Bournaud}, {Brown}, {Caputi}, {Cassata}, {Challis},
  {Chary}, {Cheung}, {Cirasuolo}, {Conselice}, {Roshan Cooray}, {Croton},
  {Daddi}, {Dav{\'e}}, {de Mello}, {de Ravel}, {Dekel}, {Donley}, {Dunlop},
  {Dutton}, {Elbaz}, {Fazio}, {Filippenko}, {Finkelstein}, {Frazer}, {Gardner},
  {Garnavich}, {Gawiser}, {Gruetzbauch}, {Hartley}, {H{\"a}ussler},
  {Herrington}, {Hopkins}, {Huang}, {Jha}, {Johnson}, {Kartaltepe},
  {Khostovan}, {Kirshner}, {Lani}, {Lee}, {Li}, {Madau}, {McCarthy},
  {McIntosh}, {McLure}, {McPartland}, {Mobasher}, {Moreira}, {Mortlock},
  {Moustakas}, {Mozena}, {Nandra}, {Newman}, {Nielsen}, {Niemi}, {Noeske},
  {Papovich}, {Pentericci}, {Pope}, {Primack}, {Ravindranath}, {Reddy},
  {Renzini}, {Rix}, {Robaina}, {Rosario}, {Rosati}, {Salimbeni}, {Scarlata},
  {Siana}, {Simard}, {Smidt}, {Snyder}, {Somerville}, {Spinrad}, {Straughn},
  {Telford}, {Teplitz}, {Trump}, {Vargas}, {Villforth}, {Wagner}, {Wandro},
  {Wechsler}, {Weiner}, {Wiklind}, {Wild}, {Wilson}, {Wuyts}, \&
  {Yun}}]{Koekemoer2011}
{Koekemoer}, A.~M., {Faber}, S.~M., {Ferguson}, H.~C., {et~al.} 2011, \apjs,
  197, 36, \dodoi{10.1088/0067-0049/197/2/36}

\bibitem[{{Kriek} {et~al.}(2009){Kriek}, {van Dokkum}, {Franx}, {Illingworth},
  \& {Magee}}]{Kriek2009}
{Kriek}, M., {van Dokkum}, P.~G., {Franx}, M., {Illingworth}, G.~D., \&
  {Magee}, D.~K. 2009, \apjl, 705, L71, \dodoi{10.1088/0004-637X/705/1/L71}

\bibitem[{{Kroupa}(2001)}]{Kroupa2001}
{Kroupa}, P. 2001, \mnras, 322, 231, \dodoi{10.1046/j.1365-8711.2001.04022.x}

\bibitem[{{Lang} {et~al.}(2014){Lang}, {Wuyts}, {Somerville}, {F{\"o}rster
  Schreiber}, {Genzel}, {Bell}, {Brammer}, {Dekel}, {Faber}, {Ferguson},
  {Grogin}, {Kocevski}, {Koekemoer}, {Lutz}, {McGrath}, {Momcheva}, {Nelson},
  {Primack}, {Rosario}, {Skelton}, {Tacconi}, {van Dokkum}, \&
  {Whitaker}}]{Lang2014}
{Lang}, P., {Wuyts}, S., {Somerville}, R.~S., {et~al.} 2014, \apj, 788, 11,
  \dodoi{10.1088/0004-637X/788/1/11}

\bibitem[{{Lee} {et~al.}(2018){Lee}, {Giavalisco}, {Whitaker}, {Williams},
  {Ferguson}, {Acquaviva}, {Koekemoer}, {Straughn}, {Guo}, {Kartaltepe},
  {Lotz}, {Pacifici}, {Croton}, {Somerville}, \& {Lu}}]{Lee2018}
{Lee}, B., {Giavalisco}, M., {Whitaker}, K., {et~al.} 2018, \apj, 853, 131,
  \dodoi{10.3847/1538-4357/aaa40f}

\bibitem[{{Leja} {et~al.}(2019){Leja}, {Carnall}, {Johnson}, {Conroy}, \&
  {Speagle}}]{Leja2019}
{Leja}, J., {Carnall}, A.~C., {Johnson}, B.~D., {Conroy}, C., \& {Speagle},
  J.~S. 2019, \apj, 876, 3, \dodoi{10.3847/1538-4357/ab133c}

\bibitem[{{Leja} {et~al.}(2018){Leja}, {Johnson}, {Conroy}, \& {van
  Dokkum}}]{Leja2018}
{Leja}, J., {Johnson}, B.~D., {Conroy}, C., \& {van Dokkum}, P. 2018, \apj,
  854, 62, \dodoi{10.3847/1538-4357/aaa8db}

\bibitem[{{Leja} {et~al.}(2017){Leja}, {Johnson}, {Conroy}, {van Dokkum}, \&
  {Byler}}]{Leja2017}
{Leja}, J., {Johnson}, B.~D., {Conroy}, C., {van Dokkum}, P.~G., \& {Byler}, N.
  2017, \apj, 837, 170, \dodoi{10.3847/1538-4357/aa5ffe}

\bibitem[{{Leja} {et~al.}(2020){Leja}, {Speagle}, {Johnson}, {Conroy}, {van
  Dokkum}, \& {Franx}}]{Leja2020}
{Leja}, J., {Speagle}, J.~S., {Johnson}, B.~D., {et~al.} 2020, \apj, 893, 111,
  \dodoi{10.3847/1538-4357/ab7e27}

\bibitem[{{Leja} {et~al.}(2022){Leja}, {Speagle}, {Ting}, {Johnson}, {Conroy},
  {Whitaker}, {Nelson}, {Dokkum}, \& {Franx}}]{Leja2021}
{Leja}, J., {Speagle}, J.~S., {Ting}, Y.-S., {et~al.} 2022, \apj, 936, 165,
  \dodoi{10.3847/1538-4357/ac887d}

\bibitem[{{Lilly} \& {Carollo}(2016)}]{Lilly2016}
{Lilly}, S.~J., \& {Carollo}, C.~M. 2016, \apj, 833, 1,
  \dodoi{10.3847/0004-637X/833/1/1}

\bibitem[{{Lintott} {et~al.}(2008){Lintott}, {Schawinski}, {Slosar}, {Land},
  {Bamford}, {Thomas}, {Raddick}, {Nichol}, {Szalay}, {Andreescu}, {Murray}, \&
  {Vandenberg}}]{Lintott2008}
{Lintott}, C.~J., {Schawinski}, K., {Slosar}, A., {et~al.} 2008, \mnras, 389,
  1179, \dodoi{10.1111/j.1365-2966.2008.13689.x}

\bibitem[{{Lower} {et~al.}(2020){Lower}, {Narayanan}, {Leja}, {Johnson},
  {Conroy}, \& {Dav{\'e}}}]{Lower2020}
{Lower}, S., {Narayanan}, D., {Leja}, J., {et~al.} 2020, \apj, 904, 33,
  \dodoi{10.3847/1538-4357/abbfa7}

\bibitem[{{Madau} \& {Dickinson}(2014)}]{Madau2014}
{Madau}, P., \& {Dickinson}, M. 2014, \araa, 52, 415,
  \dodoi{10.1146/annurev-astro-081811-125615}

\bibitem[{{Man} \& {Belli}(2018)}]{Man2018}
{Man}, A., \& {Belli}, S. 2018, Nature Astronomy, 2, 695,
  \dodoi{10.1038/s41550-018-0558-1}

\bibitem[{{Maraston} {et~al.}(2010){Maraston}, {Pforr}, {Renzini}, {Daddi},
  {Dickinson}, {Cimatti}, \& {Tonini}}]{Maraston2010}
{Maraston}, C., {Pforr}, J., {Renzini}, A., {et~al.} 2010, \mnras, 407, 830,
  \dodoi{10.1111/j.1365-2966.2010.16973.x}

\bibitem[{{Matharu} {et~al.}(2019){Matharu}, {Muzzin}, {Brammer}, {van der
  Burg}, {Auger}, {Hewett}, {van der Wel}, {van Dokkum}, {Balogh}, {Chan},
  {Demarco}, {Marchesini}, {Nelson}, {Noble}, {Wilson}, \& {Yee}}]{Matharu2019}
{Matharu}, J., {Muzzin}, A., {Brammer}, G.~B., {et~al.} 2019, \mnras, 484, 595,
  \dodoi{10.1093/mnras/sty3465}

\bibitem[{{McGrath} {et~al.}(2008){McGrath}, {Stockton}, {Canalizo}, {Iye}, \&
  {Maihara}}]{McGrath2008}
{McGrath}, E.~J., {Stockton}, A., {Canalizo}, G., {Iye}, M., \& {Maihara}, T.
  2008, \apj, 682, 303, \dodoi{10.1086/589631}

\bibitem[{{Merlin} {et~al.}(2021){Merlin}, {Castellano}, {Santini},
  {Cipolletta}, {Boutsia}, {Schreiber}, {Buitrago}, {Fontana}, {Elbaz},
  {Dunlop}, {Grazian}, {McLure}, {McLeod}, {Nonino}, {Milvang-Jensen},
  {Derriere}, {Hathi}, {Pentericci}, {Fortuni}, \& {Calabr{\`o}}}]{Merlin2021}
{Merlin}, E., {Castellano}, M., {Santini}, P., {et~al.} 2021, \aap, 649, A22,
  \dodoi{10.1051/0004-6361/202140310}

\bibitem[{{Mo} {et~al.}(2010){Mo}, {van den Bosch}, \& {White}}]{Mo2010}
{Mo}, H., {van den Bosch}, F.~C., \& {White}, S. 2010, {Galaxy Formation and
  Evolution}

\bibitem[{{Mowla} {et~al.}(2019){Mowla}, {van Dokkum}, {Brammer}, {Momcheva},
  {van der Wel}, {Whitaker}, {Nelson}, {Bezanson}, {Muzzin}, {Franx},
  {MacKenty}, {Leja}, {Kriek}, \& {Marchesini}}]{Mowla2019}
{Mowla}, L.~A., {van Dokkum}, P., {Brammer}, G.~B., {et~al.} 2019, \apj, 880,
  57, \dodoi{10.3847/1538-4357/ab290a}

\bibitem[{{Muzzin} {et~al.}(2013){Muzzin}, {Marchesini}, {Stefanon}, {Franx},
  {McCracken}, {Milvang-Jensen}, {Dunlop}, {Fynbo}, {Brammer}, {Labb{\'e}}, \&
  {van Dokkum}}]{Muzzin2013}
{Muzzin}, A., {Marchesini}, D., {Stefanon}, M., {et~al.} 2013, \apj, 777, 18,
  \dodoi{10.1088/0004-637X/777/1/18}

\bibitem[{{Nayyeri} {et~al.}(2017){Nayyeri}, {Hemmati}, {Mobasher}, {Ferguson},
  {Cooray}, {Barro}, {Faber}, {Dickinson}, {Koekemoer}, {Peth}, {Salvato},
  {Ashby}, {Darvish}, {Donley}, {Durbin}, {Finkelstein}, {Fontana}, {Grogin},
  {Gruetzbauch}, {Huang}, {Khostovan}, {Kocevski}, {Kodra}, {Lee}, {Newman},
  {Pacifici}, {Pforr}, {Stefanon}, {Wiklind}, {Willner}, {Wuyts}, {Castellano},
  {Conselice}, {Dolch}, {Dunlop}, {Galametz}, {Hathi}, {Lucas}, \&
  {Yan}}]{Nayyeri2017}
{Nayyeri}, H., {Hemmati}, S., {Mobasher}, B., {et~al.} 2017, \apjs, 228, 7,
  \dodoi{10.3847/1538-4365/228/1/7}

\bibitem[{{Nelson} {et~al.}(2012){Nelson}, {van Dokkum}, {Brammer},
  {F{\"o}rster Schreiber}, {Franx}, {Fumagalli}, {Patel}, {Rix}, {Skelton},
  {Bezanson}, {Da Cunha}, {Kriek}, {Labbe}, {Lundgren}, {Quadri}, \&
  {Schmidt}}]{Nelson2012}
{Nelson}, E.~J., {van Dokkum}, P.~G., {Brammer}, G., {et~al.} 2012, \apjl, 747,
  L28, \dodoi{10.1088/2041-8205/747/2/L28}

\bibitem[{{Nelson} {et~al.}(2016){Nelson}, {van Dokkum}, {F{\"o}rster
  Schreiber}, {Franx}, {Brammer}, {Momcheva}, {Wuyts}, {Whitaker}, {Skelton},
  {Fumagalli}, {Hayward}, {Kriek}, {Labb{\'e}}, {Leja}, {Rix}, {Tacconi}, {van
  der Wel}, {van den Bosch}, {Oesch}, {Dickey}, \& {Ulf Lange}}]{Nelson2016}
{Nelson}, E.~J., {van Dokkum}, P.~G., {F{\"o}rster Schreiber}, N.~M., {et~al.}
  2016, \apj, 828, 27, \dodoi{10.3847/0004-637X/828/1/27}

\bibitem[{{Newman} {et~al.}(2015){Newman}, {Belli}, \& {Ellis}}]{Newman2015}
{Newman}, A.~B., {Belli}, S., \& {Ellis}, R.~S. 2015, \apjl, 813, L7,
  \dodoi{10.1088/2041-8205/813/1/L7}

\bibitem[{{Newman} {et~al.}(2018){Newman}, {Belli}, {Ellis}, \&
  {Patel}}]{Newman2018}
{Newman}, A.~B., {Belli}, S., {Ellis}, R.~S., \& {Patel}, S.~G. 2018, \apj,
  862, 126, \dodoi{10.3847/1538-4357/aacd4f}

\bibitem[{{Newman} {et~al.}(2014){Newman}, {Ellis}, {Andreon}, {Treu},
  {Raichoor}, \& {Trinchieri}}]{Newman2014}
{Newman}, A.~B., {Ellis}, R.~S., {Andreon}, S., {et~al.} 2014, \apj, 788, 51,
  \dodoi{10.1088/0004-637X/788/1/51}

\bibitem[{{Newman} {et~al.}(2012){Newman}, {Ellis}, {Bundy}, \&
  {Treu}}]{Newman2012}
{Newman}, A.~B., {Ellis}, R.~S., {Bundy}, K., \& {Treu}, T. 2012, \apj, 746,
  162, \dodoi{10.1088/0004-637X/746/2/162}

\bibitem[{{Ocvirk} {et~al.}(2006){Ocvirk}, {Pichon}, {Lan{\c{c}}on}, \&
  {Thi{\'e}baut}}]{Ocvirk2006}
{Ocvirk}, P., {Pichon}, C., {Lan{\c{c}}on}, A., \& {Thi{\'e}baut}, E. 2006,
  \mnras, 365, 46, \dodoi{10.1111/j.1365-2966.2005.09182.x}

\bibitem[{{Oser} {et~al.}(2012){Oser}, {Naab}, {Ostriker}, \&
  {Johansson}}]{Oser2012}
{Oser}, L., {Naab}, T., {Ostriker}, J.~P., \& {Johansson}, P.~H. 2012, \apj,
  744, 63, \dodoi{10.1088/0004-637X/744/1/63}

\bibitem[{{Pacifici} {et~al.}(2012){Pacifici}, {Charlot}, {Blaizot}, \&
  {Brinchmann}}]{Pacifici2012}
{Pacifici}, C., {Charlot}, S., {Blaizot}, J., \& {Brinchmann}, J. 2012, \mnras,
  421, 2002, \dodoi{10.1111/j.1365-2966.2012.20431.x}

\bibitem[{{Papovich} {et~al.}(2001){Papovich}, {Dickinson}, \&
  {Ferguson}}]{Papovich2001}
{Papovich}, C., {Dickinson}, M., \& {Ferguson}, H.~C. 2001, \apj, 559, 620,
  \dodoi{10.1086/322412}

\bibitem[{{Papovich} {et~al.}(2011){Papovich}, {Finkelstein}, {Ferguson},
  {Lotz}, \& {Giavalisco}}]{Papovich2011}
{Papovich}, C., {Finkelstein}, S.~L., {Ferguson}, H.~C., {Lotz}, J.~M., \&
  {Giavalisco}, M. 2011, \mnras, 412, 1123,
  \dodoi{10.1111/j.1365-2966.2010.17965.x}

\bibitem[{{Papovich} {et~al.}(2012){Papovich}, {Bassett}, {Lotz}, {van der
  Wel}, {Tran}, {Finkelstein}, {Bell}, {Conselice}, {Dekel}, {Dunlop}, {Guo},
  {Faber}, {Farrah}, {Ferguson}, {Finkelstein}, {H{\"a}ussler}, {Kocevski},
  {Koekemoer}, {Koo}, {McGrath}, {McLure}, {McIntosh}, {Momcheva}, {Newman},
  {Rudnick}, {Weiner}, {Willmer}, \& {Wuyts}}]{Papovich2012}
{Papovich}, C., {Bassett}, R., {Lotz}, J.~M., {et~al.} 2012, \apj, 750, 93,
  \dodoi{10.1088/0004-637X/750/2/93}

\bibitem[{{Patel} {et~al.}(2013){Patel}, {van Dokkum}, {Franx}, {Quadri},
  {Muzzin}, {Marchesini}, {Williams}, {Holden}, \& {Stefanon}}]{Patel2013}
{Patel}, S.~G., {van Dokkum}, P.~G., {Franx}, M., {et~al.} 2013, \apj, 766, 15,
  \dodoi{10.1088/0004-637X/766/1/15}

\bibitem[{{Peng} {et~al.}(2002){Peng}, {Ho}, {Impey}, \& {Rix}}]{Peng2002}
{Peng}, C.~Y., {Ho}, L.~C., {Impey}, C.~D., \& {Rix}, H.-W. 2002, \aj, 124,
  266, \dodoi{10.1086/340952}

\bibitem[{{Peng} {et~al.}(2010{\natexlab{a}}){Peng}, {Ho}, {Impey}, \&
  {Rix}}]{galfit}
---. 2010{\natexlab{a}}, \aj, 139, 2097, \dodoi{10.1088/0004-6256/139/6/2097}

\bibitem[{{Peng} {et~al.}(2010{\natexlab{b}}){Peng}, {Lilly}, {Kova{\v{c}}},
  {Bolzonella}, {Pozzetti}, {Renzini}, {Zamorani}, {Ilbert}, {Knobel},
  {Iovino}, {Maier}, {Cucciati}, {Tasca}, {Carollo}, {Silverman}, {Kampczyk},
  {de Ravel}, {Sanders}, {Scoville}, {Contini}, {Mainieri}, {Scodeggio},
  {Kneib}, {Le F{\`e}vre}, {Bardelli}, {Bongiorno}, {Caputi}, {Coppa}, {de la
  Torre}, {Franzetti}, {Garilli}, {Lamareille}, {Le Borgne}, {Le Brun},
  {Mignoli}, {Perez Montero}, {Pello}, {Ricciardelli}, {Tanaka}, {Tresse},
  {Vergani}, {Welikala}, {Zucca}, {Oesch}, {Abbas}, {Barnes}, {Bordoloi},
  {Bottini}, {Cappi}, {Cassata}, {Cimatti}, {Fumana}, {Hasinger}, {Koekemoer},
  {Leauthaud}, {Maccagni}, {Marinoni}, {McCracken}, {Memeo}, {Meneux}, {Nair},
  {Porciani}, {Presotto}, \& {Scaramella}}]{Peng2010}
{Peng}, Y.-j., {Lilly}, S.~J., {Kova{\v{c}}}, K., {et~al.} 2010{\natexlab{b}},
  \apj, 721, 193, \dodoi{10.1088/0004-637X/721/1/193}

\bibitem[{{Perez-Gonzalez} {et~al.}(2013){Perez-Gonzalez}, {Cava}, {Barro},
  {Villar}, {Cardiel}, {Ferreras}, {Rodr{\'\i}guez-Espinosa}, {Alonso-Herrero},
  {Balcells}, {Cenarro}, {Cepa}, {Charlot}, {Cimatti}, {Conselice}, {Daddi},
  {Donley}, {Elbaz}, {Espino}, {Gallego}, {Gobat}, {Gonz{\'a}lez-Mart{\'\i}n},
  {Guzm{\'a}n}, {Hern{\'a}n-Caballero}, {Mu{\~n}oz-Tu{\~n}{\'o}n}, {Renzini},
  {Rodr{\'\i}guez-Zaur{\'\i}n}, {Tresse}, {Trujillo}, \&
  {Zamorano}}]{PerezGonzalez2013}
{Perez-Gonzalez}, P.~G., {Cava}, A., {Barro}, G., {et~al.} 2013, \apj, 762, 46,
  \dodoi{10.1088/0004-637X/762/1/46}

\bibitem[{{Salim} {et~al.}(2007){Salim}, {Rich}, {Charlot}, {Brinchmann},
  {Johnson}, {Schiminovich}, {Seibert}, {Mallery}, {Heckman}, {Forster},
  {Friedman}, {Martin}, {Morrissey}, {Neff}, {Small}, {Wyder}, {Bianchi},
  {Donas}, {Lee}, {Madore}, {Milliard}, {Szalay}, {Welsh}, \& {Yi}}]{Salim2007}
{Salim}, S., {Rich}, R.~M., {Charlot}, S., {et~al.} 2007, \apjs, 173, 267,
  \dodoi{10.1086/519218}

\bibitem[{{Sanders} {et~al.}(1988){Sanders}, {Soifer}, {Elias}, {Madore},
  {Matthews}, {Neugebauer}, \& {Scoville}}]{Sanders1988}
{Sanders}, D.~B., {Soifer}, B.~T., {Elias}, J.~H., {et~al.} 1988, \apj, 325,
  74, \dodoi{10.1086/165983}

\bibitem[{{Santini} {et~al.}(2015){Santini}, {Ferguson}, {Fontana}, {Mobasher},
  {Barro}, {Castellano}, {Finkelstein}, {Grazian}, {Hsu}, {Lee}, {Lee},
  {Pforr}, {Salvato}, {Wiklind}, {Wuyts}, {Almaini}, {Cooper}, {Galametz},
  {Weiner}, {Amorin}, {Boutsia}, {Conselice}, {Dahlen}, {Dickinson},
  {Giavalisco}, {Grogin}, {Guo}, {Hathi}, {Kocevski}, {Koekemoer},
  {Kurczynski}, {Merlin}, {Mortlock}, {Newman}, {Paris}, {Pentericci},
  {Simons}, \& {Willner}}]{Santini2015}
{Santini}, P., {Ferguson}, H.~C., {Fontana}, A., {et~al.} 2015, \apj, 801, 97,
  \dodoi{10.1088/0004-637X/801/2/97}

\bibitem[{{Schreiber} {et~al.}(2015){Schreiber}, {Pannella}, {Elbaz},
  {B{\'e}thermin}, {Inami}, {Dickinson}, {Magnelli}, {Wang}, {Aussel}, {Daddi},
  {Juneau}, {Shu}, {Sargent}, {Buat}, {Faber}, {Ferguson}, {Giavalisco},
  {Koekemoer}, {Magdis}, {Morrison}, {Papovich}, {Santini}, \&
  {Scott}}]{Schreiber2015}
{Schreiber}, C., {Pannella}, M., {Elbaz}, D., {et~al.} 2015, \aap, 575, A74,
  \dodoi{10.1051/0004-6361/201425017}

\bibitem[{{Scoville} {et~al.}(2017){Scoville}, {Lee}, {Vanden Bout},
  {Diaz-Santos}, {Sanders}, {Darvish}, {Bongiorno}, {Casey}, {Murchikova},
  {Koda}, {Capak}, {Vlahakis}, {Ilbert}, {Sheth}, {Morokuma-Matsui}, {Ivison},
  {Aussel}, {Laigle}, {McCracken}, {Armus}, {Pope}, {Toft}, \&
  {Masters}}]{Scoville2017}
{Scoville}, N., {Lee}, N., {Vanden Bout}, P., {et~al.} 2017, \apj, 837, 150,
  \dodoi{10.3847/1538-4357/aa61a0}

\bibitem[{{Shen} {et~al.}(2003){Shen}, {Mo}, {White}, {Blanton}, {Kauffmann},
  {Voges}, {Brinkmann}, \& {Csabai}}]{Shen2003}
{Shen}, S., {Mo}, H.~J., {White}, S. D.~M., {et~al.} 2003, \mnras, 343, 978,
  \dodoi{10.1046/j.1365-8711.2003.06740.x}

\bibitem[{{Shibuya} {et~al.}(2015){Shibuya}, {Ouchi}, \&
  {Harikane}}]{Shibuya2015}
{Shibuya}, T., {Ouchi}, M., \& {Harikane}, Y. 2015, \apjs, 219, 15,
  \dodoi{10.1088/0067-0049/219/2/15}

\bibitem[{{Speagle} {et~al.}(2014){Speagle}, {Steinhardt}, {Capak}, \&
  {Silverman}}]{Speagle2014}
{Speagle}, J.~S., {Steinhardt}, C.~L., {Capak}, P.~L., \& {Silverman}, J.~D.
  2014, \apjs, 214, 15, \dodoi{10.1088/0067-0049/214/2/15}

\bibitem[{{Spilker} {et~al.}(2016){Spilker}, {Bezanson}, {Marrone}, {Weiner},
  {Whitaker}, \& {Williams}}]{Spilker2016}
{Spilker}, J.~S., {Bezanson}, R., {Marrone}, D.~P., {et~al.} 2016, \apj, 832,
  19, \dodoi{10.3847/0004-637X/832/1/19}

\bibitem[{{Springel} {et~al.}(2005){Springel}, {Di Matteo}, \&
  {Hernquist}}]{Springel2005}
{Springel}, V., {Di Matteo}, T., \& {Hernquist}, L. 2005, \apjl, 620, L79,
  \dodoi{10.1086/428772}

\bibitem[{{Straatman} {et~al.}(2016){Straatman}, {Spitler}, {Quadri},
  {Labb{\'e}}, {Glazebrook}, {Persson}, {Papovich}, {Tran}, {Brammer},
  {Cowley}, {Tomczak}, {Nanayakkara}, {Alcorn}, {Allen}, {Broussard}, {van
  Dokkum}, {Forrest}, {van Houdt}, {Kacprzak}, {Kawinwanichakij}, {Kelson},
  {Lee}, {McCarthy}, {Mehrtens}, {Monson}, {Murphy}, {Rees}, {Tilvi}, \&
  {Whitaker}}]{Straatman2016}
{Straatman}, C. M.~S., {Spitler}, L.~R., {Quadri}, R.~F., {et~al.} 2016, \apj,
  830, 51, \dodoi{10.3847/0004-637X/830/1/51}

\bibitem[{{Strazzullo} {et~al.}(2013){Strazzullo}, {Gobat}, {Daddi}, {Onodera},
  {Carollo}, {Dickinson}, {Renzini}, {Arimoto}, {Cimatti}, {Finoguenov}, \&
  {Chary}}]{Strazzullo2013}
{Strazzullo}, V., {Gobat}, R., {Daddi}, E., {et~al.} 2013, \apj, 772, 118,
  \dodoi{10.1088/0004-637X/772/2/118}

\bibitem[{{Suess} {et~al.}(2022){Suess}, {Kriek}, {Bezanson}, {Greene},
  {Setton}, {Spilker}, {Feldmann}, {Goulding}, {Johnson}, {Leja}, {Narayanan},
  {Hall-Hooper}, {Hunt}, {Lower}, \& {Verrico}}]{Suess2022}
{Suess}, K.~A., {Kriek}, M., {Bezanson}, R., {et~al.} 2022, \apj, 926, 89,
  \dodoi{10.3847/1538-4357/ac404a}

\bibitem[{{Tacchella} {et~al.}(2016){Tacchella}, {Dekel}, {Carollo},
  {Ceverino}, {DeGraf}, {Lapiner}, {Mandelker}, \& {Primack}}]{Tacchella2016}
{Tacchella}, S., {Dekel}, A., {Carollo}, C.~M., {et~al.} 2016, \mnras, 458,
  242, \dodoi{10.1093/mnras/stw303}

\bibitem[{{Tacchella} {et~al.}(2022){Tacchella}, {Conroy}, {Faber}, {Johnson},
  {Leja}, {Barro}, {Cunningham}, {Deason}, {Guhathakurta}, {Guo}, {Hernquist},
  {Koo}, {McKinnon}, {Rockosi}, {Speagle}, {van Dokkum}, \&
  {Yesuf}}]{Tacchella2022}
{Tacchella}, S., {Conroy}, C., {Faber}, S.~M., {et~al.} 2022, \apj, 926, 134,
  \dodoi{10.3847/1538-4357/ac449b}

\bibitem[{{Tacconi} {et~al.}(2020){Tacconi}, {Genzel}, \&
  {Sternberg}}]{Tacconi2020}
{Tacconi}, L.~J., {Genzel}, R., \& {Sternberg}, A. 2020, \araa, 58, 157,
  \dodoi{10.1146/annurev-astro-082812-141034}

\bibitem[{{Tadaki} {et~al.}(2017){Tadaki}, {Genzel}, {Kodama}, {Wuyts},
  {Wisnioski}, {F{\"o}rster Schreiber}, {Burkert}, {Lang}, {Tacconi}, {Lutz},
  {Belli}, {Davies}, {Hatsukade}, {Hayashi}, {Herrera-Camus}, {Ikarashi},
  {Inoue}, {Kohno}, {Koyama}, {Mendel}, {Nakanishi}, {Shimakawa}, {Suzuki},
  {Tamura}, {Tanaka}, {{\"U}bler}, \& {Wilman}}]{Tadaki2017}
{Tadaki}, K.-i., {Genzel}, R., {Kodama}, T., {et~al.} 2017, \apj, 834, 135,
  \dodoi{10.3847/1538-4357/834/2/135}

\bibitem[{{Taniguchi} {et~al.}(2015){Taniguchi}, {Kajisawa}, {Kobayashi},
  {Shioya}, {Nagao}, {Capak}, {Aussel}, {Ichikawa}, {Murayama}, {Scoville},
  {Ilbert}, {Salvato}, {Sanders}, {Mobasher}, {Miyazaki}, {Komiyama}, {Le
  F{\`e}vre}, {Tasca}, {Lilly}, {Carollo}, {Renzini}, {Rich}, {Schinnerer},
  {Kaifu}, {Karoji}, {Arimoto}, {Okamura}, {Ohta}, {Shimasaku}, \&
  {Hayashino}}]{Taniguchi2015}
{Taniguchi}, Y., {Kajisawa}, M., {Kobayashi}, M. A.~R., {et~al.} 2015, \pasj,
  67, 104, \dodoi{10.1093/pasj/psv106}

\bibitem[{{Toft} {et~al.}(2017){Toft}, {Zabl}, {Richard}, {Gallazzi},
  {Zibetti}, {Prescott}, {Grillo}, {Man}, {Lee}, {G{\'o}mez-Guijarro},
  {Stockmann}, {Magdis}, \& {Steinhardt}}]{Toft2017}
{Toft}, S., {Zabl}, J., {Richard}, J., {et~al.} 2017, \nat, 546, 510,
  \dodoi{10.1038/nature22388}

\bibitem[{{Tojeiro} {et~al.}(2007){Tojeiro}, {Heavens}, {Jimenez}, \&
  {Panter}}]{Tojeiro2007}
{Tojeiro}, R., {Heavens}, A.~F., {Jimenez}, R., \& {Panter}, B. 2007, \mnras,
  381, 1252, \dodoi{10.1111/j.1365-2966.2007.12323.x}

\bibitem[{{Toomre}(1964)}]{Toomre1964}
{Toomre}, A. 1964, \apj, 139, 1217, \dodoi{10.1086/147861}

\bibitem[{{Valentinuzzi} {et~al.}(2010){Valentinuzzi}, {Fritz}, {Poggianti},
  {Cava}, {Bettoni}, {Fasano}, {D'Onofrio}, {Couch}, {Dressler}, {Moles},
  {Moretti}, {Omizzolo}, {Kj{\ae}rgaard}, {Vanzella}, \&
  {Varela}}]{Valentinuzzi2010}
{Valentinuzzi}, T., {Fritz}, J., {Poggianti}, B.~M., {et~al.} 2010, \apj, 712,
  226, \dodoi{10.1088/0004-637X/712/1/226}

\bibitem[{{van der Wel} {et~al.}(2011){van der Wel}, {Rix}, {Wuyts}, {McGrath},
  {Koekemoer}, {Bell}, {Holden}, {Robaina}, \& {McIntosh}}]{vanderWel2011}
{van der Wel}, A., {Rix}, H.-W., {Wuyts}, S., {et~al.} 2011, \apj, 730, 38,
  \dodoi{10.1088/0004-637X/730/1/38}

\bibitem[{{van der Wel} {et~al.}(2012){van der Wel}, {Bell}, {H{\"a}ussler},
  {McGrath}, {Chang}, {Guo}, {McIntosh}, {Rix}, {Barden}, {Cheung}, {Faber},
  {Ferguson}, {Galametz}, {Grogin}, {Hartley}, {Kartaltepe}, {Kocevski},
  {Koekemoer}, {Lotz}, {Mozena}, {Peth}, \& {Peng}}]{vanderWel2012}
{van der Wel}, A., {Bell}, E.~F., {H{\"a}ussler}, B., {et~al.} 2012, \apjs,
  203, 24, \dodoi{10.1088/0067-0049/203/2/24}

\bibitem[{{van der Wel} {et~al.}(2014){van der Wel}, {Franx}, {van Dokkum},
  {Skelton}, {Momcheva}, {Whitaker}, {Brammer}, {Bell}, {Rix}, {Wuyts},
  {Ferguson}, {Holden}, {Barro}, {Koekemoer}, {Chang}, {McGrath},
  {H{\"a}ussler}, {Dekel}, {Behroozi}, {Fumagalli}, {Leja}, {Lundgren},
  {Maseda}, {Nelson}, {Wake}, {Patel}, {Labb{\'e}}, {Faber}, {Grogin}, \&
  {Kocevski}}]{vanderWel2014}
{van der Wel}, A., {Franx}, M., {van Dokkum}, P.~G., {et~al.} 2014, \apj, 788,
  28, \dodoi{10.1088/0004-637X/788/1/28}

\bibitem[{{van Dokkum} {et~al.}(2010){van Dokkum}, {Whitaker}, {Brammer},
  {Franx}, {Kriek}, {Labb{\'e}}, {Marchesini}, {Quadri}, {Bezanson},
  {Illingworth}, {Muzzin}, {Rudnick}, {Tal}, \& {Wake}}]{vanDokkum2010}
{van Dokkum}, P.~G., {Whitaker}, K.~E., {Brammer}, G., {et~al.} 2010, \apj,
  709, 1018, \dodoi{10.1088/0004-637X/709/2/1018}

\bibitem[{{Wang} {et~al.}(2018){Wang}, {Elbaz}, {Daddi}, {Liu}, {Kodama},
  {Tanaka}, {Schreiber}, {Zanella}, {Valentino}, {Sargent}, {Kohno}, {Xiao},
  {Pannella}, {Ciesla}, {Gobat}, \& {Koyama}}]{Wang2018}
{Wang}, T., {Elbaz}, D., {Daddi}, E., {et~al.} 2018, \apjl, 867, L29,
  \dodoi{10.3847/2041-8213/aaeb2c}

\bibitem[{{Wellons} {et~al.}(2015){Wellons}, {Torrey}, {Ma}, {Rodriguez-Gomez},
  {Vogelsberger}, {Kriek}, {van Dokkum}, {Nelson}, {Genel}, {Pillepich},
  {Springel}, {Sijacki}, {Snyder}, {Nelson}, {Sales}, \&
  {Hernquist}}]{Wellons2015}
{Wellons}, S., {Torrey}, P., {Ma}, C.-P., {et~al.} 2015, \mnras, 449, 361,
  \dodoi{10.1093/mnras/stv303}

\bibitem[{{Whitaker} {et~al.}(2012){Whitaker}, {van Dokkum}, {Brammer}, \&
  {Franx}}]{Whitaker2012}
{Whitaker}, K.~E., {van Dokkum}, P.~G., {Brammer}, G., \& {Franx}, M. 2012,
  \apjl, 754, L29, \dodoi{10.1088/2041-8205/754/2/L29}

\bibitem[{{Whitaker} {et~al.}(2011){Whitaker}, {Labb{\'e}}, {van Dokkum},
  {Brammer}, {Kriek}, {Marchesini}, {Quadri}, {Franx}, {Muzzin}, {Williams},
  {Bezanson}, {Illingworth}, {Lee}, {Lundgren}, {Nelson}, {Rudnick}, {Tal}, \&
  {Wake}}]{Whitaker2011}
{Whitaker}, K.~E., {Labb{\'e}}, I., {van Dokkum}, P.~G., {et~al.} 2011, \apj,
  735, 86, \dodoi{10.1088/0004-637X/735/2/86}

\bibitem[{{Whitaker} {et~al.}(2017){Whitaker}, {Bezanson}, {van Dokkum},
  {Franx}, {van der Wel}, {Brammer}, {F{\"o}rster-Schreiber}, {Giavalisco},
  {Labb{\'e}}, {Momcheva}, {Nelson}, \& {Skelton}}]{Whitaker2017}
{Whitaker}, K.~E., {Bezanson}, R., {van Dokkum}, P.~G., {et~al.} 2017, \apj,
  838, 19, \dodoi{10.3847/1538-4357/aa6258}

\bibitem[{{Williams} {et~al.}(2009){Williams}, {Quadri}, {Franx}, {van Dokkum},
  \& {Labb{\'e}}}]{Williams2009}
{Williams}, R.~J., {Quadri}, R.~F., {Franx}, M., {van Dokkum}, P., \&
  {Labb{\'e}}, I. 2009, \apj, 691, 1879, \dodoi{10.1088/0004-637X/691/2/1879}

\bibitem[{{Williams} {et~al.}(2010){Williams}, {Quadri}, {Franx}, {van Dokkum},
  {Toft}, {Kriek}, \& {Labb{\'e}}}]{Williams2010}
{Williams}, R.~J., {Quadri}, R.~F., {Franx}, M., {et~al.} 2010, \apj, 713, 738,
  \dodoi{10.1088/0004-637X/713/2/738}

\bibitem[{{Wuyts} {et~al.}(2011){Wuyts}, {F{\"o}rster Schreiber}, {van der
  Wel}, {Magnelli}, {Guo}, {Genzel}, {Lutz}, {Aussel}, {Barro}, {Berta},
  {Cava}, {Graci{\'a}-Carpio}, {Hathi}, {Huang}, {Kocevski}, {Koekemoer},
  {Lee}, {Le Floc'h}, {McGrath}, {Nordon}, {Popesso}, {Pozzi}, {Riguccini},
  {Rodighiero}, {Saintonge}, \& {Tacconi}}]{Wuyts2011}
{Wuyts}, S., {F{\"o}rster Schreiber}, N.~M., {van der Wel}, A., {et~al.} 2011,
  \apj, 742, 96, \dodoi{10.1088/0004-637X/742/2/96}

\bibitem[{{Zavala} {et~al.}(2019){Zavala}, {Casey}, {Scoville}, {Champagne},
  {Chiang}, {Dannerbauer}, {Drew}, {Fu}, {Spilker}, {Spitler}, {Tran},
  {Treister}, \& {Toft}}]{Zavala2019}
{Zavala}, J.~A., {Casey}, C.~M., {Scoville}, N., {et~al.} 2019, \apj, 887, 183,
  \dodoi{10.3847/1538-4357/ab5302}

\bibitem[{{Zolotov} {et~al.}(2015){Zolotov}, {Dekel}, {Mandelker}, {Tweed},
  {Inoue}, {DeGraf}, {Ceverino}, {Primack}, {Barro}, \& {Faber}}]{Zolotov2015}
{Zolotov}, A., {Dekel}, A., {Mandelker}, N., {et~al.} 2015, \mnras, 450, 2327,
  \dodoi{10.1093/mnras/stv740}

\end{thebibliography}
\bibliographystyle{aasjournal}

\end{document}